\documentclass[12pt]{article}

\pdfoutput=1 

\usepackage{graphicx}
\usepackage{epstopdf}
\usepackage[centertags]{amsmath}
\usepackage{amssymb}
\usepackage{amsthm}
\usepackage{amsfonts}
\usepackage{ccaption}
\usepackage[usenames]{color}
\usepackage{mathrsfs}
\usepackage[textsize=tiny, textwidth=2cm,color=red,backgroundcolor=pink]{todonotes}
\usepackage[
      colorlinks=true,
      linkcolor=blue,  
      urlcolor=blue,    
      filecolor=black,     
      citecolor=red,
      pdfstartview=FitV,
      pdftitle={},
        pdfauthor={Mukund Rangamani, Veronika Hubeny},
        pdfsubject={},
        pdfkeywords={},
        pdfpagemode=None,
        bookmarksopen=true    
      ]{hyperref}
       
\usepackage{rotating}

\makeatletter
\@addtoreset{equation}{section}

\makeatletter
\renewcommand\section{\@startsection {section}{1}{\z@}%
                                   {-3.5ex \@plus -1ex \@minus -.2ex}
                                   {2.3ex \@plus.2ex}%
                                   {\normalfont\large\bfseries}}
\renewcommand\subsection{\@startsection{subsection}{2}{\z@}%
                                     {-3.25ex\@plus -1ex \@minus -.2ex}%
                                     {1.5ex \@plus .2ex}%
                                     {\normalfont\bfseries}}

  \captionnamefont{\bfseries}
  \captiontitlefont{\small\sffamily}
  \captiondelim{: }
  \hangcaption

\def\sec#1{\S\ref{#1}}
\def\fig#1{Fig.\,\ref{#1}}
\def\req#1{(\ref{#1})}

\def\ie{{\it i.e.}}

\def\etc{{\it etc.}}

\def\thus{\Longrightarrow}

\def\CA{{\cal A}}
\def\CB{{\cal B}}
\def\CC{{ \cal C }}
\def\CD{{ \cal D }}

\def\CF{{\cal F}}
\def\CG{{\cal G}}
\def\CH{{\cal H}}

\def\CJ{{\cal J}}

\def\CL{{\cal L}}
\def\CM{{\cal M}}
\def\CN{{\cal N}}
\def\CO{{\cal O}}

\def\CR{{\cal R}}
\def\CS{{\cal S}}

\def\CX{{\cal X}}

\def\CZ{{\cal Z}}
\def\ZZ{\mathbb{Z}}

\def\R{{\bf R}}
\def\Sp{{\bf S}}

\def\vev#1{\langle\, #1 \, \rangle}
\def\ket#1{\mid  \! #1\rangle}

\def\expval#1{{\langle \, #1  \, \rangle}}

\def\Tr#1{{\rm Tr}\left(#1\right)}

\def\AdS#1{AdS$_{#1}$}
\def\SAdS#1{Schwarzschild-AdS$_{#1}$}
\def\RAdS{R_{\rm AdS}}

\title{{\bf \Large A holographic view on physics out of equilibrium}}

\author{\normalsize 
Veronika E. Hubeny\footnote{veronika.hubeny@durham.ac.uk},   \ Mukund Rangamani\footnote{mukund.rangamani@durham.ac.uk} \\
\small \sl   Centre for Particle Theory \& Department of
Mathematical Sciences,
\\[-1.5mm]
\small \sl Science Laboratories, South Road, Durham DH1 3LE, United Kingdom. \\
}

\begin{document}

\setlength{\baselineskip}{16pt}
\begin{titlepage}
\maketitle
\begin{picture}(0,0)(0,0)
\put(360, 240){DCPT-10/21} 
\end{picture}
\vspace{-36pt}

\begin{abstract}
We review the recent developments in applying holographic methods to understand non-equilibrium physics in strongly coupled field theories. The emphasis will be on elucidating the relation between evolution of quantum field theories perturbed away from equilibrium and the dual picture of dynamics of classical fields in black hole backgrounds.  In particular, we discuss the linear response regime, the hydrodynamic regime and finally the non-linear regime of interacting quantum systems. We also describe how the duality might be used to learn some salient aspects of black hole physics in terms of field theory observables.
 \end{abstract}
\thispagestyle{empty}
\setcounter{page}{0}
\end{titlepage}

\renewcommand{\thefootnote}{\arabic{footnote}}


\tableofcontents

\section{Introduction}
\label{s:intro}

Astounding amount of progress and understanding in physics has been achieved by studying special systems in equilibrium, which are by definition `non-dynamical', i.e.\ independent of time.  One important reason for the prevalent focus on time-independent systems is that they are of course far simpler to study than dynamical ones.  Indeed, vast majority of exactly tractable systems are of this kind.  Known exact solutions are seldom fully generic and typically admit a large degree of symmetry, usually including time-translation invariance.  Nevertheless, while the study of such non-dynamical, stationary situations might seem rather looking-under-the-lamppost type strategy, it has proved to be a very successful one.  Although due to an incredible complexity of nature around us such exactly-tractable systems are at best only approximations to the real world, they are often remarkably relevant and useful.  Dynamically-evolving systems tend to equilibrate, and in absence of external forcing they typically settle down to stationary configurations.\footnote{
Implicit in this statement is an assumption  that  the systems under consideration are sufficiently ergodic.
} Thus stable stationary solutions can reveal the late-time physics of a {\it generic} system.  Moreover, many interesting physical processes such as phase transitions, which we observe occurring dynamically, can be well-studied without invoking any explicit time-dependence at all.

Nevertheless, not many would dispute that most interesting physical phenomena {\it do} involve non-trivial temporal dynamics.  Not only are dynamically evolving systems ubiquitous in nature, but they are the {\it raison d'\^ etre} for everything we see around us.  Hence the need to understand dynamics scarcely needs motivation.  However, progress so far has unfortunately been hindered by lack of adequate techniques.  Typically, one resorts to perturbative methods, though the regime of their validity places severe limitations on their applicability.  Alternately, one might try to ``put the system on a computer" and evolve numerically, but the computational cost involved is usually too astronomical to allow for convenient extraction of the physics.  In certain cases one may circumvent both of these limitations by mapping the system into a more tractable one.  We will see that all these ingredients come into play in the present Review.

So far our somewhat self-evident remarks have been rather abstract and general.  We will now specify the particular context we wish to address.  We will concentrate on exploring the dynamics of a certain class of quantum field theories, focusing especially on the strong coupling regime.   Although we have best understanding of field theories at weak coupling, strongly-coupled systems are ubiquitous in nature, ranging from typical condensed matter systems, to quark-gluon plasmas created in high-energy experiments, and in fact play a role in most areas of physics.  Apart from the evident applicability, there is also the rewarding aspect of serendipity related to the present cutting-edge experiments involving these systems.   

Which properties of strongly coupled field theories do we wish to understand, and which ones can we hope to understand?  Fortuitously there is a substantial overlap between the answers to both questions.  In particular, it is both interesting and tractable to extract certain universal features, as we will discuss below.  Conversely, we seldom can, nor want, to calculate the detailed microscopic dynamics, due to the sheer level of complexity; it is generally much more instructive to take a coarse-grained view of the system. As a result we will often focus on obtaining a low energy effective description for such strongly interacting systems. From a conventional renormalization group (RG) picture, it is then clear that one expects similar low energy physics for systems within the same universality class. One such low energy theory which we will discuss in some detail is hydrodynamics, or more generally fluid dynamics, which is expected to be a good description as long as the local fluid variables vary slowly compared to the microscopic scale, i.e.\ at long wavelengths and small frequencies.  

Hence in studying dynamics of strongly coupled field theories, we are simultaneously exploring the dynamics of fluids.  Our study is then bolstered by the insights which we have already acquired from hydrodynamics.  On the other hand, despite decades of theoretical as well as numerical, observational, and experimental scrutiny which fluid dynamics has received, there are still many deep questions which remain to be answered, especially involving dynamical evolution.    For example, one of the famous Clay Millennium Prize Problems concerns the global regularity (existence and smoothness) of the Navier-Stokes equations \cite{Fefferman:2000fk}. Intriguingly, the solutions often include turbulence, which, in spite of its practical importance in science and engineering, still remains one of the great unsolved problems in physics.

As already mentioned above, understanding dynamics in strongly coupled field theories, or their effective description in terms of fluids, is an exceedingly hard problem.  
One of the key strategies has been to focus on field theories which admit a holographic dual, and use this dual description to extract the physical properties of the field theoretic system under consideration.  The prototypical case is the AdS/CFT correspondence \cite{Maldacena:1997re, Gubser:1998bc, Witten:1998qj}
which relates the four-dimensional $\CN=4$ Super Yang-Mills  (SYM) gauge theory to a IIB string theory (or supergravity) on asymptotically AdS$_5 \times \Sp^5$ spacetime.\footnote{
The AdS/CFT correspondence is  comprehensively reviewed in the classic reviews \cite{Aharony:1999ti,DHoker:2002aw}.  See also \cite{McGreevy:2009xe} for a nice review with emphasis on `applied holography'.
} As is well known, this is a strong/weak coupling duality; the strongly-coupled field theory can be accessed via the semi-classical gravitational dual, which has obvious computational, as well as conceptual, advantages.  If we wish to know how a given strongly-coupled system behaves, we need not devise methods to calculate this in the field theory at all -- a rather daunting, if not impossible, task -- we only need to translate the system into the dual language and calculate the classical evolution using Einstein's equations.

While the $\CN =4$ SYM is fundamentally distinct from the `real-world' systems which we would ultimately like to understand, it can serve as a useful toy model to motivate and study new classes of strongly coupled phenomena. This philosophy has led to the correspondence being applied to QCD via the so called AdS/QCD approach (for reviews see \cite{Peeters:2007ab,Mateos:2007ay,Gubser:2007zz,Gubser:2009md}), and more recently to condensed matter systems, often dubbed AdS/CMT, where a variety of physical effects raging from superfluid transitions to non-Fermi liquid behaviour are being actively investigated. An excellent account of these efforts can be found in the reviews \cite{Hartnoll:2009sz,Herzog:2009xv,McGreevy:2009xe,Horowitz:2010gk}. Of course, what makes this enterprise fascinating is the degree to which the computations in the AdS/CFT framework agree with experimentally measured quantities in the real world. One such quantity which received much attention is the shear viscosity of the quark-gluon plasma; based holographic computations, \cite{Kovtun:2004de} suggest that the dimensionless ratio of viscosity to entropy density ($\eta/s$) has a universal lower bound, $\eta/s \ge 1/4\pi$, which subsequently has been subject to intense scrutiny; for the current  status in string theory see \cite{Buchel:2008vz,Sinha:2009ev}.

Perhaps the most intriguing -- and promising -- aspect of the gauge/gravity correspondence is highlighted by its holographic nature: the theories which are dual to each other are naturally formulated in different number of dimensions.\footnote{
The holographic principle was proposed much before the advent of the AdS/CFT correspondence and was motivated in part by the peculiar non-extensive nature of black hole entropy in \cite{tHooft:1993gx,Susskind:1994vu}.
} This automatically implies that the effective degrees of freedom on the two sides of the correspondence are related in a highly non-local (and hitherto rather mysterious) fashion.  This of course makes the task of extracting physics of one theory from its dual formulation even more formidable than merely doing calculations in the dual; indeed, much of the AdS/CFT-related efforts of the last decade have concentrated on elucidating the dictionary between the two sides.  Yet at the same time this very feature, which might initially appear as an unwanted complication, lies at the heart of the enormous potential the duality holds for solving the system.  The effects we seek to unravel are often very complex, emergent phenomena which cannot (in practice, or even in principle) be described in terms of the fundamental degrees of freedom.  In other words, they require a different description.  This by itself does not of course guarantee that the holographic description is the desired one, but the fact that out of such formidably complicated repackaging of the information a new simple and elegant picture arises, is at least promising.  It may well transpire that the complex emergent phenomena we seek to understand, at least those which are in some sense fundamental, appear simple when re-written in the dual language. 

One classic indication that the above hope is more that just a wishful thinking may be found in the manner in which the radial direction of the bulk emerges from the boundary field theory, sometimes referred to as the scale/radius, or UV/IR, duality \cite{Susskind:1998dq}.  From everyday experience, we are well-aware that physics likes to organize itself by energy (or length) scales; processes occurring at widely-separated scales do not interact very much.  In the holographic dual, this hierarchy of scales, mathematically expressed in terms of Wilsonian RG, is neatly packaged  in terms of an emergent radial direction of the bulk geometry (this idea is at the heart of the holographic renormalization group developed initially in \cite{deBoer:1999xf}).  But rather than just serving as a mnemonic for the energy scale, this dimension takes on a life of its own: it mixes with the other directions in a fully covariant fashion! Taking the bulk perspective, the field theory's decoupling of scales is simply a manifestation of bulk locality.  So this simple and natural feature of the bulk description has far-reaching implications for the boundary description.

Phrased more prosaically, it is not just to be hoped-for, but rather it is guaranteed, that certain complex phenomena must have a very simple and natural explanation in the dual picture.  After all, the two dual theories are just different descriptions of the same physical system: there is no absolute notion of which side is `more fundamental' than the other.  Whichever side we use as a starting point, there are certain fundamental properties or principles  underlying the theory that we understand in simple terms; yet these will be mapped into the dual framework, wherein they will appear as highly non-trivial.  If we were to come upon them from the other side -- which due to their inherent importance is perhaps not entirely unlikely -- it would appear rather magical that in the dual language they suddenly become simple. Of course, this is not to say that all complex phenomena must admit a simple description in some reformulation, but it is not so outrageous to expect that the fundamentally important ones do.

The preceding philosophical interlude was meant to motivate the use of gauge/gravity duality to understand certain field theories at strong coupling in terms of their gravitational dual, beyond the mere fact that we can usually calculate in classical gravity more easily than in the strongly-coupled field theory.  But the gauge/gravity correspondence likewise has profound implications when applied in the other direction.  The field theory, albeit strongly coupled, provides a definition of quantum gravity with asymptotically AdS boundary conditions.  Hence even in extreme regimes where classical gravity breaks down, and  where we don't yet have the tools to understand the physics within string theory directly, the field theory remains well-defined.
This means that once we achieve sufficient understanding of the dictionary between the two sides, we can elucidate the long-standing quantum gravitational puzzles by re-casting them into the field theoretic language.  

Let us briefly indicate some of the gravitational questions we ultimately hope to answer, since these have posed the underlying motivation in many of the works mentioned in this Review.  A central question of quantum gravity concerns the fundamental nature of spacetime. We have come to realize that spacetime is an emergent concept, but exactly how it emerges remains a mystery.
Happily, certain aspects of this emergence can be conveniently explored using the AdS/CFT framework, including ones pertaining to the present theme of out-of-equilibrium dynamics.

To set the stage, let us first recall several well-understood classical highlights of the correspondence.
According to the AdS/CFT dictionary, different asymptotically AdS spacetimes manifest themselves by different states in the boundary field theory.  For example, the vacuum state in the CFT corresponds to the pure AdS spacetime.  Metric perturbations which maintain the  AdS asymtopia are related to the stress-energy-momentum tensor expectation value in the CFT.   More importantly, putting a black hole in the bulk has the effect of heating up the boundary theory.  Specifically, a large\footnote{
AdS is a space of constant negative curvature, which introduces a length scale, called the AdS scale $\RAdS$, corresponding to the radius of curvature.  The black hole size is then measured in terms of this AdS scale; large black holes have horizon radius $r_+  > \RAdS$.
} Schwarzschild-AdS black hole corresponds to (approximately) thermal state in the gauge theory.  This can be easily conceptualized as the late-time configuration a generic state evolves to: in the bulk, the combined effect of gravity and negative curvature tends to make a generic large-energy configuration collapse and form a black hole which then quickly settles down to the Schwarzschild-AdS geometry.  On the other hand, in the field theory, a generic large-energy excitation will thermalize. At the level of this coarse entry in the dictionary we see that heating the field theory corresponds to black hole formation in the dual gravity, and the subsequent thermalization corresponds to the black hole settling down to a stationary state.

While appealingly simple, this level of understanding is far too coarse to allow us to extract the more interesting aspects.  We need to probe the AdS/CFT dictionary further to uncover what happens in regions where the classical description of the black hole breaks down, such as near the curvature singularity, or in the more general dynamical situations.  Ultimately, we would like to answer such questions as:  Which CFT configurations admit a dual description in terms of classical spacetime?  What types of spacetime singularities are physically allowed?  How are the disallowed singularities resolved?  How is spacetime causal structure encoded in the dual field theory?

Emergence of time is, if anything, even more mysterious than the emergence of space.  Not only do we have more satisfactory toy models of the latter than of the former, but conceptually the problem of time is one of the deepest problems in quantum gravity.  The `time' (conjugate to the Hamiltonian) which quantum mechanics uses for evolution is ingrained in the fundamental formulation of the system; it is there from the start rather than being emergent.  We can take this evolution parameter to be simply the time of the non-dynamical (and non-emergent) background on which the CFT lives.
What is then the relation between this CFT time and the notion of time in the dual bulk spacetime?  For bulk spacetimes which are globally static we tend to associate the two; but already for general dynamical spacetimes there is no natural identification even classically, since there is no uniquely-specified foliation of the bulk.  Even if we can construct geometrically defined spacelike slices through the bulk (such as volume-maximizing ones), there is a-priori no reason that the bulk events localized on a given slice should be dual to boundary events localized at the corresponding boundary time; to the contrary, we have many indications that the correspondence is much more temporally non-local.
Nevertheless, we still expect that time-dependence on the boundary will be manifested by time-dependence in the bulk, and vice-versa.  Indeed, focusing on certain characteristic features of a given time-dependence can enable us to elucidate the gauge/gravity dictionary further. Thus, our overarching motif of considering dynamical systems in strongly coupled field theories will naturally translate to studying bulk spacetime dynamics. 

We now return to our initial comments regarding time-dependence being difficult to handle.  The reader might observe that these general comments applied equally to the bulk side of the correspondence as well as to the boundary side, and might therefore wonder what have we gained by translating one hard problem into another hard problem.  
The purpose of this Review is to indicate what in fact we {\it have} gained by using a holographic dual, and to outline some of the methods that have been used to obtain further understanding of dynamical systems.   We will structure our presentation of the various approaches according to the severity of time-dependence they can handle, i.e.\ how strongly out-of-equilibrium evolutions they apply to.  

A useful starting point is to consider a well-understood global equilibrium situation, and try to understand the response of the system under small deviations from this equilibrium.  If the amplitude of such deviations is suitably small everywhere, the system may be studied using a {\it linear response theory}, which will be the focus of \sec{s:linresp} and \sec{s:lrads}. We will briefly review the basic concepts in linear response theory in \sec{s:linresp} focussing on two main aspects: the use of retarded correlation function in equilibrium to extract dynamics in the presence of small fluctuations, and  the behaviour of fast probes in an equilibrated medium as modeled by Langevin dynamics. In \sec{s:lrads} we will see how this physics can be mapped into the gravitational arena. The linear response regime is in fact simplest to understand from the AdS/CFT correspondence, for the computation of correlation functions is the best understood part of the AdS/CFT dictionary. We will also describe the behaviour of probes and their stochastic dynamics by drawing connection with semi-classical dynamics in black hole backgrounds. 

While surprisingly powerful, linear response theory requires small deviations from global equilibrium, leaving more general dynamics inaccessible.  To go beyond linear response, we will first focus on long-wavelength IR physics, where the coarse-grained description of an interacting field theory is provided  by fluid dynamics.   Fluid dynamics can of course describe fluids well out of global equilibrium, as long as the fluid variables, such as local temperature, fluid velocity, etc.,  make sense.  In particular, the spatial and temporal variations of these variables must occur on much longer scales than the microscopic ones, but the amplitudes need not be small.  Borrowing notation from kinetic theory, let us denote this microscopic scale $\ell_{\rm mfp}$, and the typical scale on which the fluid variables vary $L$.
Then the regime in which the fluid description is meaningful, or equivalently the regime where the system attains {\it local} thermal equilibrium everywhere, is given by the condition $L \gg 
\ell_{\rm mfp}$; we will refer to this as the `long-wavelength' regime.  How does the fact that we have a fluid description of a given configuration help?  As we will see in \sec{s:flugra}, it conveniently constitutes a large truncation of the relevant degrees of freedom describing the system.  Using the recently-formulated {\it fluid/gravity correspondence} \cite{Bhattacharyya:2008jc}, there is a one-to-one mapping between any such solution to fluid dynamics and a bulk solution to Einstein's equations describing a large  non-uniform and dynamically evolving black hole.  Note that material covered in sections \sec{s:corrads}-\sec{s:bhquasi} and \sec{s:flugra} has previously been reviewed in \cite{Son:2007vk} and \cite{Rangamani:2009xk,Hubeny:2010fk}, respectively. 

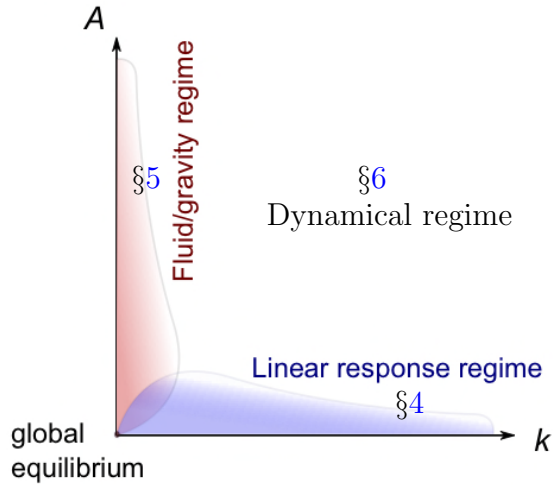
\begin{figure}
\begin{center}
\includegraphics[width=3in]{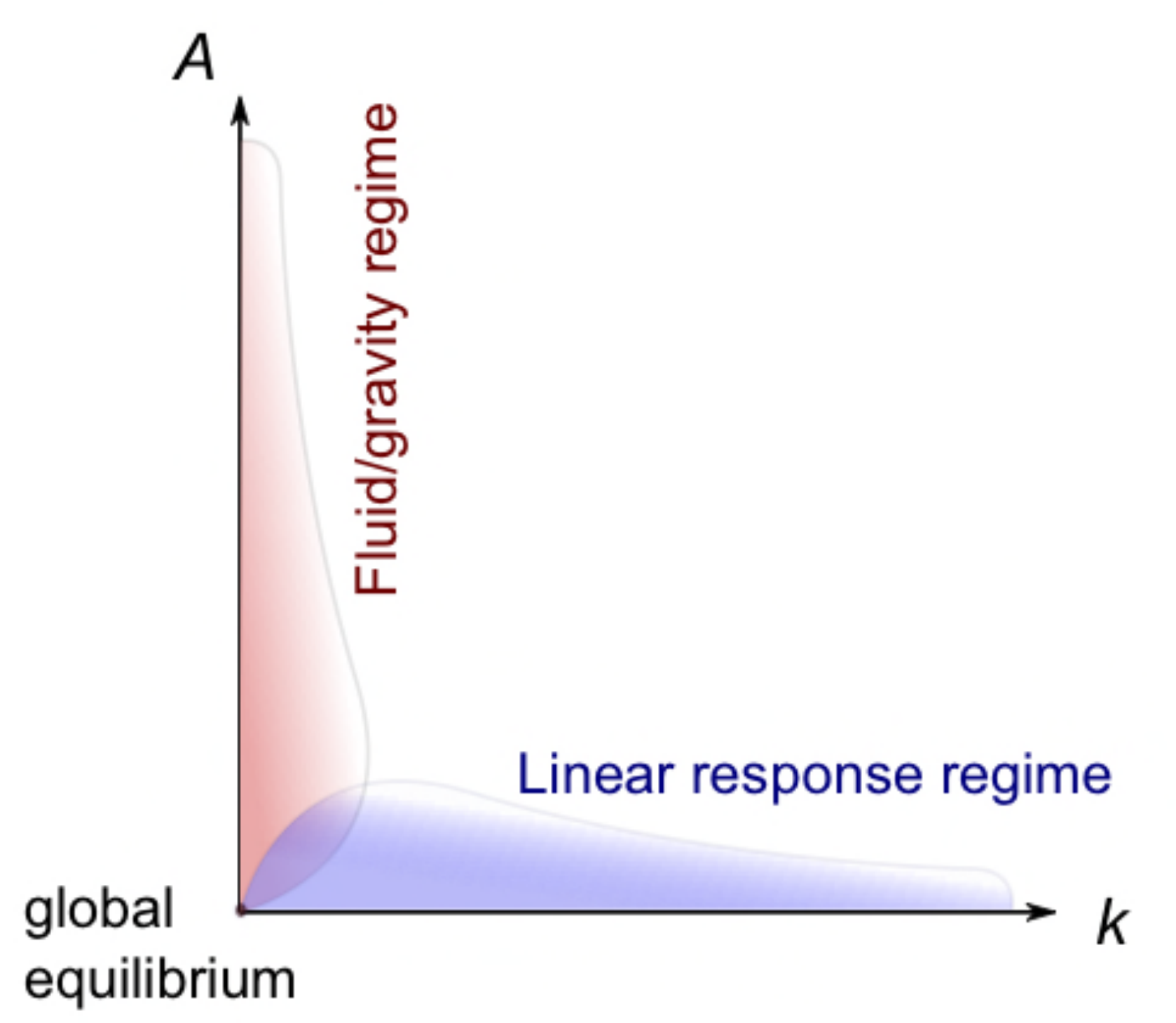}
\begin{picture}(0,0)
\setlength{\unitlength}{1cm}
\put(-2.5,1){\sec{s:lrads}}
\put(-6,4){\sec{s:flugra}}
\put(-3,4){\sec{s:dynam}}
\put(-4.2,3.5){Dynamical regime}
\end{picture}
\caption{Rough indication of the regimes of validity of the linear response theory and the fluid/gravity correspondence, in the space of perturbations from global thermodynamic equilibrium, labeled by the amplitude of perturbations $A$ and the wave number (or frequency) $k = \ell_{\rm mfp}/L$, relative to the microscopic scale. We have indicated the relevant sections of the paper where the different regimes are discussed from the holographic perspective.}
\label{fig:regimes}
\end{center}
\end{figure}

\fig{fig:regimes} illustrates the two regimes of validity discussed so far.  Linear response theory is valid for small deformations from global equilibrium, while fluid/gravity correspondence can be used in the long-wavelength regime.   The main points to note are that these two regimes are a-priori distinct, and that they still leave the more interesting region of large deviations from global equilibrium out of reach. 

The general story of understanding physics out of equilibrium then involves consideration of deformations that take us to the non-linear regime in amplitude and frequency. This translates to the full-blown study of gravitational dynamics, including the fascinating physics of black hole formation. In \sec{s:dynam} we will review recent progress in understanding various aspects of such analyses. Finally in \sec{s:diagnostics} we will take the opportunity to address interesting questions about gravity: what does our understanding of physics of strongly coupled field theories teach us about gravitational dynamics? Our focus in this brief section will be to describe various attempts in the past decade to extract features of the bulk geometry in terms of field theory observables. We conclude with a discussion in \sec{s:concl}.

\section{The AdS/CFT dictionary}
\label{s:AdSCFT}

Before we delve into the details of physics out of equilibrium, we pause to recall some salient facts about the AdS/CFT correspondence.  As already mentioned, this is well explained in the classic reviews \cite{Aharony:1999ti,DHoker:2002aw}, so accordingly we will be brief and simply collect facts that are necessary for the current discussion.

The essence of the AdS/CFT correspondence is that strongly coupled field theory dynamics is recorded in terms of string theory (or classical gravity if the field theory admits an appropriate planar limit) with appropriate asymptotically Anti-de Sitter boundary conditions.  Spacetimes which are asymptotically AdS can be
thought of as deformations of pure AdS by normalizable modes in the supergravity description. The framework is also sufficiently rich to allow for deformations of the field theory, for instance by turning on relevant operators.\footnote{
This type of deformation can be used to study field theories on curved backgrounds as described in detail in \cite{deHaro:2000xn} (see also \cite{Hubeny:2009ru,Hubeny:2009kz,Hubeny:2009rc} for recent investigations). However, from the bulk perspective, these are large deformations, involving asymptotically ``locally AdS'' geometries.
} The basic dictionary relates the field theory observables, which are gauge invariant operators (local or non-local), to their counter-parts in the string (or gravitational) description. For instance, local gauge invariant single-trace operators formed out of the fundamental fields of the gauge theory such as $\Tr{F^{\mu\nu} \, F_{\mu\nu}}$ map to single-particle states in the bulk spacetime, while non-local operators such as Wilson loops map to string or D-brane world-sheets. 

In this language, pure AdS spacetime characterizes the UV fixed point of a quantum field theory. The fixed points of interest are field theories with at least $\CN =1$ superconformal symmetry.\footnote{
While non-supersymmetric AdS compactifications also arise in string theory, they are seldom stable. Some notable exceptions which satisfy perturbative stability conditions, i.e.\ spectrum statisfying the Breitenlohner-Freedman bound \cite{Breitenlohner:1982jf}, were described in \cite{DeWolfe:2001nz}; however to the best of our knowledge non-perturbative stability of such vacua have not been established (for instance note that non-supersymmetric quotients of \AdS{5}$\times \Sp^5/\ZZ_k$ do suffer from non-perturbative instabilities \cite{Horowitz:2007pr}).
} The central charge of this CFT is given in terms of the geometry of the AdS spacetime. In the familiar examples arising from string theory one typically encounters spacetimes of the form \AdS{d+1}$\times \CX$  supported by various fluxes, where $\CX$ is generically a compact manifold (which is $(9-d)$-dimensional in string theory or analogously $(10-d)$-dimensional for M-theory solutions). 

The field theories are characterized by a dimensionless coupling constant(s) and the gauge group. We will use the 't Hooft coupling parameter $\lambda$, while the information regarding the rank of the gauge group is given by the central charge $c$ of the CFT. The latter is in turn  given in terms of the volume of $\CX$. Schematically, 
\begin{equation}
\frac{1}{16\pi\, G_N^{(10)}} \, \int d^{10} x\, \sqrt{-G} \, \left( R^{(G)} + \cdots \right) = 
\frac{\rm{Vol}(\CX)}{16\pi\, G_N^{(10)}}\, \int d^{d+1} x\, \sqrt{-g} \,  \left( R + \cdots \right) \ .
\label{}
\end{equation}	
Denoting the \AdS{d+1} curvature scale by $\RAdS$ and taking into account the basic relation between the 10-dimensional Planck and string length scales \cite{Polchinski:1998wt} 
\begin{equation}
16\pi\, G_N^{(10)} = (2\pi)^7 \, g_s^2 \, \ell_s^8 =  (2\pi)^7 \, \ell_p^8 \ ,
\label{g10def}
\end{equation}	
we obtain
\begin{equation}
\frac{\RAdS^{d-1}}{16\pi\, G_N^{(d+1)}} = \frac{{\rm Vol}(\CX) \, \RAdS^{d-1}}{ (2\pi)^7 \, g_s^2 \, \ell_s^8}\propto c \ .
\label{g5Nrel}
\end{equation}	
For instance, in the celebrated duality between $SU(N)$ $\CN =4$ SYM  and Type IIB string theory on \AdS{5} $\times \Sp^5$ one finds
\begin{equation}
c_{\CN =4} \propto \frac{\RAdS^3}{16\pi\, G_N^{(5)}} = \frac{\pi^3 \, \RAdS^8}{ (2\pi)^7 \, g_s^2 \, \ell_s^8} = \frac{N^2}{8\, \pi^2} \ .
\label{centn4}
\end{equation}	

As described above, deformations away from the UV fixed point correspond to deformations from pure AdS spacetime. Normalizable modes in the bulk spacetime would be related to deformations that are engineered by giving vacuum expectation values to the dual field theory operator. Non-normalizable modes in the bulk are non-fluctuating sources that can be used to deform  the field theory Lagrangian.\footnote{
In certain situations  one encounters a choice of boundary conditions to impose on bulk fields. A case in point is that of massive scalar fields with mass lying close to the Breitenlohner-Freedman bound \cite{Breitenlohner:1982jf,Klebanov:1999tb}. We assume that a particular choice has been made among the various possibilities and refer to normalizable and non-normalizable with respect to this choice of boundary conditions.
} Finally, in our considerations we will not allow irrelevant deformations of the CFT, as these would correspond to destroying the AdS asymptotics.

\subsection{Regimes of interest}
\label{s:regimes}

The AdS/CFT correspondence is a profound correspondence between two quantum theories; but for general values of the parameters these theories are complicated and beyond computational control.  The set of limits we focus on to gain control and insight is the following.  
\begin{itemize}
\item $c \to \infty$ or equivalently $N \to \infty$:  quantum corrections are suppressed and the gravitational theory becomes classical. This also has the added merit of suppressing string interactions since $g_s \sim 1/c$ in the planar limit.
\item $\lambda \to \infty$: stringy ($\alpha'$) corrections are suppressed, so the bulk theory is simply Einstein gravity interacting with other fields, while the boundary theory describes dynamics of local single trace operators.  At this level, the field theory dynamics still looks complicated and highly nonlocal: the single trace operator expectation value is related to the asymptotic fall-off of the corresponding bulk field. Thus although the bulk fields evolve according to the 2-derivative field equations, the dynamics on the boundary contains infinite number of (spatial and temporal) derivatives. 
\end{itemize}

Hence in the large-$\lambda$, large-$c$ regime we are essentially down to studying classical gravitational dynamics in order to elucidate aspects of the field theory at strong coupling. This is of course a drastic simplification of the problem, but nevertheless it warrants further simplification to gain tractability.   Classical gravitational dynamics in asymptotically \AdS{d+1} spacetime involves an infinite number of fields arising from Kaluza-Klein (KK) modes on the compact space $\CX$.  Fortunately, further truncation can often be achieved by the magic of {\it consistent truncation} \cite{Duff:1985jd}. The basic philosophy behind consistent truncations is to find an appropriate ansatz for Type II or M-theory fields, which can acturally be reproduced by examining the dynamics of a truncated set of fields in $d+1$ dimensions. The most familiar example of consistent truncation are the gauged supergravity theories which keep only the lightest modes under the KK reduction. More complicated examples including massive fields have been constructed recently \cite{Maldacena:2008wh,Gauntlett:2009zw} and play an important role in the applied AdS/CFT correspondence.

One hassle with consistent truncations is that the lower-dimensional theory, and hence the dynamics of the dual field theory, are model dependent, i.e., dependent on the choice of the internal manifold $\CX$ (and in general on the fluxes turned on). Although this prevents one from making universal statements, it has the opposite advantage of incorporating richer dynamics. As we describe later, such extensions are in fact necessary in order to study the behaviour of field theories in grand canonical ensembles with prescribed chemical potentials. Nevertheless, it is useful to ask whether one can further distill the essential features of the correspondence to a minimal classical gravitational Lagrangian, and use this to identify universal features that are shared by a wide variety of QFTs. This is in fact possible and is achieved in the simplest imaginable manner: by studying the dynamics of pure gravity in AdS spacetimes. Let us pause to review that argument before proceeding. 

Given any solution of the form \AdS{d+1}$ \times \CX$ one can in fact argue that there is an universal sub-sector which simply comprises pure gravitational dynamics in \AdS{d+1}, as can be seen by Kaluza Klein reducing on the compact space $\CX$.  Clearly any non-trivial solution obtained in $d+1$ dimensions uplifts to a solution of the full string theory equations of motion. As a result, we always have a consistent truncation of the complicated lower dimensional Lagrangian to just Einstein gravity with negative cosmological constant, where the only dynamical mode is the graviton.  Since in the field theory dual this graviton mode corresponds to the stress tensor expectation value, this truncation provides a decoupled sector with universal dynamics for the stress tensor.  In other words, the stress tensor obeys the same equations of motion in each of such infinite class of strongly coupled field theories.  From the field theory perspective, the different theories are simply characterized by their differing central charge. Hence apart from an overall normalization we will usually be probing dynamics across a wide class of theories.

The complete dynamics of the stress tensor, which in particular allows one to compute all the $n$-point functions of $T_{\mu\nu}$ in the prescribed state, is still a complicated and non-local system from the field theory perspective. After all, this dynamics requires one to be able to solve for the dynamics of the non-linear Einstein-Hilbert Lagrangian (with negative cosmological constant) in $d+1$ dimensions.  Therefore to simplify things further, in \sec{s:flugra} we will take one further limit:  we will focus on configurations wherein the stress tensor varies sufficiently slowly compared to the local equilibration length scale (the long-wavelength limit).  Such configurations will then be locally thermalized, allowing for a much simpler description in terms of fluid dynamics.  In other words, the equations for the stress tensor in this limit reduce to generalized Navier-Stokes equations \cite{Bhattacharyya:2008jc}.

Once one has an understanding of the dynamics of the field theory stress tensor, or equivalently the bulk dynamics of pure gravitational degrees of freedom, one can enlarge the system to include other fields. For instance, a natural extension of the canonical ensemble in field theories with conserved charges is to include chemical potentials for the said charges and examine the behaviour of the grand canonical ensemble. Since global symmetries in the field theory map to  gauge fields in the bulk spacetime, one naturally ends up studying the behaviour of Einstein-Maxell or Einstein-Yang-Mills type theories (again with negative cosmological constant of course).  In this context one can furthermore consider the behaviour of composite gauge invariant operators of the field theory carrying various quantum numbers in these ensembles. The gravitational problem then generalizes appropriately to the physics of the dual bulk fields. This general scheme has recently been applied to study the phase structure of field theories at finite temperature and density (i.e., non-zero chemical potential) and has unearthed many interesting features which share qualitative similarities with the dynamics of superconducting phase transitions, non-Fermi liquid behaviour etc., which has been reviewed in \cite{Hartnoll:2009sz,Herzog:2009xv,McGreevy:2009xe,Horowitz:2010gk}. 

Our strategy in the following will be to consider the simplest setting of just gravitational physics in the bulk, since as argued above this sector is universal across a wide variety of field theories. To this end, our bulk analysis will involve the dynamics of Einstein gravity with a negative cosmological constant, i.e., 
\begin{equation}
\CS_{\text{bulk}} = \frac{1}{16\pi\,G_N^{(d+1)}}\,\int\, d^{d+1}x\, \sqrt{-G} \,\left(R - 2\,\Lambda\right) \ .
\label{ehact}
\end{equation}	
Einstein's equations are
given by\footnote{
We use upper case Latin indices $\{M,N, \cdots\}$ to denote bulk directions, while lower case Greek indices $\{\mu ,\nu, \cdots\}$ will refer to field theory (or ``boundary") directions. Finally, we will use lower case Latin indices $\{i,j,\cdots\}$ to denote the spatial directions in the boundary.}
\begin{equation}\label{ein} \begin{split}
&E_{M N} \equiv R_{M N} - \frac{1}{2} \, R \,  G_{M N} - \frac{d(d-1)}{2\, \RAdS^2}\, G_{MN}=0\\
& \implies \ \  R_{M N} + \frac{d}{\RAdS^2}\, G_{M N}=0 \ . 
\end{split}
\end{equation}

If the bulk AdS spacetime geometry is some negatively curved $(d+1)$-dimensional Lorentzian manifold, ${\cal M}_{d+1}$, with conformal boundary $\partial {\cal M}_{d}$, then the  field theory lives on a spacetime ${\cal B}_d$ of dimension $d$ in the same conformal class as $\partial {\cal M}_{d}$. Choosing an appropriate conformal frame, one may identify ${\cal B}_d$ and $\partial {\cal M}_{d}$ and speak of the field theory as living on the AdS boundary.  From the standpoint of the bulk theory, the choice of metric on ${\cal B}_d$ fixes a boundary condition that the bulk solution must satisfy.\footnote{
In standard AdS/CFT parlance, this amounts to fixing the non-normalizable mode of the bulk graviton to obtain the desired metric on ${\cal B}_d$.
} The correspondence is simplest to state for conformal field theories in dimensions where the trace anomaly vanishes, but with appropriate care, the correspondence also holds in the presence of a trace anomaly, and it can accommodate non-conformal deformations.

In general, one could have multiple bulk spacetimes $\CM_{d+1}$ whose boundary is the spacetime $\CB_d$ on which our field theory lives. In such cases, the AdS/CFT prescription of \cite{Witten:1998qj,Witten:1998zw} requires that one view all such possibilities as saddle points for the string theory (or gravity) path integral and one is instructed to sum over all such possibilities. Of course, the saddles might exchange dominance as one changes the boundary manifold; this can be viewed as a phase transition of the field theory as one dials an external parameter (in this case the geometry of the non-dynamical spacetime which it lives on). We will shortly encounter an example of such phase transitions for field theories on compact spatial volume.

\subsection{Field theories in the canonical ensemble \& thermal equilibrium}
\label{s:canensemble}

Finite temperature physics in the field theory can be realized by coupling the system to a heat bath, or more precisely by looking at the thermal density matrix 
\begin{equation}
\rho = e^{-\beta \, H} 
\label{tdenmat}
\end{equation}	
where $H$ is the field theory Hamiltonian. The dual spacetime should have a natural thermal interpretation. It is a well-known fact going back to the seminal works of Bekenstein and Hawking \cite{Bekenstein:1973ur,Hawking:1974sw} that black hole spacetimes with non-degenerate event horizons naturally exhibit features associated with thermal physics; thereby one is led to expect that black hole spacetimes to play a role in describing the dual of a finite temperature field theory, which is further supported by the intuition mentioned in \sec{s:intro} that endpoints of generic evolutions should match in the dual descriptions. It is, however, logically possible that one also has to consider `thermal geometries' (such as thermal AdS) which just have the Euclidean time circle periodically identified. 

To understand this issue better it is useful to think of the thermal density matrix by working in Euclidean time. On the field theory side one can achieve finite temperature by putting the theory on $\CB_d = \Sp^1_\beta \times \CA_{d-1}$ where the Euclidean time circle $\Sp^1_\beta$ has period $\beta = 1/T$  and $\CA_{d-1}$ is the spatial manifold. For compact $\CA_{d-1}$, such as for $\CA_{d-1} = \Sp^{d-1}$, one does indeed have two candidate bulk spacetimes satisfying the boundary conditions \cite{Witten:1998zw}. The first is the so-called thermal AdS spacetime which is \AdS{d+1} with periodically identified Euclidean time coordinate,
\begin{equation}
ds^2 = -\left(1+\frac{r^2}{\RAdS^2}\right) \, dt^2 + \frac{dr^2}{\left(1+\frac{r^2}{\RAdS^2}\right) } + r^2\, d\Omega_{d-1}^2 \ , \qquad \tau_E = -i\, t \ , \;\; \tau_E \simeq \tau_E + \beta \ .
\label{thermads}
\end{equation}	
The other saddle point is the static, spherically symmetric \SAdS{d+1} spacetime,
\begin{equation}
ds^2 = -f_g(r) \, dt^2 + \frac{dr^2}{f_g(r)} + r^2\, d\Omega_{d-1}^2 \ , \qquad f_g(r) = 1+\frac{r^2}{\RAdS^2}- \frac{r_+^{d-2}}{r^{d-2}}\, \left(1+\frac{r_+^2}{\RAdS^2}\right) \ ,
\label{globads}
\end{equation}	
for which the Hawking temperature $T$ is related to the size of the horizon $r_+$, as
\begin{equation}
T = \frac{d\, r_+^2 + (d-2)\, \RAdS^2}{4 \pi\, r_+ \, \RAdS^2} \ .
\label{bhtemp}
\end{equation}

These two geometries, \req{thermads} and \req{globads}, exchange dominance at $r_+ = \RAdS$ \cite{Witten:1998zw}: at low temperature, the thermal ensemble is dual to the thermal AdS spacetime and has a free energy of $\CO(1)$, while at high temperature, the correct dual is the \SAdS{d+1} which has a free energy of $\CO(c)$. This phase transition is referred to as the Hawking-Page transition \cite{Hawking:1982dh} and is best thought of as a confinement-deconfinement transition (since the jump in the free energy is large at large central charge $c$ as required for the planar limit). There is furthermore strong evidence that the transition persists to  weak coupling where it has been identified as a Hagedorn transition \cite{Aharony:2003sx}.

For most of our discussion, however, we are going to be interested in the dynamics of field theories on non-compact spacetimes; our focus will typically be on Minkowski spacetime $\R^{d-1,1}$. In this case there is no phase transition: the flat space limit is essentially the same as the high temperature limit for conformal field theories. As a result, the relevant geometry dual to the thermal density matrix \req{tdenmat} in the field theory is the planar \SAdS{d+1} black hole (where WLOG\footnote{
Note that in addition to the choice of AdS length scale, the presence of full $SO(d-1,1)$ invariance of the solution combined with the underlying $SO(d,2)$ isometry of \AdS{d+1} allows us to rescale the coordinates and set the horizon to be located at $r = 1$.  
} we have fixed $\RAdS = 1$ and $r_+ = 1$):
\begin{equation}
ds^2 = r^2 \, \left( -f(r) \, dt^2 + \sum_i (dx^i)^2 \right) + {dr^2 \over r^2 \, f(r)} \ , \qquad f(r) \equiv 1- \frac{1}{ r^d}
\label{SAdSmet}
\end{equation}	
The causal structure of this solution is easily determined:  the spacetime has a spacelike curvature singularity at $r=0$, cloaked by a regular event horizon at $r=1$, and a timelike boundary at $r= \infty$. This simple solution is of course static and translationally invariant in the boundary directions parameterized by $(t,x^i)$. One can in fact generate a $d$-parameter family of solutions by boosting in $\R^{d-1,1}$ with normalized $d$-velocity $u^\mu$ and scaling $ r$. This generates stationary black holes whose horizon size is given (after scaling) by $r_+$ and the boost velocity $u^\mu$ enters into the Killing generator of the horizon via 
\begin{equation}
\xi^a = u^\mu  \left(\frac{\partial}{\partial x^\mu}\right)^{\! a} \ . 
\label{}
\end{equation}	

It will turn out that a more convenient form of the metric is one which is manifestly regular on the horizon as well as being boundary-covariant.  We can obtain such a form by starting from \req{SAdSmet}, then
change to ingoing Eddington coordinates to avoid the coordinate singularity on the horizon:
$v = t + r_{\ast}$ where $dr_{\ast} = {dr \over r^2 \, f(r)}$, and finally  `covariantize' by boosting:  $v \to u_{\mu} \, x^{\mu}$, $x_i \to P_{i \mu} \, x^{\mu}$, where $P_{\mu\nu}$ is the spatial projector,
$P_{\mu\nu} =  \eta_{\mu\nu} + u_{\mu} \,  u_{\nu}$. This leads to the form of the metric we will use later:
\begin{equation}
ds^2 =-2\, u_{\mu}\, dx^{\mu} \,  dr 
+ r^2\, \left[ \eta_{\mu\nu} + \left(1-f(r/r_+)\right) \, u_{\mu}\, u_{\nu} \right] \, dx^{\mu} \,  dx^
{\nu}  \ , 
\label{SAdSzero}
\end{equation}
The event horizon is now at $r= r_+$, which in turn is related to the temperature via the large $r_+$ limit of \req{bhtemp},
\begin{equation}
T = \frac{d}{4\pi}\, r_+ \ .
\label{planarbhtemp}
\end{equation}

Once the bulk black hole solution is determined, it is straightforward to use the holographic prescription of \cite{Henningson:1998gx,Balasubramanian:1999re} to compute the boundary stress tensor. To perform the computation we regulate the asymptotically \AdS{d+1} spacetime at some cut-off hypersurface $r  = \Lambda_\text{c}$ and consider the induced metric on this surface, which (up to a scale factor involving $\Lambda_{\text{c}}$) is our boundary metric $g_{\mu\nu}$. The holographic stress tensor is given in terms of the extrinsic curvature $K_{\mu\nu}$ and metric data of this  cut-off hypersurface. Denoting the unit outward normal to the surface by $n^\mu$ we have 
\begin{equation}
K_{\mu\nu} = g_{\mu\rho}\,\nabla^\rho n_\nu
\label{extrinsicc}
\end{equation}	
For example, for asymptotically \AdS{d+1} spacetimes, the prescription of \cite{Balasubramanian:1999re} gives
\begin{equation}
T^{\mu\nu} = \lim_{\Lambda_\text{c} \to \infty}\; \frac{\Lambda_\text{c}^{d-2}}{16\pi \, G_N^{(d+1)}} \, \left[ K^{\mu\nu} - K \, g^{\mu\nu} - (d-1)\, g^{\mu\nu} - \frac{1}{d-2}\,  \left(R^{\mu\nu} -\frac{1}{2}\, R \, g^{\mu\nu}\right)\right] 
\label{bdysten}
\end{equation}	
where $K^{\mu\nu}$ is the extrinsic curvature of the boundary. Implementing this procedure for the metric \req{SAdSzero} we learn that the AdS/CFT correspondence maps this bulk solution to an ideal fluid characterized by temperature $T$  and fluid velocity $u^\mu$.   In particular, the induced stress tensor on the boundary is
\begin{equation}
T^{\mu \nu} = \CN_d\, \pi^d \, T^d \, \left(\eta^{\mu \nu} + d \, u^\mu \, u^\nu \right) \ ,
\label{TzeroO}
\end{equation}
where $\CN_d$ is an overall normalization that is proportional to the central charge $c$ of the field theory. 
Note that this stress tensor is traceless, $T_{\ \mu}^{\mu} =0$, as indeed is expected for a CFT.

\subsection{Equilibrium in grand canonical ensembles}
\label{s:gcensemble}

A natural extension of the above framework is to consider systems at finite density by generalizing \req{tdenmat} to
\begin{equation}
\rho = e^{-\beta (H - \mu_I\,Q^I)} 
\label{}
\end{equation}	
where $\mu_i$ are chemical potentials for the conserved charges $Q^I$. The choice of chemical potentials is of course restricted by the global symmetries of the field theory. If we insist on restricting attention to the universal dynamics of the stress tensor alone, then the only possible generalization is to consider chemical potentials for rotations $\Omega_I$. For a field theory on $\CA_{d-1} = \Sp^{d-1}$ one in general has $\lfloor \frac{d-1}{2}\rfloor$ independent rotations and each of these can be given a non-zero angular velocity $\Omega_I$. 

However, if we are willing to add matter fields in the bulk, i.e.\ extend \req{ehact} to include additional (non-gravitational) degrees of freedom, then we can generalize the discussion to include more interesting chemical potentials. First of all, we note that in order to incorporate chemical potentials, the field theory must admit conserved currents. Conserved currents in the field theory map to gauge symmetries in the bulk spacetime.\footnote{
This is in fact reminiscent of the theorem ``No global symmetries in string theory'', which follows from the observation that conserved currents in spacetime lead to world-sheet vertex operators of weight $(1,0)$ and $(0,1)$ respectively, which  in turn imply massless particle states in spacetime. 
} For every conserved field theory current $J^\mu_I$ one therefore has a bulk gauge field $A_{I\mu}$. The bulk gauge field obeys some equations of motion in the \AdS{d+1} spacetime and its non-normalizable mode in the near-boundary expansion corresponds to the boundary chemical potential $\mu_I$. 

Typically, large-$N$ field theories with holographic duals have conserved charges; for instance $\CN=4$ SYM has a $SO(6) \simeq SU(4)$ global symmetry and one can consider the grand canonical ensemble with $\mu_I$(s) corresponding to Cartan subgroup of this non-abelian global symmetry group. In fact, for a wide variety of $\CN=1$ superconformal field theories in $d=4$ one has three conserved $U(1)$ charges which geometrically can be related to the fact that these field theories arise from compactifications of Type IIB supergravity on toric Sasaki-Einstein manifolds.  A simple example to keep in mind is the Einstein-Maxwell Lagrangian (with perhaps Chern-Simons terms in odd bulk dimensions) which can be used to study the grand canonical ensemble with $U(1)$ charge chemical potentials, which can be obtained via a consistent truncation of gauged supergravity theories.

Given an appropriate definition of the field theory grand canonical ensemble, the equilibrium solution corresponding to this ensemble can be found by looking for static black holes carrying the appropriate charges. For Einstein-Maxwell theory these would be just the Reissner-Nordstrom-AdS (RNAdS) black holes, and the thermodynamic properties of these black holes correspond to the equilibrium thermodynamics of the field theory. 

In the presence of matter, one could have interesting phase transitions over and above the analog of the  Hawking-Page transition discussed in \sec{s:canensemble}. For pure Maxwell field interacting with gravity, these include charge redistribution as discussed originally in \cite{Gubser:2000ec,Gubser:2000mm}, or the more recent examples involving Chern-Simons dynamics \cite{Nakamura:2009tf}. Inclusion of charged scalars (and sometimes neutral scalars) can also lead to interesting phase transitions as originally pointed out by \cite{Gubser:2008zu}, which plays an important role in the physics of holographic superconductors \cite{Hartnoll:2008vx,Hartnoll:2008kx}.

\section{Linear response theory}
\label{s:linresp}

Having reviewed the basic framework of the AdS/CFT correspondence we now turn to describing the essential features of linear response theory. This will set the stage for our discussion of deviations away from equilibrium in \sec{s:lrads}.

Consider a generic quantum system characterized by a unitary Hamiltonian $H$ acting on a Hilbert space $\CH$ in equilibrium. The system could be in a pure state $\ket{\Psi}\in \CH$ or more generally in a density matrix $\rho$; the only requirement is that the system be in a stationary state. We now wish to perturb the system away from equilibrium and analyze the dynamics. A general deformation can be thought of as a change in the evolution operator $H \to H + H_{\text{pert}}$ and one is left with having to examine the dynamics with respect to this new Hamiltonian. Things however simplify if we can focus on deviations which are small in amplitude -- this is the regime of linear response theory, which we now briefly review. For a beautiful account see the original derivation by Kubo \cite{Kubo:1957fk,Kubo:1957lr} and the review \cite{Kubo:1966fk}.

In the linear response regime, one imagines the system being perturbed by a weak external force. To wit, one can write $H_{\text{pert}} = - \CA \, \CF(t)$ where $\CF(t)$ is the external force (with time dependence explicitly indicated) and $\CA$ is the canonical conjugate operator to the force. In the linear response regime the amplitude of the force is constrained to be small, $|\CF(t) |\simeq \CA \ll 1$, while we allow arbitrary temporal dependence, which in particular could involve high frequency modes being excited; see \fig{fig:regimes}. We essentially wish to do time dependent perturbation theory (being careful of causality in relativistic theories) for such deviations away from equilibrium.

\subsection{Response functions for deviations from equilibrium}
\label{s:responsef}

To monitor the departure from the stationary state, we can pick some observable, say the expectation value of a local operator $\CO$ in the theory.\footnote{
One could also consider non-local operators and more exotic observables such as entanglement entropy; we will discuss the latter in \sec{s:diagnostics}.
} For concreteness, we will assume that the system was in a density matrix $\rho$ before we perturbed it. Stationarity of this state demands that $[H,\rho] =0$. Turning on the perturbation $H_\text{pert}$ will cause a deformation to the density matrix. Suppose that we  encounter a new density matrix $\widetilde{\rho}$, which now has to  satisfy Hamiltonian evolution with respect to $H + H_\text{pert}$, i.e.,
\begin{equation}
i\,\frac{\partial}{\partial t} \tilde{\rho} = [H+H_\text{pert}, \widetilde{\rho}]
\label{rhoev}
\end{equation}	
For small amplitude perturbations, we can assume that $\widetilde{\rho} = \rho + \delta \rho$ and solve for $\delta \rho$ formally in terms of a time ordered integral expression:
\begin{equation}
\delta \rho(t) = i\, \int_{-\infty}^{t} \, dt' \, e^{-i(t-t')\, H} \; [\CA,\rho]\;
 e^{i(t-t')\, H}\, \CF(t') 
\label{delrho}
\end{equation}	
which is derived by solving the linearized version of the evolution equation \req{rhoev}.

Once we have the change in the density matrix, it is easy to compute the change in any measurement occurring due to the perturbation. We can estimate the response of the system by examining the expectation value of some observable $\CO$, which can be obtained directly as:
\begin{eqnarray}
&& \vev{\CO(t)} = \Tr{\widetilde{\rho} \, \CO} \nonumber \\
&& \thus \delta\vev{\CO(t)} = i\, \Tr{\int_{-\infty}^t\, dt' \, [\CA,\rho]\;
\CO(t-t')\, \CF(t') }
\label{delOexp}
\end{eqnarray}	
It is conventional to define the {\it response function} $\CR$ (sometimes called the after-effect function) as the change in the operator expectation value for a delta function perturbation; i.e., for $F(t) = \delta (t)$. From \req{delOexp} one recovers 
\begin{equation}
\CR_{\CO}^\CA(t)  = i\, \Tr{[\CA , \rho]\, \CO(t)} \equiv -i\, \Tr{\rho \,[\CA,\CO(t)]}
\label{respfn}
\end{equation}	
where we have tried to make clear in the notation the idea that one measures the response of $\CO$ to a perturbation caused by the deformation due to a force conjugate to $\CA$. By taking a Fourier transform of the response function, one arrives at the {\it admittance}:
\begin{equation}
\chi^\CA_\CO (\omega) = \lim_{\epsilon\to 0^+} \, \int_0^\infty\, \CR_\CO^\CA(t) \, e^{-i\,\omega\, t - \epsilon \, t} \, dt 
\label{}
\end{equation}	
It should be clear from the above discussion that this formalism is sufficiently general to accommodate arbitrary linear changes in a given dynamical system. Non-linear corrections can be explored in perturbation theory, generalizing familiar ideas from time dependent perturbation theory in quantum mechanics.

%

\subsection{Retarded correlators \& Kubo formulae}
\label{s:kubo}

The perturbations discussed so far are explicit perturbations on the system caused by some external force $\CF$. One can as well envisage a perturbation driven purely by thermal fluctuations, which are not a-priori related in any obvious way to external forces acting on the system. However, thermal fluctuations can be measured by looking at the response of the extensive thermodynamic variables to variations  in the local energy or charge densities. These in fact can also be treated in linear response and lead to the famous Kubo formulae for the transport coefficients. We now give a brief overview of these concepts valid in any field theory; in \sec{s:lrads} we will demonstrate how these formulae can be applied in the AdS/CFT context.

Let us return to the response function given in \req{respfn} and extract some essential features that we will use in later analysis. While that discussion was sufficiently general, it is worth translating this into more familiar language. Physically we wish to monitor the behaviour of the expectation value of some operator $\CO_a$ when the system is subject to some perturbation. We can imagine this occurring via a direct coupling of the operator to some sources $\CJ_a$, whose effect is to change the action via:
\begin{equation}
\CS = \CS_0 + \int \, d^dx\, \CJ_a(x) \, \CO_a(x)
\label{actionchange}
\end{equation}	
The response of the system due to this change can be obtained by rewriting the result \req{respfn}  slightly. Since the system will respond only after the perturbation, causality demands that we simply convolve the retarded Green's function of the operator $\CO_a$ to the sources $\CJ_b$ that causes it to deviate from stationarity. In particular, we have 
\begin{equation}
\vev{\CO_a(x)} = -\int d^dy \, G_{ab}^R(x,y)\, \CJ_b(y)
\label{}
\end{equation}	
where the retarded correlation function is defined as usual via
\begin{equation}
G^R_{ab}(x,y) =-i\, \theta(x^0-y^0) \, \vev{[\CO_a(x),\CO_b(y)]}
\label{}
\end{equation}	
Therefore given the retarded correlation functions, one can immediately infer, in the linear response regime, the manner in which the system under consideration reacts to the external disturbance caused by the sources. One will often be interested in the behaviour of the Green's functions in momentum space. To this end we define the Fourier transform of the retarded correlators via
\begin{equation}
G_{ab}^R(\omega, {\bf k}) = -i\, \int  d^d x\, e^{-i\,k\cdot x} \, \theta(t) \, G^R_{ab}\,(x,0)
\label{}
\end{equation}	
where we have assumed translational invariance and defined the $d$-vector $k^\mu = (\omega, {\bf k})$.

A class of observables that are interesting to examine are those corresponding to conserved currents in the theory, viz., the energy-momentum tensor $T_{\mu\nu}$ or generic conserved global charges $J_\mu$.  If we consider systems in thermal equilibrium, where deviations from thermality are engineered by density or charge fluctuations, then one is naturally led to studying the retarded Green's functions of these conserved currents: 
\begin{eqnarray}
G_{\mu\nu,\alpha\beta}(x,y) &=& -i\, \Tr{\rho\, [T_{\mu\nu}(x) , T_{\alpha \beta}(y) ]}  \equiv  -i\,\theta(x^0-y^0) \, \vev{ [T_{\mu\nu}(x) , T_{\alpha \beta}(y) ]}_\beta \nonumber \\
G_{\mu\nu, \alpha }(x,y)  &=& -i\, \Tr{\rho\, [T_{\mu\nu}(x) , J_{\alpha}(y) ]}  \equiv  -i\,\theta(x^0-y^0) \, \vev{[T_{\mu\nu}(x) , J_\alpha(y) ]}_\beta \nonumber \\
G_{\mu,\alpha}(x,y) &=& -i\, \Tr{\rho\, [J_\mu(x) , J_\alpha(y) ]}  \equiv  -i\,\theta(x^0-y^0) \, \vev{ [J_\mu(x) , J_\alpha(y) ]}_\beta
\label{retcorr}
\end{eqnarray}	
where we have assumed that the density matrix in question is appropriate to the ensemble under consideration. The correlators indicated on the RHS in \req{retcorr} are therefore the  thermal correlators, with perhaps fixed chemical potentials. 

Interesting physical quantities characterizing the system can be extracted by examining the momentum dependence of the retarded correlation functions of the conserved currents. The momentum space correlators have non-trivial analytic behaviour, with poles in the complex $k$ plane. One can read off the dispersion relation for the associated modes, by solving for $\omega({\bf k})$.  Since thermodynamic systems typically incorporate dissipative effects, these dispersion relations typically have imaginary pieces which capture the rate at which the system relaxes back to equilibrium. Stability of the quantum system demands that the perturbations damp out exponentially in time. This translates to the poles of the retarded Green's functions lying entirely in the lower half plane of the complex frequency space.
We will see shortly that these poles of the retarded Green's functions are in fact associated with the quasinormal modes of black hole geometries (which describe how a black hole settles back to its quiescent equilibrium state) via the AdS/CFT correspondence.

To be specific, let us consider a four dimensional conformal field theory and examine the stress tensor retarded correlator. It is useful to take into account the stress tensor conservation equation which  implies a Ward identity $k^\mu G_{\mu\nu,\alpha\beta} = 0$.\footnote{
We are here ignoring contact terms which can appear in the Ward identities. In momentum space these will give rise to analytic pieces and will therefore be irrelevant to our discussion of of poles.
} It is in fact convenient to pick a direction in momentum space, say ${\bf k} = k \, {\hat {\bf e}}_z$, and describe modes as being longitudinal or transverse to this choice of momentum. One can then show that the transverse components of the stress tensors have the behaviour
\begin{eqnarray}
\left\{G_{tx,tx}(\omega,k), G_{tx,xz}(\omega,k), G_{xz,xz}(\omega,k)\right\} &=&\frac{1}{2}\,
\left\{k^2 ,- \omega\, k , \omega^2 \right\} \, \frac{\CG_1(\omega,k)}{\omega^2-k^2}  \nonumber \\
G_{xy,xy}(\omega, k) &=& \frac{1}{2}\, \CG_3(\omega, k)
\label{transTT}
\end{eqnarray}	
while the longitudinal modes behave as  
\begin{eqnarray}
\left\{G_{tt,tt}(\omega,k), G_{tt,tz}(\omega,k), G_{tz,tz}(\omega,k)\right\} &=&\frac{2}{3}\,
\left\{k^2 ,- \omega\, k , \omega^2 \right\} \, \frac{k^2 \, \CG_2(\omega,k) }{(\omega^2-k^2)^2} \nonumber \\
&&
\label{longTT}
\end{eqnarray}	
The correlators are thus completely described by three scalar functions of momenta, $\CG_i(k)$ for $i=1,2,3$,  which exhibit the aforementioned poles.

Out of the set of poles of the retarded Green's function, of special interest are those that capture the late time behaviour. These necessarily involve small imaginary parts in $\omega({\bf k})$ (since the modes with large imaginary parts damp out quickly and therefore have short half-lives). These special set of poles are the {\it hydrodynamic poles}; they capture, and in fact provide a basis for, a complete description of the interacting quantum system via linearized hydrodynamics.\footnote{
Note that hydrodynamics is a good approximation to any physical system close to equilibrium, for fluctuations that are sufficiently long in wavelength. We discuss this in detail in \sec{s:flugra}.
} In the low-frequency regime, i.e., for $\omega, k \ll T$,  it turns out that the function $\CG_3(\omega,k)$ defined in \req{transTT} is non-singular, while the other correlation functions exhibit poles. In fact,  at low frequencies only the pieces of the correlation functions that correspond to conserved quantities can be singular, for it is these modes which due to the conservation law take a long time to relax back to equilibrium. 

The hydrodynamic poles of the system are characterized by having dispersion relations which satisfy the constraint $\omega(k) \to 0$ as $k\to0$. Fluctuations transverse to the direction of momentum flow,  captured by $\CG_1(\omega, k)$, give rise to dispersion relation of the form 
\begin{equation}
\text{Diffusive mode:} \qquad 
\omega = - \, i\, D\, k^2  , \label{}
\end{equation}	
which is characteristic of a diffusive mode, with diffusion constant $D$. For energy-momentum transport this is the shear mode, with the diffusion constant $D$ being related to the shear viscosity $\eta$ of the system via $D = \frac{\eta}{\epsilon + P}$, where $\epsilon$ and $P$ are the equilibrium energy density and pressure, respectively.  The longitudinal component, given by $\CG_2(\omega, k)$, has poles at locations 
\begin{equation}
\text{Sound mode:} \qquad \omega = \pm v_s \, k - i\, \Gamma_s\, k^2
\label{}
\end{equation}	
which describe sound propagation in the medium with velocity $v_s$ and attenuation $\Gamma_s$. Note that the sound mode is the only propagating mode in the hydrodynamic limit.

While one can examine the analytic structure of the retarded correlators in momentum space and extract  interesting transport properties, it is useful to obtain direct formulae for them. These are the famous Kubo formulae. For instance to find the shear viscosity of the system, one simply takes an appropriate zero frequency limit of the retarded correlator, i.e.,
\begin{equation}
\eta = -\lim_{\omega \to 0} \, \frac{1}{\omega}\, \text{Im}\left(G^R_{xy,xy}(\omega, {\bf 0})\right) .
\label{}
\end{equation}	
Similar expressions can be written down for the charge conductivity, etc..

\subsection{Brownian motion of probes and Langevin dynamics}
\label{ss:brown}

Our discussion thus far has been anti-chronological in a historical sense, for we have used general notions of quantum field theories and statistical mechanics to arrive at the response of the system to external perturbations. Historically, these ideas were first explored in kinetic theory, where it was realized that one could systematically account for the deviations away from purely thermal behaviour. We will refrain from repeating these ideas here in the context of departures from equilibrium of the system as a whole, but instead use these concepts to describe the physics of a single probe particle in a thermal medium. We have in mind a projectile that moves through some plasma medium. The medium itself will be taken to be in a thermal ensemble and we will be interested in the manner in which the probe particle loses energy to the medium. Moreover, it is also well known that even if the particle attains equilibrium with the plasma, it will continue to be buffeted by thermal fluctuations from the medium and undergo random motion. This is the famous stochastic Brownian motion, which is best described in the limit of a heavy probe particle in a thermal medium.

Let us therefore consider the Langevin equation, which is the simplest model
describing a non-relativistic Brownian particle of mass $m$ in one
spatial dimension:
\begin{align}
 \dot p(t)&=-\gamma_0 \,p(t)+R(t) \ .
\label{simpleLE}
\end{align}
Here $p=m \, \dot x$ is the (non-relativistic) momentum of the
Brownian particle at position $x$ and time $t$, and $\dot{}\equiv d/dt$.  The
two terms on the right hand side of \eqref{simpleLE} represent friction and random force, respectively, and $\gamma_0$ is a
constant called the friction coefficient.  One can think of the
particle as losing energy to the medium due to friction,
and concurrently getting a random kick from the thermal bath,
modeled by the random force.  We assume the latter to be  simply white noise
with
\begin{align}
 \vev{R(t)}&=0,\qquad \vev{R(t)R(t')}=\kappa_0 \, \delta(t-t'),
 \label{RRcorrMarkovian}
\end{align}
where $\kappa_0$ is a constant.
Note that the separation of the force into frictional and random parts on the
right hand side of \eqref{simpleLE} is merely a phenomenological
simplification -- microscopically, the two forces have the same origin, namely collision with the fluid constituents.

Assuming equipartition of energy at temperature $T = \vev{m\, \dot x^2}$, one can derive the following time evolution for the square of the 
displacement $s(t)$ \cite{Uhlenbeck:1930zz}:
\begin{align}
 \vev{s(t)^2}\equiv
 \vev{[x(t)-x(0)]^2}
 ={2\,D\over\gamma_0}\, \left(\gamma_0 \,t-1+e^{-\gamma_0 \,t}\right)
 \approx
 \begin{cases}
  \displaystyle {T\over m}\,t^2 & \displaystyle \Bigl(t\ll {1\over\gamma_0}\Bigr) \\[3ex]
  \displaystyle 2\,D\,t           & \displaystyle \Bigl(t\gg {1\over\gamma_0}\Bigr)
 \end{cases}
 \label{s^2_simple}
\end{align}
where the diffusion constant $D$ is related to the friction coefficient
$\gamma_0$ by the Sutherland-Einstein relation:
\begin{align}
 D&={T\over\gamma_0 \,m} \ .
 \label{diff_const_def}
\end{align}
We can see that in the ballistic regime, $t\ll 1/\gamma_0$, the particle moves
inertially ($s\sim t$) with the velocity determined by equipartition,
$\dot x\sim \sqrt{T/m}$.  On the other hand, in the diffusive regime, $t\gg
1/\gamma_0$, the particle undergoes a random walk ($s\sim\sqrt{t}$).  The transition is not instantaneous
because the Brownian particle must collide with a certain number of fluid
particles to get substantially diverted from the direction of its initial velocity.
The crossover time between the two regimes is the relaxation time
\begin{align}
 t_{\text{relax}}&\sim {1\over\gamma_0}\ ,
\label{t_relax}
\end{align}
which characterizes the time scale for the Brownian particle to forget
its initial velocity and thermalize.
One can also derive the important relation between the friction coefficient
$\gamma_0$ and the size of the random force $\kappa_0$:
\begin{align}
 \gamma_0&={\kappa_0\over 2\,m\,T}\ ,
 \label{gamma_and_kappa}
\end{align}
which is the simplest example of the fluctuation-dissipation theorem and
arises precisely because the frictional and random forces have the
same origin.

In $n$ spatial dimensions, the momentum $p$ and force $R$ in \eqref{simpleLE} are
generalized to $n$-component vectors, and \eqref{RRcorrMarkovian} is then naturally
generalized to
\begin{align}
 \vev{R_i(t)}&=0\ ,\qquad \vev{R_i(t)R_j(t')}=\kappa_0\, \delta_{ij}\, \delta(t-t')\ ,
\label{RRcorr_gendim}
\end{align}
where $i,j=1,\dots, n$.  In the diffusive regime, the displacement squared
scales as 
\begin{equation}
\vev{s(t)^2}\approx 2 \, n \, D \, t \ .
\label{s^2_n}
\end{equation}	
On the other hand, the Sutherland-Einstein relation
\eqref{diff_const_def} and the fluctuation-dissipation relation \eqref{gamma_and_kappa} are
independent of $n$.

Let us now return to the case with one spatial dimension ($n=1$).  The
Langevin equation \eqref{simpleLE}, \eqref{RRcorrMarkovian} captures
certain essential features of physics, but nevertheless is too simple to describe realistic systems, since it assumes that the friction is instantaneous and that there is no correlation between
random forces at different times  \eqref{RRcorrMarkovian}. If the
Brownian particle is not infinitely more massive than the fluid
particles, these assumptions are no longer valid; friction will depend
on the past history of the particle, and random forces at different
times will not be fully independent.  We can incorporate these effects
by generalizing the simplest Langevin equation \eqref{simpleLE} to the
so-called generalized Langevin equation \cite{Kubo:1966fk, Mori:1965yq},
\begin{gather}
 \dot p(t)=-\int_{-\infty}^t dt'\, \gamma(t-t')\, p(t')+R(t)+K(t) \ .
 \label{genLE}
\end{gather}
The friction term now depends on the past trajectory via the memory
kernel $\gamma(t)$, and the random force is taken to satisfy
\begin{align}
 \vev{R(t)}=0\ ,\qquad
 \vev{R(t) \, R(t')}=\kappa(t-t') \ ,
 \label{RRcorr}
\end{align}
where $\kappa(t)$ is some function.  We have in addition introduced an external
force $K(t)$ that can be applied to the system.

Let us briefly indicate how, faced with such a system with only $K$ under our control, we would extract the physical information, namely $\gamma(t)$ and $\kappa(t)$, as well as the characteristic timescales involved in collision and relaxation.  The latter will offer more direct insight into the nature of the medium under consideration.
To analyze the physical content of the generalized Langevin equation, it is convenient to first Fourier transform \eqref{genLE},  obtaining
\begin{align}
 p(\omega)={R(\omega)+K(\omega)\over \gamma[\omega]-i\omega} \ ,
 \label{genLE_omega}
\end{align}
where $p(\omega),R(\omega),K(\omega)$ are Fourier transforms, {\it
e.g.},
\begin{align}
 p(\omega)=\int_{-\infty}^\infty dt\, p(t)\,e^{i\omega t} \ ,
 \label{Fourier_trfm}
\end{align}
while
$\gamma[\omega]$ is the Fourier--Laplace transform:
\begin{align}
 \gamma[\omega]=\int_{0}^\infty dt\, \gamma(t) \,e^{i\omega t} \ .
 \label{FL_trfm}
\end{align}

If we take the statistical average of \eqref{genLE_omega}, the random force vanishes
because of the first equation in \eqref{RRcorr}, and we obtain
\begin{align}
 \vev{p(\omega)}=
 \mu(\omega)K(\omega),\qquad
 \mu(\omega)\equiv{1\over \gamma[\omega]-i\omega} \ ,
\label{adm_def}
\end{align}
where $\mu(\omega)$ is called the admittance.  The strategy, then, is to first determine the
admittance $\mu(\omega)$, and thereby $\gamma[\omega]$, by
measuring the response $\vev{p(\omega)}$ to an external force.  For example, if the external force is
\begin{align}
 K(t)=K_0 \,e^{-i\omega t},\label{monochr_force}
\end{align}
then $\vev{p(t)}$ is simply
\begin{align}
 \vev{p(t)}&=\mu(\omega)\, K_0\, e^{-i\omega t} \ .
 \label{presponse0}
\end{align}

For a quantity $\CO(t)$, we define the power spectrum $I_\CO(\omega)$ by
\begin{align}
  I_\CO(\omega)=\int_{-\infty}^\infty dt\,\vev{\CO(t_0)\,\CO(t_0+t)}\,e^{i\omega t} \ .
 \label{pwrspctr_def}
\end{align}
Note that $\vev{\CO(t_0) \, \CO(t_0+t)}$ is independent of $t_0$ in a
stationary system.  The knowledge of the power spectrum is equivalent to that
of 2-point function, because of the Wiener--Khintchine theorem:
\begin{align}
 \vev{\CO(\omega) \, \CO(\omega')} =
 2\pi\delta(\omega+\omega') \, I_\CO(\omega) \ .
 \label{WienerKhintchine}
\end{align}
Now consider the case without an external force, {\it i.e.},
$K=0$. In this case, from \eqref{genLE_omega},
\begin{align}
 p(\omega)={R(\omega)\over \gamma[\omega]-i\omega} \ .
\end{align}
Therefore, the power spectrum of $p$ and that for $R$ are related by
\begin{align}
 I_p(\omega)={I_R(\omega)\over |\gamma[\omega]-i\omega|^2} \ .
 \label{Ip_and_IR}
\end{align}
Hence, combining \eqref{presponse0} and \eqref{Ip_and_IR}, one can determine
both $\gamma(t)$ and $\kappa(t)$ appearing in the Langevin equation
\eqref{genLE} and \eqref{RRcorr} separately.  However, these two quantities are not
independent but are related to each other by the fluctuation-dissipation
theorem, generalizing the relation
\eqref{gamma_and_kappa}, cf.,  \cite{Kubo:1966fk}.

For the generalized Langevin equation, the analog of the
relaxation time \eqref{t_relax} is given by 
\begin{align}
 {t_{\text{relax}}}&=\left[\int_0^\infty dt\,\gamma(t)\right]^{-1}
 ={1\over \gamma[\omega=0]}=\mu(\omega=0) \ .
\label{t_relax_gen}
\end{align}
If $\gamma(t)$ is sharply peaked around $t=0$, we can ignore the
retarded effect of the friction term in \eqref{genLE} and write
\begin{align}
 \int_0^\infty dt'\,\gamma(t-t')\,p(t')
 \approx
 \int_0^\infty dt'\,\gamma(t')\cdot p(t)
 = {1 \over t_{\text{relax}}}\,p(t) \ .
\end{align}
The generalized Langevin equation \req{genLE} then reduces to the simple Langevin equation
\req{simpleLE}, so that $t_{\text{relax}}$ corresponds to
the thermalization time for the Brownian particle.

Another physically relevant time scale, the microscopic (or collision
duration) time $t_{\text{coll}}$, is defined to be the width of the
random force correlator function $\kappa(t)$.  Specifically, let us
define
\begin{align}
 t_{\text{coll}} = \int_0^{\infty}\!\! dt \,{\kappa(t)\over \kappa(0)} \ .
 \label{def_t_coll}
\end{align}
If $\kappa(t)=\kappa(0) \, e^{-t/t_{\text{coll}}}$, the right hand side of
this precisely gives $t_{\text{coll}}$.  This $t_{\text{coll}}$
characterizes the time scale over which the random force is correlated,
and thus can be thought of as the time elapsed in a single process of
scattering.  In many cases, it is natural to expect that 
\begin{align}
 t_{\text{relax}}\gg t_{\text{coll}} \ ,
\label{tr>>tc}
\end{align}
since, after all, we indicated that it takes a heavy probe many collisions to thermalize.
Typical examples for which \eqref{tr>>tc} holds include settings where
the particle is scattered occasionally by dilute scatterers as described by kinetic theory, and
settings where a heavy particle is hit frequently by much smaller
particles \cite{Kubo:1966fk}.  However, as we will discuss in \sec{s:brownian}, for the
Brownian motion dual to AdS black holes, the field theories in question are
strongly coupled CFTs and in fact \eqref{tr>>tc} does not necessarily
hold. There is also a third natural time scale $t_{\text{mfp}}$ given by
the typical time elapsed between two collisions.  In the kinetic
theory, this mean free path time is typically between the single-collision and relaxation time scales, $t_{\text{coll}} \ll
t_{\text{mfp}} \ll t_{\text{relax}}$; but again, this hierarchy is not expected to hold beyond perturbation theory.

The basic message to take away from this discussion is that the linear response regime is accessible once one understands the dynamics in equilibrium. The response functions are simply given in terms of Green's functions evaluated in the stationary configuration.

\section{Linear response from AdS/CFT: Probes of thermal plasma}
\label{s:lrads}

We now proceed to put together the toolkits we presented in the preceding two sections.
In \sec{s:linresp} we have described the basic methods employed in non-equilibrium statistical mechanics to understand the physics of systems  out of equilibrium, viz., linear response theory. One of the fundamental tenets in this approach is that for small-amplitude deviations, one can compute relevant observables by computing appropriate correlation functions in the equilibrium ensemble. Since the time-evolution part of the problem has been effectively dealt with in this manner, one is left with a much easier task in general. 

However, the computation of equilibrium  correlation functions is not as trivial as it sounds, especially in circumstances where the underlying quantum system is intrinsically strongly coupled. One therefore requires some further insight to deal with such situations. Fortunately, for a class of field theories which have holographic duals, the gauge/gravity correspondence comes to rescue, as indicated  in \sec{s:AdSCFT}. In fact, one the earliest developed technologies within the correspondence was the recipe to compute correlation functions of gauge invariant local operators in field theories using their dual gravity picture. 
In the present context this means that the gauge/gravity correspondence provides an efficient way to compute the correlation functions relevant for the linear response theory directly in terms of classical computations in an asymptotically AdS spacetime. 

We will begin by a brief review of the techniques employed to compute correlation functions in the AdS/CFT correspondence, followed by a discussion of lessons learnt by examining the linear response regime. Finally, we will discuss how one can monitor the behaviour of probe motion (both ballistic and stochastic) in a strongly coupled plasma medium.

\subsection{Computing correlation functions in AdS/CFT}
\label{s:corrads}

Let us consider a local gauge-invariant single trace operator $\CO(x)$ with conformal dimension $\Delta$ in the boundary CFT.  To compute correlation functions of this operator, one would deform the CFT action by adding a term $\int d^dx \, \phi_0(x) \, \CO(x)$ and obtain the generating function of the correlators as a functional of the sources $\phi_0(x)$. The requirement that the deformation term be dimensionless implies that $\phi_0(x)$ has scaling dimension $d- \Delta$.  

In the AdS/CFT correspondence, a given boundary operator maps to a bulk field whose spin $s$ is determined by the Lorentz transformation property of the operator $\CO$ in question. The bulk field $\phi(x,r)$, with $r$ being the radial coordinate in \AdS{d+1}, has mass $m^2(\Delta, s)$. For scalar operators one has the relation \cite{Witten:1998qj}\footnote{
This is not required to be true for $m_{BF}^2 \le m^2 \le m^2_{BF} +1$ where $m^2_{BF} = -\frac{d^2}{4\, R^2}$ is the Brietenlohner-Freedman mass, providing the lower bound on the mass of a scalar field in \AdS{d+1}. In the said range, one can equally well associate bulk field of mass $m^2$ with a boundary operator of dimension $\Delta =\frac{d}{2} - \sqrt{\frac{d^2}{4} + m^2 \, R^2}$ as discussed in \cite{Klebanov:1999tb}, which corresponds to an alternative quantization of fields in \AdS{d+1} \cite{Breitenlohner:1982jf}.
}
\begin{equation}
\Delta =\frac{d}{2} + \sqrt{\frac{d^2}{4} + m^2 \, R^2} \ .
\label{delm}
\end{equation}	
Similarly, for  a $p$-form operator on the boundary, the relation between the conformal dimension of the operator $\Delta$  and the mass of the bulk field $m$ is given as:
\begin{equation}
(\Delta + p)\,(\Delta+ p-d) = m^2 \ .
\label{delmp}
\end{equation}	

Given this map between fields and operators we can go ahead and use the AdS/CFT correspondence to compute the generating function of correlation functions,
\begin{equation}
W[\phi_0] = -\log\left(\expval{ e^{\int d^4x\, \phi_0\, \CO}}\right) \ ,
\label{corgen}
\end{equation}	
where we have schematically indicated the path integral over the quantum fields of the CFT. 

The statement of the AdS/CFT correspondence asserts that this generating functional is given by the partition function $\CZ_\text{string}$ of the string theory with the fields $\phi(r,x)$ prescribed to take on the boundary values $\phi_0(x)$ at the boundary of \AdS{d+1}. In particular, in the limit when classical gravity is a good approximation, the string partition function simply reduces to the on-shell action of gravity evaluated on the solution to the field equation. The on-shell action is usually divergent since we are turning on  mode that is non-normalizable to act as a source. To ensure that we capture the correct physics, we can regulate the AdS spacetime at $r=\epsilon^{-1}$ and demand that we satisfy the boundary condition there, i.e., demand $\phi(r,x) \to \phi_0(x) $ as $r \to \epsilon^{-1}$. Thus one arrives at the relation derived in \cite{Gubser:1998bc,Witten:1998qj}:  
\begin{equation}
\expval{e^{\int_{\partial \text{AdS}}\, \phi_0(x) \, \CO(x)} }_\text{CFT} = \CZ_\text{string} \left[\phi(r=\epsilon^{-1},x) = \phi_0(x)\right]
\label{adscft}
\end{equation}	
or in the limit where classical gravity in the bulk is a good approximation
\begin{equation}
W[\phi_0] = -\log \CZ_\text{string}
 \simeq \text{extremum}_{\phi(r=\epsilon^{-1}) = \phi_0} \; I_\text{sugra}(\phi_0) 
\label{gcorr}
\end{equation}	

The general scheme we have outlined above works well for computing Euclidean correlation functions in asymptotically AdS spacetimes, but there are certain subtleties to keep in mind while computing retarded correlation functions. A clear prescription was initially given in \cite{Son:2002sd} for which a nice supporting argument based on Schwinger-Keldysh contours was provided in \cite{Herzog:2002pc}. More recently, these arguments have been revisited in \cite{Marolf:2004fy} and a compact expression for computing two-point functions was provided in \cite{Iqbal:2008by,Iqbal:2009fd}. 
Formal studies of these correlation functions from a holographic renormalization scheme and general contour prescriptions for higher-point functions were discussed in \cite{Skenderis:2008dh,Skenderis:2008dg}; recently three-point functions at finite temperature were computed in \cite{Barnes:2010jp}.

Since we will be primarily interested in addressing issues in linear response theory, let us record here the prescription derived in \cite{Iqbal:2008by}, relating the retarded Green's function to a simple ratio involving the field and its conjugate momentum. In particular, for massive sscalar fields in asymptotically AdS spacetimes,
\begin{equation}
G_R(k) =\text{finite}\bigg\{ \lim_{r\to \infty} \, r^{2(\Delta -d)}\, \frac{\Pi(r,k)}{\phi_\text{sol}(r,k)} \bigg|_{\phi_0 =0} \bigg\} .
\label{ilcorr}
\end{equation}	
Here $\Pi$ is the canonical momentum conjugate to the field $\phi$ (which itself is dual to the operator $\CO$ under consideration) under radial evolution in AdS. Furthermore, $\phi_{\text{sol}}(r,k)$ is the on-shell solution to the appropriate wave equation subject to the boundary conditions that it be regular in the interior\footnote{
In the case of spacetimes with horizons, this amounts to demanding that the field be  infalling at the horizon, as we explain in \sec{s:bhquasi}.
}  and approaching some chosen boundary value $\phi_0$ at the boundary of the \AdS{} spacetime. The constraints $\phi_0=0$ (i.e.\ switching off the source) and the limit $r\to \infty$ are necessary to obtain the boundary observable as  
one anticipates from \req{gcorr}. We should note that this formula has been written down after taking into account the intricacies of the holographic renormalization and hence one is instructed to extract the finite part of the bulk calculation. For details on these techniques we refer the reader to \cite{Bianchi:2001kw,Skenderis:2002wp}.

\subsection{Retarded correlators and black hole quasinormal modes}
\label{s:bhquasi}

Given the utility of the AdS/CFT correspondence in computing correlation functions, let us now return to the issue of linear response around a given equilibrium configuration.  As we have seen in \sec{s:canensemble} and \sec{s:gcensemble}, physics of thermal equilibrium is captured by stationary black hole spacetimes in the dual geometric description. Therefore, linear response behaviour in the field theory translates directly to the behaviour of linearized fluctuations of bulk fields on AdS black hole backgrounds.

One of the first steps in this direction was taken in \cite{Horowitz:1999jd}, who pointed out the connection between AdS black hole quasinormal modes and the rate at which disturbances away from equilibrium re-equilibrate. Since this connection underpins much of the linear response theory we are about to describe, we will pause to recall the basics of the quasinormal mode spectrum in black hole spacetimes.

 Physically, quasinormal modes correspond to the late-time ``ringing'' of the black hole geometry. In particular, perturbations of the black hole undergo damped oscillations, whose frequencies and damping times are entirely fixed by the geometry and the nature of the propagating field, i.e., the modes are determined by the linearized wave operator and  are independent of the initial perturbation.  In fact,  it is  well understood that black hole spacetimes, owing to the presence of an event horizon into which the fields can dissipate, act as open systems; 
the corresponding spectrum of fluctuating modes is complex.
The reason for this behaviour is intuitively easy to understand. In classical general relativity, the event horizon acts as a one-way membrane; fields fall into the black hole but do not emerge out. Mathematically, this translates to an infalling boundary condition  on fields at the horizon in this black hole background. These same fields are also required to be normalizable near the AdS boundary, for one wishes to retain the AdS asymptotics (and therefore in the field theory side retain the  UV fixed point CFT unperturbed by relevant or irrelevant operators). Quasinormal modes for a classical field $\Phi$ (suppressing Lorentz indices) are defined as eigenfunctions of the linearized fluctuation operator which acts on $\Phi$ in the black hole background, satisfying these boundary conditions i.e, ingoing at the horizon and normalizable at infinity.\footnote{ An excellent summary of known spectra of quasinormal modes for various fields in diverse black hole backgrounds, together with their implication for both astrophysical black holes and in the AdS/CFT context can be found in \cite{Berti:2009kk}. See also \cite{Kokkotas:1999bd} for an earlier discussion of black hole quasinormal modes.}
 
As initially described in \cite{Horowitz:1999jd}, the quasinormal modes of AdS black holes capture the rate at which the field theory, when perturbed away from thermal equilibrium, returns back to the quiescent equilibrium state. In this context, one usually concentrates on the lowest set of modes, as these dominate the long-time behaviour.  Nevertheless, it is possible to give a clear interpretation to the entire quasinormal mode spectrum. As was pointed out in \cite{Birmingham:2001pj} in the context of 1+1 dimensional boundary CFTs and asymptotically \AdS{3} BTZ black holes, the entire quasinormal mode spectrum maps to the poles of the retarded Green's functions of operators in the canonical ensemble. This was extended to higher dimensions in the seminal work of \cite{Son:2002sd} and further elaborated upon in \cite{Kovtun:2005ev}.

While the relation between quasinormal modes and poles of retarded Green's functions holds in general for any operator in the dual field theory, it takes on interesting hues for the case when the dual operator corresponds to a conserved current. As discussed in \sec{s:kubo}, the analytic structure in the retarded Green's functions of the stress tensor which corresponds to the hydrodynamic modes of the system, has complex dispersion relations $\omega(k)$ characterized by the long-wavelength behaviour $\omega(k) \to 0$ as $k\to0$. 
This was explicitly verified by the computation of gravitational quasinormal modes in planar \AdS{d+1} black hole backgrounds, which have translationally invariant horizons and allow for arbitrarily long-wavelength modes. On the contrary, global \AdS{d+1} black holes have horizons of spherical topology and correspond to field theories living in finite volume on $\text{ESU}_d = \R \times \Sp^{d-1}$. One then encounters IR effects coming from the finiteness of spatial volume which precludes the existence of quasinormal modes with vanishing frequencies. In order to see the hydrodynamic behaviour one has to scale the curvature to zero as well, which reduces the problem to the planar case; one can systematically account for the curvature corrections as we indicate in \sec{s:flugra}.

The fact that black hole quasinormal mode spectrum admits modes with hydrodynamic dispersion relation leads one to suspect that one can use the gravity analysis to compute properties of the fluid description. Indeed as we have sketched in \sec{s:kubo}, one can use the behaviour of the retarded Green's functions at zero momentum to learn about transport coefficients like viscosity. This was first carried out in the AdS/CFT context in \cite{Policastro:2002se,Policastro:2002tn}. The analysis was ground-breaking in that it not only verified the general intuition that one can relate the classical dynamics in a black hole background to the physics of strongly coupled plasmas, but it also paved the way for what is perhaps the most famous conjecture in the subject, viz., the bound on the ratio of shear viscosity to entropy density, $\eta/s \ge \frac{1}{4\pi}$, \cite{Kovtun:2004de}. For a large class of two-derivative theories of gravity one finds by direct computation that this bound is in fact saturated, which prompted \cite{Kovtun:2004de}. Understanding its implications and its {\it raison d'\^ etre}  has been the focus of a large body of literature, which we cannot do justice to here and point the reader to the excellent review article \cite{Son:2007vk} for developments till a couple of years ago.
 
 This rather small value of shear viscosity obtained in the holographic computations has been instrumental in forging connections with ongoing experimental efforts to understand the state of matter, the quark-gluon plasma (QGP),  produced in heavy-ion collisions and RHIC and soon at LHC. Fits to data  from the STAR detector at RHIC suggests that the QGP behaves close to the deconfinement transition in QCD as a nearly-ideal fluid with very low viscosity (see \cite{Schafer:2009dj} for a discussion of near perfect fluidity in physical systems). This has prompted a concentrated effort in the literature and spurred the growth of the AdS/QCD enterprise; we refer the reader to the reviews \cite{Mateos:2007ay, Gubser:2009md} for these developments.

Recently, this bound has been shown to be violated in higher-derivative theories of gravity: there are example toy-models such as Gauss-Bonnet gravity \cite{Brigante:2008gz, Brigante:2007nu} and other higher-derivative theories \cite{deBoer:2009pn,Camanho:2009vw,Buchel:2009sk, deBoer:2009gx, Camanho:2009hu, Myers:2010jv} and also some string theory inspired constructions of large-$N$ superconformal theories \cite{Kats:2007mq, Sinha:2009ev}. The general consensus at the stage of writing this review seems to be that the bound, whilst robust in the two-derivative approximation (which corresponds to the strong coupling, large-$N$ theory), could in general be violated by finite-$N$ (string interactions) and also perhaps by $\alpha'$ (large finite coupling) effects. 

The quasinormal mode analysis can also be used to go to higher orders in the hydrodynamic expansion. After all,  for a planar black hole one has a spectrum of poles of the retarded Green's function which can be used to extract an exact non-linear dispersion relation $\omega(k)$ beyond the leading long-wavelength $k\to0$ approximation. Such techniques were first employed in the analysis of \cite{Baier:2007ix} who computed certain second-oder transport coefficients. Similar analyses were also carried out in \cite{Natsuume:2007ty,Natsuume:2008iy}. In an nice calculation \cite{Amado:2008ji} described the behaviour of  low lying quasinormal modes (the modes closest to the real axis) and displayed how it exchanges dominance with the next quasinormal mode at some finite value of momentum. This in particular indicates the regime of validity of the linear hydrodynamic approximation; for the higher quasinormal modes, while still giving poles of the retarded Green's functions, are not part of the effective hydrodynamic theory. 

In summary, there is a direct relation between the physics of black hole quasinormal modes and the retarded Green's functions of local gauge invariant operators. For a given operator $\CO$ on the boundary,  one identifies the corresponding  bulk field $\phi$ and computes its quasinormal mode spectrum to infer the location of the poles of the retarded Green's function. From the retarded Green's functions of  conserved currents one learns that the long-wavelength behaviour of the interacting CFTs which fall within the purview of the AdS/CFT correspondence is described by linearized hydrodynamics. Thus black hole quasinormal modes provide a powerful computational technique to learn about the dynamics of strongly coupled gauge theories and their relaxation back to equilibrium.  They also confirm the intuition that the thermal behaviour of strongly coupled field theories can be captured in effective field theory by viewing the system as a plasma medium.

\subsection{Probes in the plasma: Dissipation and stochastic motion}
\label{s:probes}

Thus far, we have discussed the behaviour of retarded correlation functions of local, gauge invariant operators using the AdS/CFT correspondence. These analyses allow us to picture the interacting, thermal field theory as a plasma medium. As discussed in \sec{ss:brown}, it is useful to ask how do probe particles introduced into such plasmas behave? This question is not only interesting from a theoretical viewpoint, but also from a pragmatic standpoint. For instance, in the case of the QGP, one is interested in knowing how much energy is lost by a quark produced in the deep interior of the plasma as it traverses outward. There, one does not have reliable computational methods owing to the strongly coupled nature of the plasma, but  as we shall see, in the AdS/CFT framework one can again distill this question to a simple classical computation. In this section we will explore the various attempts in the literature aimed at addressing this question, starting with the ballistic motion of quarks in the plasma and then turning to a discussion of the stochastic Brownian motion of stationary probes. 

\subsubsection{Energy loss and radiation of moving projectiles}
\label{s:projectiles}

To understand the energy loss of probe particles in plasma medium, we introduce an external probe in the form of an external quark or meson (for non-abelian plasmas) into the medium. Such probes are holographically modeled by an open string in the bulk geometry; here the geometry of interest is an asymptotically AdS black hole, which as we have described above provides the thermal medium.  Let us understand the set-up in more detail.  One of the open string end-points is pinned on the boundary of the AdS spacetime. Since this end-point carries the usual Chan-Paton index, it corresponds to the external quark we have introduced into the system. Heuristically, the external quark has a flux tube attached to it; in the holographic description one can view this flux tube as the open string world-sheet which extends into the bulk spacetime. Monitoring the motion of the external quark through the plasma thus amounts to studying the classical dynamics of string world-sheet in a black hole background. For mesonic probes we consider open strings with both end-points stuck on the boundary of \AdS{d+1}. One could also consider other probes such as monopoles or baryons (which are heavy in the large-$N$ limit); these would correspond to D-branes living in the bulk. 

The first steps to understand the energy loss for  probes in plasmas holographically were carried out in the  seminal
papers \cite{Herzog:2006gh, Liu:2006ug, Gubser:2006bz, Herzog:2006se, CasalderreySolana:2006rq, Gubser:2006nz, Liu:2006he, CasalderreySolana:2007qw}, by considering the dynamics of probe strings as described above.
A brief summary of these accounts can be found for instance in \cite{Mateos:2007ay}. The general
philosophy in these discussions was to use the probe dynamics to extract the rates of energy loss and transverse momentum broadening in the medium, which bear direct relevance to the physical problem of motion of
quarks and mesons in the quark-gluon plasma.

\begin{figure}
\begin{center}
\includegraphics[width=2.5in]{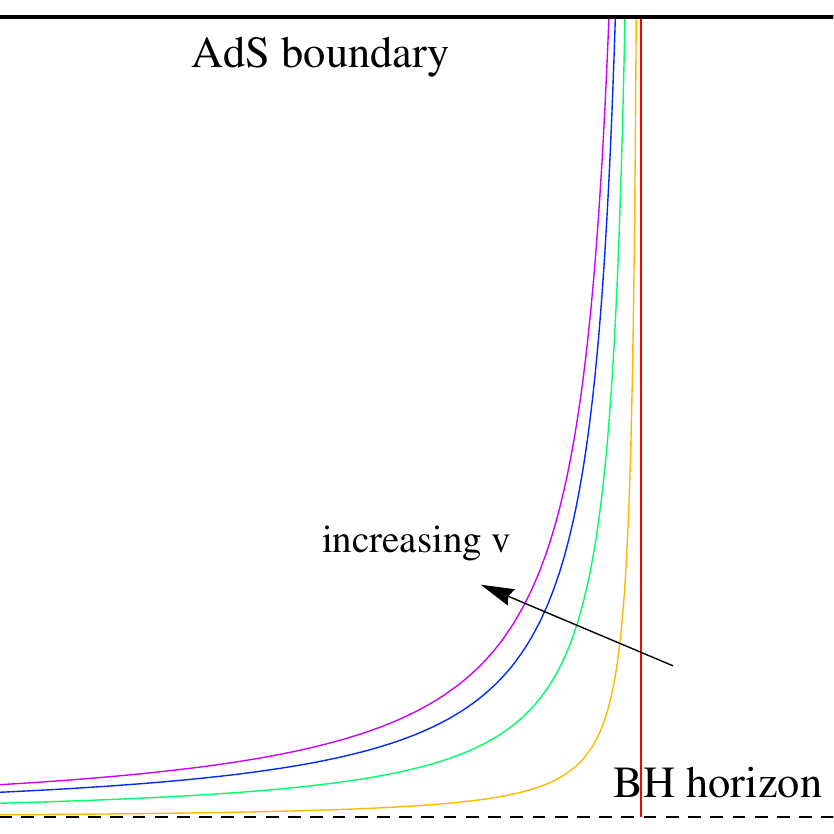}
\caption{The trailing string solution in \SAdS{4} spacetime. The curves are drawn for differing values of the quark velocity $v$ (specifically from right to left, $v=0, 0.15, 0.5, 0.8$, and $0.99999$) on the boundary.}
\label{fig:trstring}
\end{center}
\end{figure}

The simplest computation of the energy loss was originally considered in \cite{Herzog:2006gh,Gubser:2006bz}. The idea was to examine a probe in the  boundary moving with a constant velocity $v$ under the influence of an external force, applied so as to compensate for the frictional force acting on the probe and to maintain a steady state.  From this steady state speed one can then recover the friction constant $\gamma$. On the bulk side, the problem reduces to a classical solution of  Nambu-Goto action for the string, with one end-point being pulled with velocity $v$ along the AdS boundary. The constancy of velocity at the boundary is again maintained by an external force, in this case a constant electric field that drags the string end-point. By examining the classical solution of the Nambu-Goto action with these boundary conditions, one finds that while the quark moves forward with constant velocity, the bulk string world-sheet trails behind, see \fig{fig:trstring}.
Since there is no natural place for the string world-sheet to end in the bulk spacetime, it simply dips into the horizon.\footnote{
This is however not captured by the constant-$t$ snapshot of the dragged string represented by \fig{fig:trstring} where only the static region outside the horizon is visible, so the string looks like it `freezes' onto the horizon, analogously to the ``frozen star" picture of a collapsing black hole.
}  In particular, this implies that the classical world-sheet of the string itself has an induced horizon (which for non-zero velocity will always be outside the spacetime event horizon), a fact that will be important when we address the stochastic motion of the boundary end-point in \sec{s:brownian}.

 To see the construction in more detail, consider the planar \SAdS{d+1} black hole \req{SAdSmet}; we are looking for a classical solution to the Nambu-Goto action
\begin{equation}
\CS_{\text{NG}} = -\frac{1}{2\pi \alpha'} \, \int d^2\sigma\, e^{\frac{1}{2}\; \phi} \, \sqrt{-\text{det}(g_\text{induced})}  \ ,
\label{ngact}
\end{equation}	
with $\sigma^\alpha$ being the world-sheet coordinates $(\tau,\sigma)$.  The computation is easily carried out in static gauge $\tau = t$ and $\sigma = r$. In order to track the motion of a quark on the boundary, it is convenient to use the ansatz
\begin{equation}
x^1(t,r) = v\, t + \xi(r) \ , \qquad x^i = 0 \;\; \forall\;i \neq 1 \ ,
\label{xansatz}
\end{equation}	
resulting in the Lagrangian density:
\begin{equation}
\CL = -\sqrt{1+r^4\, f(r)\, \xi'(r)^2 - \frac{v^2}{f(r)}} \ ,
\label{efflagNG}
\end{equation}	
with $f(r)$ given by \req{SAdSmet}.
The solution is obtained straightforwardly by noting that the conjugate momentum to $\xi$, $\pi_\xi = \frac{\partial\CL}{\partial \xi'}$ is conserved and constant, which allows one to solve $\xi(r)$ via quadratures,\footnote{
The physically correct sign of the square root corresponds to the bulk piece of the string world-sheet trailing the quark on the boundary.}
\begin{equation}
\xi'(r) = \frac{\pi_\xi}{r^2\, f(r)} \, \sqrt{\frac{f(r) -v^2}{r^4\, f(r) - \pi_\xi^2}} \ .
\label{xiprime}
\end{equation}	
 In fact, requiring that the induced metric be timelike and non-degenerate fixes $\pi_\xi$ completely. Noting that $0\le f(r)<  1$ for $r \in [r_+,\infty)$ and $v \le 1$ implies that the numerator of \req{xiprime} can vanish somewhere in the bulk of the spacetime. At this point, it must be that the denominator also vanishes. This then constraints $\pi_\xi$ to take on the value: 
\begin{equation}
\pi_\xi = r_+^2\, \frac{v}{(1-v^2)^{2/d}} \ .
\label{conjmom}
\end{equation}	

Given the classical string configuration obtained by solving \req{xiprime}, one can compute the rate of energy loss by looking at the momentum flow along the string world-sheet. One simply has to compute the net flux of this world-sheet momentum down the string. This results in \cite{Herzog:2006gh,Gubser:2006bz,deBoer:2008gu}
\begin{equation}
\frac{dp^1}{dt} =  -\frac{8\pi \,\RAdS^2\, T^2}{d^2\,\alpha'}
 \, \frac{v}{(1-v^2)^{2/d}} \ ,\qquad  p^1={m\,v\over \sqrt{1-v^2}} \ .
\label{momentumloss}
\end{equation}	
where we have translated the result in terms of the temperature $T$ which is related to $r_+$ via \req{planarbhtemp} and reinstated the \AdS{d+1} length scale.  In the non-relativistic limit, $v\ll 1$, this means that the friction constant is
\begin{align}
 \gamma_{0}^{\text{AdS$_{d+1}$}}=
\frac{8\pi \,\RAdS^2 \,T^2}{d^2\,\alpha' \,m} \ .
\label{gamma_AdSd}
\end{align}
If we use the Sutherland--Einstein relation \eqref{gamma_and_kappa},\footnote{
Note that, as explained around
\eqref{RRcorr_gendim}, the relation \eqref{gamma_and_kappa} does not
depend on $d$.
}  we obtain the diffusion constant
\begin{align}
 D_{{\rm AdS}_d}&=\frac{d^2 \,\alpha'}{ 8\pi\, \RAdS^2 \, T}   \ .
 \label{diff_const_AdSd}
\end{align}

The calculation described above breaks down at very large velocities: physically, the force required to keep the quark moving sufficiently fast becomes so large that the quark anti-quark production is unsuppressed \cite{CasalderreySolana:2007qw}.
In \cite{Herzog:2006gh} the more involved analysis of actually trying to see a moving quark slow down explicitly was also undertaken. This involves solving the time dependent equations arising from the Nambu-Goto action \req{ngact} with the initial conditions of non-zero velocity. 

In this context, there is an interesting puzzle: a-priori one would have expected that any particle  moving through a plasma will lose energy to the medium and slow down. While we have seen that this occurs for quarks, described by the trailing string configuration \req{xiprime}, there is no mechanism for energy loss in the case of a meson moving through a plasma. A meson, modeled as a quark anti-quark pair corresponds to a configuration of an open string with both ends stuck to the boundary; in such situations one finds `no-drag' solutions where the string simply dangles down into the bulk independently of the direction of $v$. More pertinently, for small separations between the quark and anti-quark (such that the meson remains in its bound state), the string  stays above the horizon and as a result does not lose energy to the black hole.  This feature was observed in different contexts in the analysis of \cite{Peeters:2006iu, Liu:2006nn,Chernicoff:2006hi}. Likewise it was also noticed in \cite{Chernicoff:2006yp} that baryonic particle also propagate without being subject to  drag (these authors also derived the drag force on k-quarks and gluons). These probes which seemingly do not suffer from dissipation are to our knowledge rather poorly understood.\footnote{
In fact, the gravitational dual is analogously puzzling.  A dragged string above a static black hole is equivalent to a static string above a boosted black hole, since only the relative velocity matters.  For the latter configuration, the boost of the black hole produces an ergoregion above the horizon, and it is easy to see that the string cannot dip into this ergoregion.  However, one might naively expect the frame-dragging effect to extend past the ergosurface and affect the string, bending it in the direction of the boost.  Yet, as manifest from the solutions, no such effect takes place!
}  Analysis of the velocity dependence of the screening length in the plasma for mesons and baryons was explored in \cite{Liu:2006nn, Chernicoff:2006hi, Liu:2006he, Athanasiou:2008pz}. Another mechanism for quark energy loss in the medium based on Cherenkov radiation was proposed recently in \cite{CasalderreySolana:2009ch}.\footnote{
We should also remark that there have been attempts to study linearly accelerating particles and their associated radiation in  the AdS/CFT context; see \cite{Chernicoff:2008sa, Xiao:2008nr,Paredes:2008cr, Chernicoff:2009re,Chernicoff:2009ff, Caceres:2010rm,Chernicoff:2010wg}.  
}

Note that deceleration of the quark provides yet another mechanism for its energy loss, induced by radiation due to the quark's deceleration.  The interesting question of interplay between the two distinct energy loss mechanisms -- namely medium-induced (i.e.\ drag) and acceleration-induced -- has been explored in \cite{Chernicoff:2008sa,Fadafan:2008bq}, suggesting that these two effects may interfere destructively.   More specifically, the authors consider a quark undergoing a circular motion at constant angular velocity.  One advantage of such setup is that it can be treated as approximately stationary configuration while nevertheless incorporating acceleration (and in fact the classical world-sheet calculation for circular acceleration remains valid well into the acceleration-dominated regime), in contrast to the above-mentioned cases of linear motion.  Interestingly, \cite{Fadafan:2008bq} find that depending on the angular velocity $\omega$ and radius $v/\omega$ of the quark's circular trajectory, the energy loss is dominated by either by drag force acting as though the quark were moving in a straight line, or by the radiation due to the circular motion as if in absence of any plasma, whichever effect is larger, with continuous crossover between these regimes occurring at $\omega \approx \pi T (1-v^2)^{3/4}$.

Once one has an understanding of the basic mechanism for energy loss in the holographic set-up, one can attempt to study the detailed response of the projectile in the plasma medium. One would naturally expect that a moving projectile produces a wake behind it; at large distances from the projectile one can employ hydrodynamics to study the behaviour of this disturbance, but in general, high-frequency modes will be excited near the source of the perturbation. A convenient way to probe the physics of a moving projectile is to monitor the expectation value of the energy-momentum tensor:
\begin{equation}
\Delta\vev{T^{\mu\nu}(x)} = \vev{T^{\mu\nu}(x)}_{\text{with projectile}} - \vev{T^{\mu\nu}(x)}_{\text{without projectile}} \ .
\label{vevT}
\end{equation}	
The projectile as before is modeled as an external quark in the field theory and corresponds to the motion of the string in the black hole background. Now in addition to the classical motion of the string we also want to compute the back-reaction on the geometry due to this external string and its influence on the boundary values of the metric (which in turn are related to the stress tensor expectation value). The first steps in this direction were taken in \cite{Friess:2006aw} where the expectation value of the operator dual to the dilaton was considered. This was then extended to the calculation of the stress tenor expectation value \req{vevT} in a series of works,
\cite{Friess:2006fk,Gubser:2007nd,Gubser:2007xz,Yarom:2007ni,Gubser:2007xz,Chesler:2007an,Gubser:2007ga,Chesler:2007sv}.  For a good review of this development we refer the reader to \cite{Gubser:2009sn}. 

The essential idea here is to look at the perturbations about the trailing string configuration and ascertain from this the source of the bulk  graviton. One then solves the (linearized) bulk Einstein's equations including the back-reaction of this source. Using the asymptotic behaviour of the solution one can finally read off the boundary stress tensor. 
This stress tensor exhibits many characteristic features that are expected for a projectile moving through a dissipative medium.  For example, when the projectile moves faster than the speed of sound through the medium, it produces a Mach cone at a particular angle, leaving behind it a wake in the plasma (see \cite{Gubser:2009sn} for further details).

\begin{figure}
\begin{center}
\includegraphics[width=2.5in]{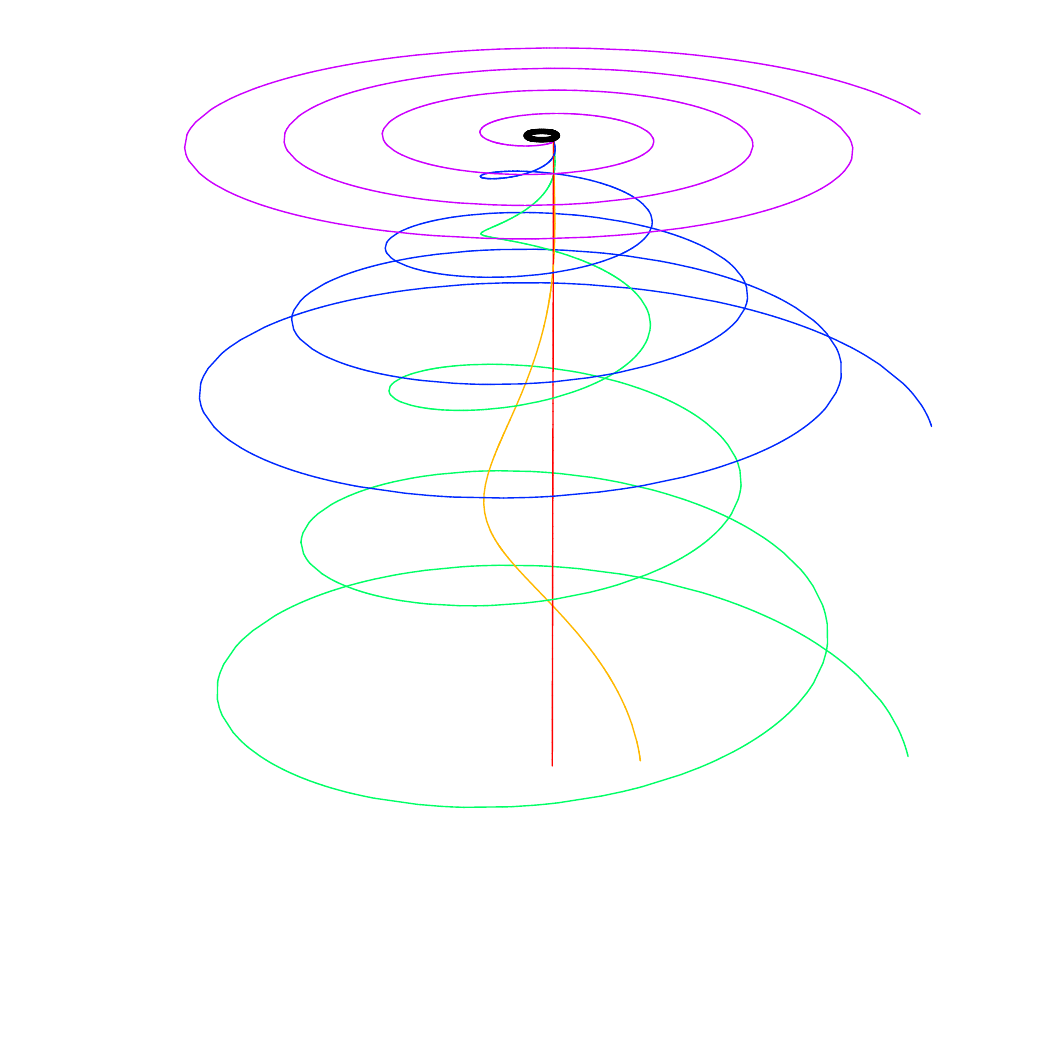}
\caption{The spiral string solution in AdS$_5$ spacetime corresponding to quark undergoing a circular motion. The curves are drawn for differing values of the quark velocity $v$ on the boundary (values of $v$ and color-coding same as in \fig{fig:trstring}).  The small black circle on the top is the quark's trajectory on the AdS boundary; the vertical direction corresponds to the bulk radial direction.}
\label{fig:spirstring}
\end{center}
\end{figure}

Using the setup of \cite{Fadafan:2008bq} describing quark in circular motion with constant angular velocity, \cite{Athanasiou:2010pv} examine (part of) the corresponding boundary stress tensor induced by metric deformation due to the trailing string in the bulk.  In particular, they compute the energy density and angular distribution of the power radiated by the quark.  Unlike the previously-mentioned cases, they focus on the plasma medium at zero temperature, so that the bulk dual is described by a string trailing in pure AdS. A snapshot of a string is plotted in \fig{fig:spirstring} for various values of the quark velocity. The string solution looks like a rigidly-rotating spiral flaring out into the bulk, which in fact induces a horizon on the world-sheet despite no horizon being present in the underlying geometry (such horizon generation was previously discussed in \cite{Chernicoff:2008sa} in more general context).\footnote{
The corresponding world-sheet temperature $T_{WS}$ is related to the Unruh temperature $T_U$ of the quark as $T_{WS} = T_U \, \sqrt{1-v^2}$.
}

  Rather curiously, \cite{Athanasiou:2010pv}  discover that this strong-coupling calculation gives very similar results to those at weak coupling, and closely resembles synchrotron radiation produced by an electron in circular motion in classical electrodynamics.  Specifically, the radiation is emitted in narrow beam along its velocity vector, with opening angle $\sim \sqrt{1-v^2}$.  Surprisingly, despite strong coupling, the pulses of radiation propagate without broadening.  While this is all explicitly derived from the bulk description, even on the gravity side it remains rather puzzling that the metric perturbation due to the trailing string remains so sharply localized.  Indeed, this is rather counter to the naive UV/IR intuition, which would lead one to expect that the part of the string situated deep in the bulk should produce rather diffuse boundary stress tensor.

The absence of broadening in the radiation pattern at zero temperature is in sharp contrast to the corresponding behaviour in the finite temperature case, where we expect that the beam of radiation propagating through thermal plasma slows down to the speed of sound and ultimately thermalizes.  This type of behaviour was indeed observed in e.g.\ \cite{Chesler:2008wd,Chesler:2008uy} in the context of light quarks and mesons being released from rest.  Here the authors start with initial state corresponding to a highly energetic quark anti-quark pair and follow its time evolution\footnote{
Unlike heavy quarks, whose energy loss rate depends only on their velocity and the characteristics of the medium (i.e.\ the Õt Hooft coupling $\lambda$ and the temperature $T$ of the plasma through which the quark is moving), light quarks are in addition influenced by details of their initial conditions, specified by the gauge field configuration.  However, a more universal quantity, which is almost insensitive to the initial profile of the string, is the maximum distance $\Delta x_{\rm max}$ traveled by a quark with initial energy $E$.  
} into two jets. Whereas at zero temperature the jets travel and spread forever, at finite temperature they stop within a finite distance $\Delta x_{\rm max}$, which scales with the cube root of the energy,
$\Delta x_{\rm max} \approx \frac{0.526}{T} \left(\frac{E}{T\sqrt{\lambda}} \right)^{\! 1/3}$.

A completementary approach to computing the energy lost by moving quarks in a non-abelian plasma is to relate the energy loss to the so called ``jet-quenching parameter'' \cite{Wiedemann:2009sh,Kovner:2003zj}. This parameter can be extracted from the expectation value of a light-like Wilson loop:
\begin{equation}
\vev{W(\CC)} \simeq \exp\left(-\frac{1}{4\,\sqrt{2}}\, \hat{q} \, L^- \, L^2\right)
\label{}
\end{equation}	
where the contour $\CC$ is taken to be a rectangle of coordindate length $L^-$ in the light-like direction and $L$ in the transverse spatial directions with $L^- \gg L$.\footnote{Note that the physical proper length along a light-like direction is of course always zero.} The computation of the expectation value of Wilson loops is well known from the early days of AdS/CFT \cite{Maldacena:1998im,Rey:1998ik} and requires computing the area of a string world-sheet ending on an appropriate contour in the boundary of AdS. The first calculations of the jet-quenching parameter in $\CN=4$ SYM were done in \cite{Liu:2006ug}, where the authors found that 
\begin{equation}
\hat{q}_{\CN =4 \, \text{SYM} } =  \frac{\pi^\frac{3}{2} \Gamma\left(\frac{3}{4}\right)}{\Gamma\left(\frac{5}{4}\right)} \, \sqrt{\lambda}\, T^3 \ .
\label{}
\end{equation}	
There is by now a large literature exploring aspects of jet quenching in various set-ups and we refer the reader to the original papers for more details.

The discussions so far have focussed on the dissipative aspects of the non-abelian plasma. As we have seen the AdS/CFT correspondence allows one to extract the basic properties of such media; specifically, by examining the classical energy loss by probes into black holes one recovers the frictional characteristics of such plasmas.

\subsubsection{Brownian motion in AdS/CFT}
\label{s:brownian}

Thus far we have focused on the classical motion of a probe particle in a thermal plasma medium. As expected, the particle loses its kinetic energy to the plasma and slows down. In the dual gravitational description of this frictional motion, this is mimicked by energy loss into the black hole horizon. However, given that we have a particle in a thermal medium, one should expect that it not only slows down, but also undergoes random motion due to the thermal fluctuations of the plasma. This is understood as Brownian motion and, as we have explained in 
\sec{ss:brown}, can be modeled in terms of a Langevin equation, which arises naturally in the context of kinetic theory. It is therefore inviting to ask whether these random fluctuations can be captured in the gravitation description, and if so, what is their physical origin? 

This question is not only interesting for academic reasons, but as described above, the holographic models provide useful toy examples to study the physics of the QGP formed in heavy-ion collisions. In that context,  the motion of an external quark in the QGP is assumed to be described by a relativistic Langevin equation \cite{Moore:2004tg}.  As reviewed in \sec{ss:brown}, in its most basic form the Langevin equation is parametrized by two constants: the friction (drag force) coefficient $\gamma$ and the magnitude of the random force
$\kappa$. However, in the relativistic case, the random force has different magnitudes $\kappa_T$ and $\kappa_L$ in the directions transverse and longitudinal to the quark momentum $p$, which in the non-relativistic limit $p\to 0$ become equal,
$\kappa_L=\kappa_T$.  The parameters $\gamma$ and $\kappa_L$ are related
to each other by the Einstein relation, under the assumption that the
Langevin dynamics holds and gives the J\"uttner distribution $e^{-\beta
E}$.  On the other hand, $\kappa_T$ is an independent parameter
\cite{Moore:2004tg, Gubser:2006nz}. Furthermore,  random force is usually
assumed to be white noise, i.e, with a delta function auto-correlation. 

One would therefore like to understand to what extent this picture works in the holographic setting. In early works on the subject, \cite{CasalderreySolana:2006rq, Gubser:2006nz,
CasalderreySolana:2007qw} computed the random force strength $\kappa_T$. The calculations of \cite{CasalderreySolana:2006rq, CasalderreySolana:2007qw} were carried out by expressing the change in the density matrix associated with the probe (and therefore $\kappa$) in terms of a Wilson loop average. This Wilson loop was taken to lie along a Schwinger-Keldysh contour and corresponds to a source of the random force operator. On the other hand, \cite{Gubser:2006nz} computes the random force strength by looking for fluctuations around the trailing string solution discussed in \sec{s:projectiles}. 

In \cite{deBoer:2008gu,Son:2009vu} the complete story for the holographic Brownian motion was worked out in detail. This was further generalized in \cite{Giecold:2009cg} to the relativistic Langevin equation. As we have alluded to earlier, classical string solutions in the black hole backgrounds have an induced world-sheet metric which has a horizon. In general, a world-sheet horizon could correspond to any of the following:\footnote{
Implicit in these statements is an assumption that one is working in static gauge, where the world-sheet time coordinate is identified with the timelike Killing field of the stationary asymptotically \AdS{d+1} black hole.
}
\begin{itemize}
\item to the bulk spacetime horizon, which is the case for static strings in static black hole spacetimes
\item to a bulk ergosurface, which occurs for a stationary string solution like the trailing string or in the case of a stationary black hole; the ergosurface occurs at the location where the asymptotic timelike Killing field has a vanishing norm, i.e. $g_{tt} =0$,
\item or simply when the string is forced to move too fast, which occurs for instance when it flares out while rotating at constant angular velocity as in \cite{Athanasiou:2010pv}.
\end{itemize}
The key physical result of \cite{deBoer:2008gu} was that the fluctuations of the classical string due to the curved spacetime particle production, i.e., Hawking radiation, associated with the world-sheet horizon was responsible for the random force on the probe particle. In fact, this connection was also noted in some of the earlier works on the subject, notably \cite{Gubser:2006nz} and \cite{Myers:2007we}.  Explicit expressions of this were worked on by assuming a thermal spectrum of the Hawking quanta in the case of BTZ black hole in \cite{deBoer:2008gu}, which had the advantage that the mode functions could be analytically obtained. Higher dimensional examples were also discussed in \cite{deBoer:2008gu} and  \cite{Son:2009vu}, the latter working directly with the maximal analytic extension of an eternal \SAdS{d+1} black hole. The extension of these results to the trailing string solution was described in \cite{Giecold:2009cg,CasalderreySolana:2009rm}. In addition \cite{Giecold:2009wi} examined the motion of a heavy quark in a toy model of a dynamical black hole, the so called conformal soliton geometry,\footnote{
The conformal soliton spacetime was first described in \cite{Friess:2006kw} and is simply the global \SAdS{d+1} spacetime considered in a single Poincar\'e patch. From the field theory point of view, it corresponds to a conformal mapping of  a thermal stress tensor on $\Sp^{d-1} \times \R$ to $\R^{d-1,1}$. We will have occasion to discuss this spacetime in the context of the fluid/gravity correspondence in \sec{ss:fast}.
} which we will revisit in \sec{ss:fast}. Similar studies for the case of an accelerating quark in the vacuum was recently considered in \cite{Caceres:2010rm}. Futhermore, in the authors of \cite{Gursoy:2009kk,Gursoy:2010aa} have examined the details of Langevin dynamics under the improved holographic QCD framework; in particular, these works compute the full Langevin correlator in non-conformal models, with the aim of using these results for studies of QGP. An important caveat to bear in mind for applications to QGP is that one generically has to consider the entire correlator and can not simplify the dynamics to the basic Langevin dynamics discussed in \sec{ss:brown}.

The simplest setting to study the stochastic motion is to monitor the fluctuations about the static straight string. For simplicity let us consider the planar \AdS{d+1} black hole \req{SAdSmet} and take the string world-volume to be $x^i = 0$. Expanding the fluctuations of the string to quadratic order about this solution, we obtain from the Nambu-Goto action \req{ngact}:
\begin{equation}
\CS_{NG} \approx -\frac{1}{4\pi\,\alpha'} \, \int d^2 \sigma \, \sqrt{-\gamma} \, \gamma^{\mu\nu} \, g_{ij}\, \frac{\partial x^i}{\partial \sigma^\mu} \, \frac{\partial x^j}{\partial \sigma^\nu}  
\label{}
\end{equation}	
where the induced metric $\gamma_{\mu\nu}$ in static gauge is
\begin{equation}
ds^2_\text{induced} = -r^2\, f(r) \, dt^2 + \frac{dr^2}{r^2\,f(r)} \ , \qquad f(r) = 1- \frac{r_+^d}{r^d} 
\label{}
\end{equation}	
and the kinetic terms of the scalars $x^i$ are determined by $g_{ij} = r^2\, \delta_{ij}$. The equation of motion for the transverse scalars on the probe string world-sheet then takes the form:
\begin{equation}
-\partial_t^2 x^i + f(r) \, \partial_r \, \left(r^4\, f(r) \, \partial_r x^i \right) = 0 
\label{}
\end{equation}	

This wave-equation for the non-minimally coupled transverse scalar $x(t,r)$ is solved by standard mode expansion $x^i(t,r) = e^{-i\,\omega t} \, \xi^i(r)$; in general $d$ this is somewhat complicated by the fact that the explicit mode solutions are not known. However, for $d = 3$, i.e., for BTZ black holes one can solve this wave equation explicitly. Analysis of the wave equation in the near-horizon region $r\simeq r_+$ reveals that the linearly independent solutions can be taken to be the ingoing and outgoing waves $\xi^{i(\pm)}(r)$. Thus one can generally write the solutions after implementing the appropriate boundary conditions on the horizon and at infinity in terms of a mode expansion:
\begin{eqnarray}
x^i(t,r) &=& \sum_{\omega >0} \, \left[a^i_\omega \, u^i_\omega(t,r) + (a^i_\omega )^\dagger\, u^i_\omega(t,r)^\star \right]\nonumber \\
u^i_\omega(t,r) &=& \CA \, \left(\xi^{i(+)}(r) + \CB \, \xi^{i(-)}(r)\right) \, e^{-i\,\omega\, t}
\label{}
\end{eqnarray}	
with the creation and annihilation operators $a_\omega$ and $a_\omega^\dagger$ satisfying the standard commutation relations: 
\begin{equation}
[a^i_\omega, a^j_{\omega'}] = [(a^i_\omega )^\dagger, (a^j_{\omega'} )^\dagger] = 0 , \qquad 
[a^i_\omega,(a^j_{\omega'})^\dagger] =  \delta_{\omega \omega'} \, \delta^{ij} 
\label{}
\end{equation}	

So far we have outlined how to write down the canonical quantized description of the fluctuations of the transverse excitations of the static string solution. The essential feature we now wish to exploit is that the asymptotic behaviour of the fluctuations (considering a regulated \AdS{} spacetime with cut-off at $r = R_c$)
\begin{equation}
x^i_\infty(t) = \lim_{r \to R_c} \, x^i(t,r) \ , 
\label{}
\end{equation}	
captures the motion of the probe quark in the boundary plasma. Monitoring the dynamics of $x^i_\infty(t)$ allows one to compute the details of the stochastic motion of the probe. In particular, one has a precise mapping between the correlation function of the quark in the boundary plasma and the correlators of the quantum operators $a_\omega$, i.e., 
\begin{equation}
\vev{a_{\omega_1}a_{\omega_2}^\dagger\dots}
\qquad \longleftrightarrow \qquad
\vev{x_\infty(t_1)\, x_\infty(t_2)\dots}
\label{}
\end{equation}	
This mapping in particular implies that, were we able to make a very precise measurement of the correlators of the end-point of the string, then one could in principle make a precise measurement of the state of the radiation emanating from the black hole (for after all the correlation function in the bulk depends on the state of the theory we are considering). 

However, if we are interested in understanding the stochastic process at a semi-classical level, all one needs to do is to evaluate the bulk correlation function in the natural Hartle-Hawking state for the fluctuating fields $x^i(r,t)$. It is a well known fact that the Hartle-Hawking state corresponds to thermal equilibrium and it therefore follows that the outgoing mode correlators are determined by the thermal density matrix
\begin{equation}
\rho = e^{-\beta \, H} \ , \qquad H = \sum_{\omega>0} \, \omega \, (a^i_\omega)^\dagger\, a^i_\omega
\label{}
\end{equation}	
Using this one can compute the mean squared displacement $\vev{:[x_\infty(t) -x_\infty(0)]^2:}$ (for simplicity monitoring only one of the scalar fields $x^i$, say $x^1 =x$). This allows one to extract the random force correlation function $\kappa(\omega)$. Similarly, one can compute the admittance $\mu(\omega)$ defined in \req{adm_def} by examining the behaviour of the string end-point when subject to an external force. As one expects the results confirm the fluctuation-dissipation theorem.  We refer the reader to \cite{deBoer:2008gu} for the precise expressions. It is more useful to record here the various time-scales obtained from the holographic computation:
\begin{equation}
 t_{\text{relax}} \sim \frac{\alpha'}{\RAdS^2} \, \frac{m}{ T^2}\ , \qquad 
t_{\text{coll}} \sim \frac{1}{T} \ , \qquad     t_{\text{mfp}} \sim 
\frac{1}{\sqrt{\lambda} \, T}  
\label{}
\end{equation}	
which in particular implies that 
\begin{equation}
 \frac{t_{\text{relax}}}{t_{\text{coll}}} \sim\frac{\alpha'\, m}{\RAdS^2 T}\sim {m\over\sqrt{\lambda}\,T}
 \ , \qquad 
\frac{t_{\text{coll}}}{t_{\text{mfp}}} \sim \sqrt{\lambda} \gg 1
\label{}
\end{equation}	
which clearly demonstrates the deviation from the hierarchy mentioned in \sec{ss:brown}. One  can understand this as a clear signal of the strongly coupled nature of the plasma dual to the black hole geometry, which obviates the wide separation of time-scales pertaining to kinetic theory. 

In \cite{Son:2009vu} the stochastic motion described above was derived using an extension of the Schwinger-Keldysh formalism. Instead of working with the static region of the Schwarzschild-AdS black hole, one considers maximally extended spacetime in Kruskal coordinates, and utilizes the Lorentzian AdS/CFT prescription of \cite{Herzog:2002pc} to obtain the retarded Green's functions. For instance the random force correlator $\kappa(t)$  is given in terms of the retarded Green's function:
\begin{equation}
\kappa(\omega)  = -(1+ 2\, n(\omega)) \, \text{Im} G_R(\omega) \ , \qquad 
n(\omega) = \frac{1}{1-\exp(\beta \omega)}
\label{}
\end{equation}	
This calculation leads to pretty much the same physical picture of the random motion as the one described above. There are some essential differences in the derivation, since the real time formalism employed in \cite{Son:2009vu} is intrinsically tied to the Lorentzian geometry, whereas the Hartle-Hawking state used in the derivation of \cite{deBoer:2008gu} is essentially an equilibrium thermal computation (which therefore can be carried out in the Euclidean black hole geometry). 

The description of the stochastic process described so far corresponds to the behaviour of the quark end-point on the boundary of the AdS spacetime (rather on a regulated boundary). An interesting question is to ask whether one can use this picture to further the membrane paradigm picture of the black hole.  The main philosophy of the membrane paradigm \cite{Thorne:1986iy} is that, as far as an observer staying outside a black hole horizon is concerned, physics can be effectively described by assuming that the
objects outside the horizon are interacting with an imaginary membrane,
which is endowed with physical properties, such as temperature and
resistance, and is sitting just outside the mathematical horizon.

In \cite{deBoer:2008gu, Son:2009vu} the stochastic properties of this fiducial membrane were explored, mainly by `integrating out' the string world-sheet all the way down to a stretched horizon located at $r = r_+ + \epsilon$. It was found that the heavy quark diffusion constant on this membrane was exactly the same as the asymptotic value \req{diff_const_AdSd}. This is in fact similar to the results on the hydrodynamic properties of the membrane (which we will revisit in \sec{s:flugra}) as discussed for instance in \cite{Iqbal:2008by}.

A crucial question about the membrane paradigm is to understand the microscopic structure of this stretched horizon in the context of string theory. 
In \cite{Susskind:1993ws, Halyo:1996vi}, Susskind and collaborators put
forward a provocative conjecture that a black hole is made of a
fundamental string covering the entire horizon.  Although this picture
must be somewhat modified \cite{Horowitz:1996nw} since we now know that
branes are essential ingredients of string theory, it is still an
attractive idea that, in the near horizon region where the local
temperature becomes string scale, a stringy ``soup'' or ``cloud'' of
strings and branes is floating around, covering the entire horizon. Ref \cite{deBoer:2008gu} and more recently \cite{Atmaja:2010uu} have attempted to find precise characteristics of this membrane by trying to use the characteristics of the stochastic motion of strings ending on the stretched horizon. In particular, one could imagine the string IR end-point interacting with the membrane halo of the black hole.  Any such model requires that we be able to match the mean free time between the interactions of the string end-point with the halo with the results derived above for the stochastic motion. The calculations in  \cite{deBoer:2008gu, Atmaja:2010uu} seem to suggest a non-trivial dependence of the interaction probability on $g_s$ and $\ell_s$. 

\section{Fluid/gravity correspondence}
\label{s:flugra}

Thus far our understanding of physics out of equilibrium has been restricted to the linear response regime. In the holographic set-up this regime is captured by the study of linearized fluctuations about AdS black hole backgrounds. We have in particular seen that the retarded correlators of local gauge invariant operators can be efficiently computed in the classical gravity approximation and the results agree with general expectations from the field theory. 

A particular application of the linear response story is the use of Kubo formulae to extract transport coefficients of strongly coupled gauge theory plasmas. As described earlier, these studies led  to the fascinating bound on the ratio $\eta/s$. A natural question is whether one can derive non-linear hydrodynamics from the holographic set-up, by relaxing the constraints of linearized fluctuations. A-priori this sounds hard, for the problem surely translates to understanding non-linear fluctuations around a black hole background. Nevertheless, as we now shall describe, this is indeed tractable and leads us to a natural relation between Einstein's equations and a relativistic generalization of the Navier-Stokes equations, which has come to be known as the {\it fluid/gravity correspondence}, \cite{Bhattacharyya:2008jc}.

The key feature we will exploit in making this connection is the general idea that 
the systems in local thermal equilibrium should in a suitable infra-red limit admit a hydrodynamic description.  As familiar from everyday experience, this framework can accommodate systems with large time-dependence, as long as they are in {\it local} thermodynamic equilibrium; proximity to global thermodynamic equilibrium is no longer required.

More specifically, fluid dynamics is the continuum effective description of any (interacting) microscopic quantum field theory. In order to meaningfully describe the system in terms of the fluid variables, the fluid description requires that the system achieves local thermodynamic equilibrium.  This means that the regime of validity where such a description is valid requires that the scale of variation of the dynamical degrees of freedom, $L$, be much larger than the microscopic scale $\ell_{\mathrm{mfp}}$, typically set by the temperature, $T$ (or the local energy density).  In this {\it long-wavelength approximation}, local equilibrium then demands that $ L \, T \equiv \frac{1}{\epsilon} \gg 1$.

To keep the discussion and formulas as clean as possible, we will restrict attention to the simplest case of uncharged conformal fluid on four-dimensional Minkowski spacetime ${\bf R}^{3,1}$ (though when sufficiently compact, we will quote the $d$ dimensional results).\footnote{
The most comprehensive discussion of conformal fluids in arbitrary dimensions can be found in \cite{Bhattacharyya:2008mz}.}
Several generalizations to this setup will be mentioned in \sec{flugra_generaliz}.  For basic details on fluid dynamics we refer the reader to \cite{Landau:1965uq}. Relativistic fluids are described in \cite{Andersson:2007gf},  aspects of the fluid/gravity correspondence are reviewed in \cite{Rangamani:2009xk,Hubeny:2010fk} and applications of relativistic fluids to heavy-ion collisions in \cite{Romatschke:2009im}. 

Before proceeding, we make few remarks on notation: the bulk metric will be denoted by $g_{MN}$ with the capital Latin indices taking values over the $d+1$ bulk dimensions; we will separate the coordinates into the radial coordinate $r$ and the remaining `boundary coordinates' $x^{\mu}$, where the $\mu$ index ranges over the $d$ boundary directions (which includes time).   The stress tensor in the boundary theory is denoted $T^{\mu\nu}$, and in writing its conservation ($\nabla_\mu \, T^{\mu \nu} = 0$), the  $\nabla_\mu$ is the covariant derivative with respect to the boundary metric $\eta_{\mu\nu}$.\footnote{
Oftentimes the reader will encounter an ordinary partial derivative $\partial_\mu$ in the following: this is because in Cartesian coordinates of flat spacetime, $\nabla_\mu = \partial_\mu$.
However, this is convenient more generally as well: since the long-wavelength approximation engenders an ultra-locality in the equations one solves (along the boundary directions), we can use boundary derivatives as just ordinary partials; however, our expressions can of course be easily `covariantized' by replacing $\partial_\mu$ by $\nabla_\mu$. We hope this does not cause much confusion.}

\subsection{Background} 
\label{s:fgbackground}

We begin  with a brief review of conformal fluid dynamics, proceed to discuss the dual gravitational solutions, and then motivate the construction of the explicit mapping between them.  

\subsubsection{Conformal fluid dynamics} 

A conformal fluid is characterized by a traceless symmetric stress tensor, which in $d$ spacetime dimensions has $\frac{d(d+1)}{2} -1$ degrees of freedom, along with a collection of charge currents (which for simplicity we have set to zero).  In a fluid dynamical characterization of the same system, the number of basic degrees of freedom is drastically reduced.  The conformal invariance fixes the equation of state, thereby determining the pressure in terms of the energy density, which can in turn be expressed in terms of the temperature.
Hence the basic variables are the local temperature $T(x)$ and velocity $u_\mu(x)$ (unit-normalized so that $u_\mu \, u^\mu = -1$) , which constitute just $d$ degrees of freedom.  

The equations of fluid dynamics are then simply the equations of local conservation of the stress tensor (as well as the charge currents in more general situations), supplemented by constitutive relations that express these currents as functions of the fluid dynamical variables. As fluid dynamics is a long wavelength effective theory, such constitutive relations are usually specified in a derivative expansion, like in any effective field theory. At any given order, thermodynamics plus symmetries determine the form of this expansion up to a finite number of undetermined coefficients. In general, the coefficients can be obtained either from measurements or from microscopic computations.  However, as we will see, in the present framework these coefficients are fully determined by the gravity side (which in a sense knows about the microscopics of the boundary field theory).

Purely  based on the symmetries, we can then write down an expression for the stress tensor of a $d$-dimensional conformal fluid, which is a local functional of the temperature and velocity fields:
\begin{equation}
T^{\mu \nu} = \alpha \, T^d\, \left(\eta^{\mu\nu} + d \, u^\mu \,u^\nu\right)  + \pi^{\mu \nu}_{\mathrm{dissipative}} \ .
\label{schematicTmunu}
\end{equation}	
The first two terms describe the ideal conformal fluid stress tensor, while $\pi^{\mu \nu}_{\mathrm{dissipative}}$ incorporates all the dissipative terms. $\alpha$ here sets the overall normalization of the stress tensor. As variations of $T(x)$ and $u_\mu(x)$  are small, we can expand $\pi^{\mu \nu}$ in a derivative expansion 
$\partial_\mu \equiv \frac{\partial}{\partial x^{\mu}}$ in the boundary directions; the leading term will turn out to be proportional to the shear viscosity. The dynamical content of the fluid equations is encoded in the conservation of the stress tensor
\begin{equation}
\nabla_{\! \mu} \, T^{\mu \nu} = 0 \ .
\label{Tconserv}
\end{equation}
Fluid dynamics viewed in this derivative expansion constructs an effective field theory for the slowly varying modes $T(x)$ and $u_\mu(x)$, analogously to the 
chiral Lagrangian for pions.

In general one should write down all possible terms at a given order in the derivative expansion consistent with the symmetries of the problem (and modulo lower order conservation equations). There are scalar functions of the thermodynamic variables multiplying these operators: these are the transport coefficients. For instance, we have indicated in \req{schematicTmunu} the explicit terms occurring at zeroth order (the ideal fluid terms). At the next order, there is one term for a conformal fluid (proportional to the shear viscosity), while a non-conformal fluid would have two. At higher orders we get more terms; a discussion of terms allowed at second order was first undertaken in \cite{Baier:2007ix} in four dimensions and  higher dimensional conformal fluids were discussed in \cite{Haack:2008cp,Bhattacharyya:2008mz} and a nice account of non-conformal fluids can be found in \cite{Romatschke:2009kr}. We should note that conformal fluids are especially simple, as Weyl covariance provides a very useful tool to classify the possible terms in the fluid stress tensor. Using a manifestly Weyl-covariant formalism \cite{Loganayagam:2008is}, many of the expressions derived in \cite{Bhattacharyya:2008jc} simplify. We refer the reader to 
\cite{Bhattacharyya:2008mz} for compact expressions in diverse dimensions. 

\subsubsection{Gravity in the bulk} 

We now turn to the gravitational solutions in asymptotically AdS spacetime. Motivated by the AdS/CFT correspondence, we will consider two-derivative theories of gravity with an AdS$_{d+1}$ ``vacuum'', such as the IIB SUGRA on AdS$_5 \times \Sp^5$.  As mentioned in \sec{s:regimes}, the solution space has a universal sub-sector, pure gravity with negative cosmological constant, for which the bulk field equations are simply Einstein's equations,
\begin{equation}
 E_{MN} \equiv R_{MN} - \frac{1}{2} \, R \, g_{MN} + \Lambda  \, g_{MN} =0 \ . 
 \label{Eeq}
 \end{equation}
(Note that taking $R_{\mathrm{AdS}} = 1$ sets $\Lambda = -6$ in five dimensions.)
We will focus on this sub-sector in the long-wavelength limit.
Apart from the pure AdS$_5$ solution, there is a 4-parameter family of solutions representing asymptotically-AdS$_5$ boosted planar black holes.
We will use these solutions to construct general dynamical spacetimes characterized by fluid-dynamical configurations.

Roughly speaking, the fluid/gravity construction may be thought of as a ``collective coordinate method" for black hole horizons. Recall that the isometry group of AdS$_5$ is $SO(4, 2)$. The Poincare algebra plus dilatations form a distinguished sub-algebra of this group (one that preserves the boundary). Out of these, the  $SO(3)$ rotations and translations in world-volume ${\bf R}^{3,1}$ leave the static planar AdS black hole invariant, but the remaining symmetry generators, dilatations and boosts, act nontrivially on this solution, generating a 4-parameter family of boosted planar black holes, parameterized by the temperature $T$ and the velocity $u^{\mu}$ of the brane. The  construction effectively promotes these parameters to ÔGoldstone fieldsÕ (or collective coordinate fields) $T(x)$ and  $u^{\mu}(x)$, and determines their dynamics, order by order in the boundary derivative expansion.  Note that this is distinct from linearization: we make no assumptions about the amplitudes of these slow variations. 

\subsubsection{The fluid/gravity map} 

Before proceeding to sketch the construction in more detail, we pause to stress an important point in mapping these long-wavelength gravity solutions to corresponding fluid configurations.  A well-known procedure of holographic renormalization (see e.g.\ \cite{Balasubramanian:1999re} and \cite{deHaro:2000xn}) links the boundary stress tensor to the behaviour of the bulk metric near the AdS boundary.
Given any asymptotically AdS spacetime, we can read-off the induced stress tensor on the boundary, since the latter is related to the normalizable modes of the gravitational field in AdS.  In particular, expanding the bulk metric in the Fefferman-Graham form near the boundary $z=0$,
$$ ds^2 = \frac{dz^2 + (\eta_{\mu\nu} + \alpha \, z^d \, T_{\mu\nu}) \, dw^\mu \, dw^\nu}{z^2} \ , $$
the stress tensor is simply given by $T_{\mu\nu}$.
Conversely, given a boundary stress tensor, there is a procedure to holographically reconstruct the bulk metric in a radial expansion around the boundary.  

Naively, this might seem puzzling:  as mentioned above, a conformally invariant stress tensor in $d$ dimensions has $\frac{d(d+1)}{2}-1$ degrees of freedom.  If any such stress tensor yielded a regular bulk spacetime, we would have a discrepancy between the fluid side which has only $d$ degrees of freedom and whose dynamics is correspondingly specified by only $d$ equations, and the gravity side that would seemingly allow more degrees of freedom.  In other words, passing from a generic quantum conformal field theory stress tensor to the stress tensor of its effective description in terms of fluid dynamics constitutes a drastic reduction in the number of degrees of freedom required to specify the spacetime.  How is this manifested in the bulk?  The answer lies in regularity.  As a series expansion around the boundary, the holographic reconstruction cannot guarantee that the metric does not become nakedly singular at some finite radial value in the bulk.  In fact, for a generic stress tensor it will.  The fluid/gravity construction demonstrates that the regular solutions are given  precisely by such stress tensors which are fluid dynamical.  Moreover, we claim that the gravity solutions thus constructed are the most general regular long-wavelength\footnote{
Note, however, that there are regular solutions which do not fall into the long-wavelength category, such as small black holes in AdS, which correspondingly are not described by fluid configurations.
} solutions to EinsteinÕs equations with negative cosmological constant.  They typically correspond to deformed and dynamical black holes; i.e.\ the solutions admit a regular event horizon which shields a curvature singularity.

The heuristic picture of a generic evolution, on the two sides of the fluid/gravity correspondence, is as follows.
Suppose we start with some generic high energy initial conditions.  On the CFT side, the system quickly settles down to local thermodynamic equilibrium, whose bulk dual is described by a dynamical, non-uniform (planar) black hole.  On both sides, such configuration is described by local velocity and temperature fields which exhibit slow variation in the boundary directions.
The subsequent evolution is described by equations of fluid dynamics on the boundary, which originate from Einstein's equations governing the bulk evolution.  Finally, at late times, the system relaxes to global thermal equilibrium, given by a stationary state parameterized  by a constant temperature and velocity.  In the bulk, this is one of the well-known stationary solutions describing a planar black hole in AdS mentioned in the previous subsection and given explicitly below.

The fluid/gravity construction specifically utilizes the fact that fluid dynamics is a long-wavelength effective theory. One writes Einstein's equations as a perturbative expansion in boundary derivatives (however keeping the exact radial dependence) to emphasize the expansion at small momenta.  This allows us to solve the equations order by order in this boundary derivative expansion.  In turns out that Einstein's equations at a given order implement the fluid stress tensor conservation equations at lower order.  Therefore, order by order, we can use the lower-order fluid dynamical solution to construct the bulk metric, and then read off the corrected fluid dynamical stress tensor.  In \cite{Bhattacharyya:2008jc}, the boundary stress tensor $T^{\mu\nu}$ and corresponding bulk metric $g_{MN}$ were constructed to second order in the boundary derivative expansion.  This yields a map between fluid dynamics and gravity, which we now proceed to sketch in more detail.

\subsection{Iterative construction of bulk metric and boundary stress tensor}
\label{hubConstruction}

The iterative procedure starts with the zeroth order configuration, corresponding to the global equilibrium, given in (\ref{SAdSzero}).  As will shortly become clear we will implement a perturbation expansion about this solution. In order for perturbation theory to make sense, it is important that we start with a seed metric in manifestly regular coordinates.

A general fluid configuration in local, but not global, equilibrium can be described by promoting the $4$ parameters $T$ and $u^\mu$ to physical fields dependent on the boundary coordinates $x^\mu$, i.e., to $T(x)$ and $u^\mu(x)$.  If these fields vary slowly compared to the microscopic scale $\ell_{\mathrm {mfp}}$, i.e.\ if
$$ \frac{\partial_\mu  \log T}{T} \sim \mathcal{O}(\epsilon) \ , \qquad \frac{\partial_\mu u}{T} \sim \mathcal{O}(\epsilon) $$ 
for small $\epsilon$,
the fluid configuration still satisfies the conditions of local equilibrium. In each local domain of slow variation, which we refer to as tube, the bulk gravitational solution is approximately that of a uniform black brane.  Remarkably, the bulk solution can be constructed by patching together these tubular domains!  Of course, if we just replace $u_\mu$ and $T$ in the metric (\ref{SAdSzero}) by $T(x)$ and $u^\mu(x)$, the resulting metric (call it $g^{(0)}_{MN}$) will no longer solve Einstein's equations  (\ref{Eeq}).  Instead, the metric $g^{(0)}_{MN}$ will need to be corrected by higher-order piece ($g^{(1)}_{MN}$, etc.), which we can obtain iteratively as an expansion in $\epsilon$.  We will find that the resulting corrected metric can be constructed systematically to any desired order, and is valid well inside the event horizon, thus allowing verification of its regularity.  It is worthwhile to stress that the success of such a procedure rests on the fact the our seed metric $g^{(0)}_{MN}$ is manifestly regular on the horizon, since otherwise the expansion would break down near the coordinate singularity at horizon.

To implement the construction algebraically, we express the line element in a boundary derivative expansion of the fields $u_\mu(x)$ and $T(x)$, and use $\epsilon$ as a book-keeping parameter (counting the number of $x^{\mu}$ derivatives):
\begin{equation}
 g_{MN} = \sum_{k=0}^\infty \, \epsilon^k \, g^{(k)}_{MN} \ , \qquad
 T =\sum_{k=0}^\infty \, \epsilon^k \, T^{(k)}  \ , \qquad
u_{\mu} =\sum_{k=0}^\infty \, \epsilon^k \, u_{\mu}^{(k)} \ . 
\label{expand}
\end{equation}
The term $g^{(k)}_{MN}$ corrects the metric at the  $k^{{\rm th}}$ order, such that  Einstein's equations will be satisfied to ${\mathcal O}(\epsilon^k)$ provided the functions $T(x)$ and $u^\mu(x)$ obey a certain set of equations of motion, which turn out to be precisely the stress tensor conservation equations of boundary fluid dynamics at ${\mathcal O}(\epsilon^{k-1})$.

Specifically, we can obtain the equations for $g^{(k)}_{MN}$ by substituting the expansion (\ref{expand}) into Einstein's equations (\ref{Eeq}), and extracting the coefficient of  ${\mathcal O}(\epsilon^k)$.  Schematically, these take the form
\begin{equation}
{\mathcal H}\left[g^{(0)}(u^{(0)}_\mu, T^{(0)})\right] g^{(k)}(x^\mu ) = s_k  
\label{schemEeq}
\end{equation}
where ${\mathcal H}$ is a second-order linear differential operator in the variable $r$ alone and $s_k$ are regular source terms which are built out of $g^{(n)}$ with $n \le k-1$.
 Since $g^{(k)}(x^\mu ) $ is already of ${\mathcal O}(\epsilon^k)$, and since every boundary derivative appears with an additional power 
of $\epsilon $, ${\mathcal H}$ is an ultralocal operator in the field theory directions. 
 Moreover, at a given $x^{\mu}$, the precise form of this operator ${\mathcal H}$ depends only on the local values of $T$ and $u^\mu$ but not on their derivatives at $x^{\mu}$.  Furthermore, the operator ${\mathcal H}$ is independent of $k$; we have the same homogeneous operator at every  order in perturbation theory. 
 This allows us to find an explicit solution of (\ref{schemEeq}) systematically at any order.
The source term $s_k$ however gets more complicated with each order, and reflects the nonlinear nature of the theory. 

Bit more explicitly, the equations of motion split up into two kinds:
Constraint equations, $E_{r \mu} = 0$ which implement stress-tensor conservation (at one lower order), and Dynamical equations $ E_{\mu\nu} = 0$ and $E_{rr} =0$  which allow determination of $g^{(k)}$.
We solve the dynamical equations
$$g^{(k)} = {\rm particular}(s_k) + {\rm homogeneous}({\mathcal H})$$ 
subject to regularity in the interior and asymptotically AdS boundary conditions.  
Using the rotational symmetry group of the seed solution (\ref{SAdSzero}) it turns out to be possible to make a judicious choice of variables such that the operator ${\mathcal H}$ is converted into a decoupled system of first order differential operators. It is then simple to solve the equation (\ref{schemEeq}) for an arbitrary source $s_k$ by direct integration.   For the details of the procedure, as well as discussion of convenient gauge choice,
 etc.,  we refer the reader to the original work \cite{Bhattacharyya:2008jc}  or the review \cite{Rangamani:2009xk}.

Instead, here we simply quote the result for the bulk metric and boundary stress tensor, corrected to first order in $\epsilon$.
To first order the bulk metric takes the form
\begin{eqnarray}
ds^2 &=&-2\, u_{\mu}\, dx^{\mu} dr 
+ r^2\, \left( \eta_{\mu\nu} + [1-f(r/\pi T)] \, u_{\mu}\, u_{\nu} \right) \, dx^{\mu}dx^{\nu} \nonumber \\
&+& 2r \left[ { r \over \pi T} \, F(r/\pi T)\, \sigma_{\mu\nu} +{1\over 3} \, u_{\mu}u_{\nu} \,\partial_{\lambda} u^{\lambda}  -  {1\over 2}\, u^{\lambda}\partial_{\lambda}\left(u_\nu u_{\mu}\right)\right] \, dx^{\mu} dx^{\nu} ,
\label{metfirstO}
\end{eqnarray}
where $T(x)$ and $u_\mu(x)$ are any slowly-varying functions which satisfy the conservation equation (\ref{Tconserv}) for the zeroth order ideal fluid stress tensor (\ref{TzeroO}), the function $F(r)$ is given by 
$$
F(r) \equiv \int_r^{\infty}\, dx \,\frac{x^2+x+1}{x (x+1) \left(x^2+1\right)} ={1\over 4}\, \left[\ln\left(\frac{(1+r)^2(1+r^2)}{r^4}\right) - 2\,\arctan(r) +\pi\right] ,
$$
and $\sigma^{\mu\nu}$ is the transverse traceless symmetric part of $\partial^{\mu} u^{\nu}$ called shear, i.e.\
$$\sigma^{\mu\nu}= P^{\mu \alpha} P^{\nu \beta} \, 
\, \partial_{(\alpha} u_{\beta)}
-\frac{1}{3} \, P^{\mu \nu} \, \partial_\alpha u^\alpha \ . $$
Note that the first line of (\ref{metfirstO}) is simply the zeroth order (boundary-derivative-free) solution (\ref{SAdSzero}), whereas each of the terms in the second line have exactly one boundary derivative.\footnote{
Note that (\ref{metfirstO}) does not have any $\partial_{\mu} T$ terms appearing explicitly, since by implementing the zeroth order stress tensor conservation, we have expressed the temperature derivatives in terms of the velocity derivatives.}

The induced fluid stress tensor on the boundary, which can be easily obtained\footnote{
One can compute the stress tensor using the general formula \req{bdysten}, which can be reduced in the current case  to the simpler formula: 
$$T^\mu_\nu = -2 \, \lim_{r \to \infty} \, r^4 \, (K^\mu_\nu - \delta^\mu_\nu ) \ .$$}
 from  the bulk metric (\ref{metfirstO}), is given by 
\begin{equation}
T^{\mu \nu} =\pi^4 \, T^4
\left( 4\,  u^\mu u^\nu +\eta^{\mu \nu}
\right)  - 2 \,\pi^3 \, T^3 \, \sigma^{\mu \nu} .
\label{TfirstO}
\end{equation}
Here the first two (derivative-free) terms describe a perfect fluid with pressure (or negative free energy density) $\pi^4 \, T^4$, and correspondingly (using thermodynamics)  entropy density  $s=4 \, \pi^4 \, T^3$.  The shear viscosity $\eta$ of this fluid may be read off from the coefficient of $\sigma^{\mu\nu}$ and is given by $\pi^3 \, T^3$. 
Notice that $\eta/s = 1/(4 \pi)$, in agreement with the well-known result of \cite{Policastro:2001yc}.

\subsection{Solution at second order}
\label{s:Solution}

In the previous section we have illustrated at first order how the iterative procedure can be implemented (in principle systematically to any order) to construct a generic long-wavelength solution.  Such a procedure was carried out in \cite{Bhattacharyya:2008jc}, where the bulk metric and boundary stress tensor was calculated explicitly to second order in the boundary derivative expansion.  In this section we will discuss the new physics which can be extracted from such a construction.

Note that already at first order, the bulk metric (\ref{metfirstO}) was a much lengthier expression than the boundary stress tensor (\ref{TfirstO}).  This remains true in general; in fact, already at second order the expression for the metric is far too unwieldy to write down here.  In the following subsections, we will therefore only write the second order boundary stress tensor explicitly but indicate the bulk metric only schematically.

\subsubsection{The 4-dimensional conformal fluid from AdS$_5$} 
\label{ss:stressT}

The second order stress tensor obtained from the gravity analysis is best expressed in terms of Weyl invariant operators.  To do so it is useful to  classify all the operators which are Weyl invariant at various orders in the derivative expansion. At first  two orders in derivatives the set of  symmetric traceless tensors which transform homogeneously under Weyl rescalings are given to be\footnote{
Various papers in the literature seem to use slightly different conventions for the normalization of the operators. We will for convenience present the results in the normalizations used initially in  \cite{Baier:2007ix}. This is the  source of the  factors of $2$ appearing in the definition of the tensors ${\mathfrak T}_i$, see \cite{Bhattacharyya:2008jc} for a discussion.
The derivative $\CD_\mu$ is a Weyl covariant derivative introduced in \cite{Loganayagam:2008is} which makes it easy to keep track of conformal transformation properties of various tensors.}
\begin{equation}
\begin{aligned}
\text{First order}: \;\;&  \sigma^{\mu\nu}  &\nonumber \\
\text{Second order}: \;\; & {\mathfrak T}_1^{\mu\nu} =2\, u^\alpha \, \CD_\alpha \sigma_{\mu\nu} \ , \quad 
 {\mathfrak T}_2^{\mu\nu} =C_{\mu\alpha\nu\beta}\,u^\alpha \,u^\beta \ , \nonumber \\
 \qquad&{\mathfrak T}_3^{\mu\nu}  =4\,\sigma^{\alpha\langle\mu}\, \sigma^{\nu\rangle}_{\ \alpha}  \ , \quad 
  {\mathfrak T}_4^{\mu\nu} =  2\, \sigma ^{\alpha\langle\mu}\, \omega^{\nu\rangle}_{\ \alpha}  \ , \quad  {\mathfrak T}_5^{\mu\nu}=   \omega ^{\alpha\langle\mu}\, \omega ^{\nu\rangle}_{\ \alpha} 
\end{aligned}
\label{winv2der}
\end{equation}	
where we have introduced a notation for the second derivative operators which will be useful to write compact expressions for the stress tensor below. The quantities involved in the operators above are constructed from the velocity derivatives, such as the acceleration $a^\mu$, shear $\sigma^{\mu \nu}$, etc.. These quantities are defined  using the decomposition of the 4-velocity gradient $\nabla^\nu u^\mu$ into transverse, traceless and trace parts, 
$$
\nabla ^{\nu}u^{\mu} 
= - a^{\mu} \, u^{\nu} 
+ \sigma^{\mu \nu} 
+ \omega^{\mu \nu} 
+ \frac{1}{3} \, \theta \, P^{\mu\nu} \ ,
$$
where expansion, acceleration, and vorticity, are respectively defined as:
$$
\theta = \nabla_{\mu}u^{\mu} \ , \qquad
a^{\mu} = u^\nu\nabla_\nu u^{\mu}  \ , \qquad
\omega^{\nu\mu} = P^{\mu \alpha} \, P^{\nu \beta} \, \nabla_{[\alpha} u_{\beta]} \ .
$$
In addition in four spacetime dimensions we also have the `curl' of the velocity field
\begin{equation}
\begin{split}
\l^{\mu} &=  u_{\alpha}\,\epsilon^{\alpha \beta \gamma \mu}
\nabla_\beta u_{\gamma} ,
 \end{split}
 \label{fderdefs}
\end{equation}

 Armed with this data  we can immediately  write down the general contribution to the stress tensor as:
\begin{eqnarray}
\Pi_{(1)}^{\mu\nu} &=& -2\, \eta\,\sigma^{\mu\nu} \nonumber \\
\Pi_{(2)}^{\mu\nu} &=&  \tau_\pi\,\eta\, {\mathfrak T}_1^{\mu\nu} + \kappa\,{\mathfrak T}_2^{\mu\nu} + \lambda_1\, {\mathfrak T}_3^{\mu\nu} + \lambda_2\, {\mathfrak T}_4^{\mu\nu} + \lambda_3\, {\mathfrak T}_5^{\mu\nu} \  .
\label{fld2}
\end{eqnarray}	
There are therefore have six transport coefficients $\eta$, $\tau_\pi$, $\kappa$, $\lambda_i$ for $i = \{1,2, 3\}$, which characterize the flow of a non-linear viscous fluid.

For a fluid with holographic dual using the  result of the gravitation solution (the procedure to construct which is described in \sec{hubConstruction}) one finds explicit values for the transport coefficients. In particular, for the $\CN=4$ SYM fluid one has\footnote{The result for generic $\CN =1$ superconformal field theories which are dual to gravity on \AdS{5} $\times \CX_5$ are given by simply replacing $\frac{N^2}{8 \pi^2}$ by the corresponding central charge of the SCFT.} \cite{Baier:2007ix,Bhattacharyya:2008jc}
\begin{equation}
\begin{aligned}
&\eta = \frac{N^2}{8 \pi} \, \left(\pi  T\right)^3 \;\;  \thus \;\;\frac{\eta}{s} = \frac{1}{4\pi}  \ , \\
&\tau_\pi = \frac{2 - \ln 2}{2\pi\, T} \ , \qquad \kappa = \frac{\eta}{\pi \,T} \\
& \lambda_1 = \frac{\eta}{2\pi\, T} \ , \qquad
\lambda_2 = \frac{\eta\, \ln 2}{\pi\, T}\ , \qquad \lambda_3=0 \ . 
\end{aligned} 
\label{transp4d}
\end{equation}	
where in the first line we have used the standard entropy density for thermal $\CN =4$ SYM at strong coupling $s = \frac{\pi^3}{2}\,N^2 \, T^3$ to exhibit the famous ratio of shear viscosity to entropy density \cite{Kovtun:2004de}.  

So far we have only discussed the behaviour of four dimensional relativistic fluids which have bulk duals as asymptotically \AdS{5} black hole spacetimes. The story is readily generalized to other dimensions $d \ge 3$. The general analysis in \AdS{d+1} of carried out in \cite{Haack:2008cp, Bhattacharyya:2008mz} in fact allows us to write down the transport coefficients described above in arbitrary dimensions in nice closed form.\footnote{For $d=3$ the general results were initially derived in \cite{VanRaamsdonk:2008fp}. See also \cite{Natsuume:2008iy} for determination of some of these coefficients in $d =3$ and $d =6$.} The transport coefficients for conformal fluids in $d$-dimensional boundary $\CB_d$ are:
\begin{equation}
\begin{aligned}
&\eta = \frac{1}{16 \pi\, G_N^{(d+1)}} \, \left(\frac{4\pi}{d}\,   T\right)^{d-1} \;\;  \thus \;\;\frac{\eta}{s} = \frac{1}{4\pi}  \ , \\
&\tau_\pi = \frac{d}{4\pi\, T} \, \left[1 + \frac{1}{d}\, \text{Harmonic}\left(\frac{2}{d} -1\right) \right] , \qquad \kappa = \frac{d}{2\pi\, (d- 2)} \,\frac{\eta}{T} \\
& \lambda_1 = \frac{d}{8\pi}\, \frac{\eta}{T} \ , \qquad
\lambda_2 = \frac{1}{2\pi}\,  \text{Harmonic}\left(\frac{2}{d} -1\right) \, \frac{\eta}{T}\ , \qquad \lambda_3=0 \ . 
\end{aligned} 
\label{transpgend}
\end{equation}	
where $\text{Harmonic}(x)$ is the harmonic number function.\footnote{The harmonic number function may be in fact be re-expressed in terms of the digamma function, or more simply as 
$$\text{Harmonic}(x) = \gamma_e + \frac{\Gamma'(x)}{\Gamma(x)} \ , $$ 
where $\gamma_e$ is Euler's constant.}

\subsubsection{The spacetime geometry dual to fluids} 
\label{s:flugeom}

\begin{figure}
\begin{center}
\includegraphics[width=3in]{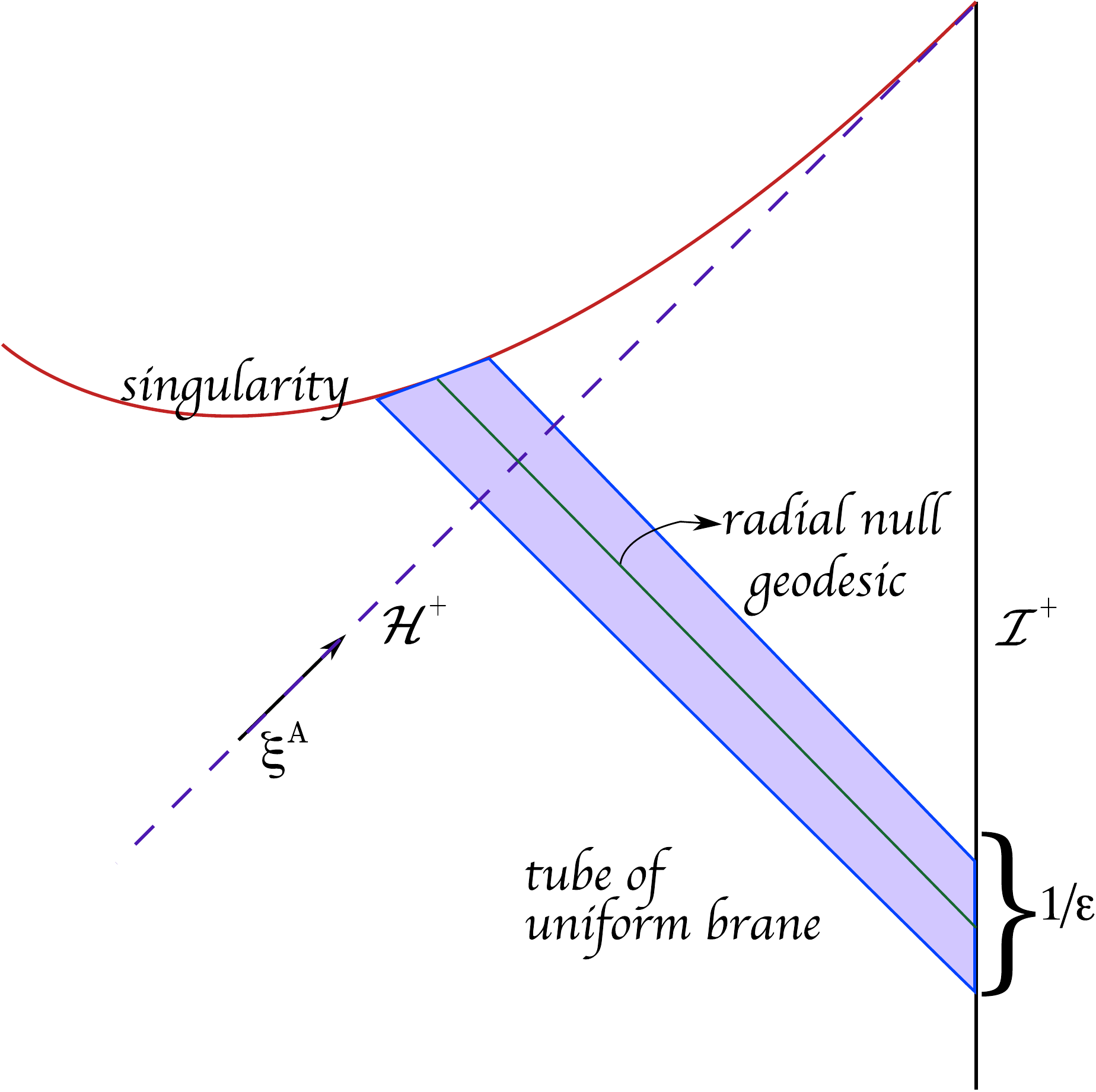}
\caption{The causal structure of the spacetimes dual to fluid mechanics illustrating the tube structure. The dashed line denotes the future event horizon $\CH^+$ generated by $\xi^A$, while the shaded tube indicates the region of spacetime over which the solution is well approximated by a tube of the uniform black brane.}
\label{f:tubes}
\end{center}
\end{figure}

Let us now turn to discuss the bulk geometry obtained at the second order in boundary derivative expansion (whose first order part is given by  (\ref{metfirstO})).
As mentioned previously, this bulk solution is `tubewise' approximated by a planar black hole. This means that in each tube, defined by a small neighborhood of given $x^{\mu}$, but fully extended in the radial direction $r$, the radial dependence of the metric is approximately that of a boosted planar black hole at some temperature $T$ and horizon velocity $u^\mu$.  These parameters vary from one position $x^{\mu}$ to another in a manner consistent with fluid dynamics.   Our choice of coordinates is such that each tube extends along an ingoing radial null geodesic; see \fig{f:tubes}. Apart from technical advantages, this is conceptually rather pleasing, since it suggests a mapping between the boundary and the bulk which is natural from causality considerations. 

It is worth stressing that although we refer to the metric written in (\ref{metfirstO}) and its second order extension presented in \cite{Bhattacharyya:2008jc} as ``a solution" in the singular form rather than the plural, these expressions actually correspond to not just a single solution or even a finite family of solutions, but rather a continuously infinite family of solutions, specified by the four functions $T(x)$ and $u^{\mu}(x)$ of four variables.  The flip side of the coin is that while very general, such a metric is not fully explicit:  in order to be so, we need to use a given solution to fluid dynamics as input.

However, even in the absence of the explicit functional dependence of $T(x)$ and $u^{\mu}(x)$, it is possible to extract certain salient features of any such geometry.
The most important feature of our geometry is the presence of an event horizon.  In \cite{Bhattacharyya:2008xc} we have demonstrated explicitly that the event horizon is regular, and determined its location in terms of the functions $T(x)$ and $u^{\mu}(x)$.  Here we only schematically motivate these results.  Intriguingly, it turns out that the location of the event horizon $ r_H(x^\mu)$ in the bulk is determined locally by the fluid dynamical data at a point $x^\mu$ (within the derivative expansion), rather than globally as usual in general relativity.

To motivate this physically, within each tube characterized by a given $x^\mu$, the position of the horizon is approximately at  $r_H \approx \pi \, T $ corresponding to that tube.  Since $T$ varies as a function of $x^\mu$, so will the horizon position $r_H(x^\mu)$.  In the corrected solution, the surface $ r  = \pi \, T(x)$ is not the event horizon (for example, it is not a null surface in general), but if the variation $T(x)$ is slow, the deviation from the true event horizon is likewise small.

One can determine the position of the event horizon within our perturbation scheme using  the fact that the solution settles down at late times to a uniformly boosted planar black hole.
In particular, if we expand the horizon location as a series in boundary derivatives, tagged as before by $\epsilon$, 
$$ r = r_H(x) = \pi \, T(x) + \sum_{k=1}^\infty \, \epsilon^k \, r_{(k)}(x) \ ,$$
then the coefficient functions $r_{(k)}(x)$ are determined algebraically by demanding that the surface given by $r= r_H(x)$ be null.

A very simple toy model which captures the gist of this argument is given by a time-dependent but spherically symmetric black hole, the Vaidya spacetime:
$$ ds^2 = - \left( 1 - \frac{2\,m(v)}{r} \right) \,dv^2 + 2\, dv \,dr + r^2 \,d\Omega^2 \ .$$
This metric describes a four-dimensional asymptotically flat black hole accreting null dust, so that the mass $m(v)$ increases with time.
Assuming that at late times the black hole settles down to Schwarzschild, $m(v) \to m_f$ as $v \to\infty$, and denoting the location of event horizon by $r=r_H(v)$,  we can find its position by demanding that it describes the null surface which at late times approaches the correct event horizon $r_H \to 2 m_F$ as $v \to \infty$.  Note that the normal $n$ to a null surface will be simultaneously tangent to that surface and likewise null.  For Vaidya, the normal  1-form $n = dr - \dot{r} \, dv$ (where $\dot{} \equiv \frac{d}{dv}$) is null when 
$$r_H(v) = 2\, m(v) + 2\, r_H(v)\,  \dot{r}_H(v) \ .$$
Of course, the exact solution to this equation yields the horizon $r_H(v)$ non-locally in terms of $m(v)$, requiring the knowledge of $m(v)$ for all $v<\infty$.  However, when $m(v)$ varies slowly, so that $\dot{m}={\mathcal O}(\epsilon)$, $m \,  \ddot{m} = {\mathcal O}(\epsilon^2) $, etc.,  we can determine this location in terms of an $\epsilon$ expansion. For the ansatz
$$r_H(v) = 2\, m(v) + a\, m(v) \,\dot{m}(v) + b\, m(v) \,\dot{m}(v)^2 + c\, m(v)^2 \,\ddot{m}(v) + \ldots $$
this expansion gives $a=8$, $b=64$, $c=32$, ...
This toy model illustrates that in spite of the event horizon being defined globally (as the boundary of the causal past of the future null infinity) and therefore requiring knowledge of the mass $m(v)$ for all time $v < \infty$, for slowly varying $m(v)$ we can nevertheless express $r_H(v_0)$ solely in terms of $m$ its derivatives at $v=v_0$.

Returning to the problem of interest, we can similarly locate the event horizon $r_H(x^\mu)$ in our dynamical non-uniform planar black hole geometry in terms of $T$, $u^\mu$, and all their derivatives, at the given point $x^\mu$.  At first order, the position of the horizon is unchanged, whereas at second order it is corrected by terms which scale with square of the shear and vorticity (see \cite{Bhattacharyya:2008xc}, \cite{Rangamani:2009xk} for explicit expressions.)

Once we identify the position of the event horizon in our geometry, it is easy to check that this horizon is regular.  In fact, our construction manifestly guarantees regularity: the only curvature singularity of the seed metric is at $r=0$, and the source terms which appear in correcting the metric order by order do not introduce any additional singularities.  The final issue to check is the regime of validity of our expansion, and this can be seen to extend well inside the event horizon.

Therefore we  have an explicit one-to-one map relating conformal fluid configurations on ${\bf R}^{3,1}$ to asymptotically AdS$_{5}$ inhomogeneous black brane solutions having regular event horizons. This is a remarkable statement about gravity, suggesting that fluid dynamical configurations within the AdS/CFT correspondence naturally uphold Cosmic Censorship.

To extract further physics from the position of the event horizon, let us consider the proper area of its spatial slices.  By the second law of black hole mechanics, the horizon area cannot decrease with time; or equivalently, the expansion of the horizon generators must be non-negative.  A well-known identification with thermodynamics translates this statement to that of the entropy increasing, or more locally, the entropy current having non-negative divergence.  Having obtained the event horizon for our geometry explicitly in terms of the metric functions $u^\mu(x)$ and $T(x)$, we can verify these statements, and identify the entropy current naturally induced on the boundary.  

To obtain the boundary entropy current $J^\mu_S$ from the bulk geometry, we can pull-back the area form $A$ on the event horizon to the boundary.  We perform this pull-back along a tube of constant $x^\mu$, i.e.\ along ingoing radial null geodesics.    This yields the expression 
\begin{eqnarray}
(4\pi\,\eta)^{-1}\,J^\mu_S = &\left[1+b^2\left(A_1 \,\sigma_{\alpha\beta}\,\sigma^{\alpha\beta}+A_2 \,\omega_{\alpha\beta}\,\omega^{\alpha\beta} +A_3 \,\mathcal{R}\,\right)\right]u^\mu \nonumber \\
&\quad  + b^2 \,\left[ B_1 \,\mathcal{D}_\lambda \sigma^{\mu\lambda} + B_2 \,\mathcal{D}_\lambda \omega^{\mu\lambda} \right] \nonumber \\
&\quad + C_1\ b\ \l^\mu + C_2\ b^2 u^\lambda \,\mathcal{D}_\lambda \l^\mu  +\ldots 
\label{entcur}
\end{eqnarray}
with $b\equiv \frac{1}{\pi T}$ and 
\begin{eqnarray}
A_1 &=&\frac{1}{4}+\frac{\pi}{16}+\frac{\ln 2}{4} \ ,\qquad A_2 =-\frac{1}{8} \ ,\qquad A_3 =\frac{1}{8} ,\nonumber \\
B_1 &=&\frac{1}{4} \ , \qquad B_2=\frac{1}{2} \ , \qquad
C_1 =C_2 = 0 
\label{entcurcoeffs}
\end{eqnarray} 
which can be easily verified to have non-negative divergence.  At leading order, the divergence is proportional to $\sigma_{\mu\nu} \, \sigma^{\mu\nu}$ which is manifestly positive for any non-zero shear.  In case of zero shear, the relations $ B_1=2 \, A_3 $ and $C_1+C_2=0  $ guarantee non-negativity.  We can readily see that these relations are indeed satisfied for the entropy current given by our gravity dual construction, (\ref{entcurcoeffs}).  

The expression (\ref{entcur}) was written suggestively in a most general form yielding a Weyl-covariant entropy current for any set of constant coefficients $A_i,B_i,C_i$.  In general, to the order we work, the seven coefficients get reduced to five independent ones allowing for  Weyl-covariant entropy current with non-negative divergence.  From the field theory side, this would therefore suggest a 5-parameter ambiguity in constructing a sensible entropy current, purely based on the symmetries and the requirement that it correctly reproduces equilibrium physics.  This is at first sight puzzling, since our gravity construction seemingly fixed all these parameters.  In fact, this was not quite the case: there is still an ambiguity even on the gravity side, corresponding to the freedom to add total derivative terms without changing the horizon area, and the pull-back being ambiguous to boundary diffeomorphisms.  However, at second order this results in a 2-parameter ambiguity for Weyl covariant current with positive divergence, so that some mismatch remains.\footnote{
Although including higher orders may remedy this discrepancy, \cite{Romatschke:2009kr}.}

\subsection{Generalizations} 
\label{flugra_generaliz}

The simple analysis presented above was carried out in \cite{Bhattacharyya:2008jc}, but this was quickly generalized in a number of interesting directions.  Here we list some of the most important ones.
\begin{itemize}
\item 
One of the earliest such generalizations involved extending the correspondence to other dimensions, relating a $d$-dimensional conformal fluid to asymptotically AdS$_{d+1}$ black hole (see \cite{VanRaamsdonk:2008fp} for the interesting case of $d=3$ and \cite{Haack:2008cp,Bhattacharyya:2008mz} for general\footnote{
The $d=2$ case is rather trivial: the most general solution consists of left and right movers, so the fluids cannot locally equilibrate, relatedly the fluid dynamics in $1+1$ dimensions is equivalent to conservation of general conformal stress tensor; also $\sigma_{\mu\nu} = \omega_{\mu\nu} =0$.  From the bulk standpoint, any $2+1$ dimensional solution to Einstein's equations with negative cosmological constant is locally AdS$_3$.
} $d$).  
This was already mentioned in \sec{ss:stressT} (cf.\ \req{transpgend}); here we collect the result in more convenient form.
The most general analysis is presented in \cite{Bhattacharyya:2008mz} who give the second order stress tensor (and corresponding transport coefficients), the dual bulk metric along with its event horizon and the corresponding entropy current for these fluid flows, for a fluid on slowly-varying curved background; see below.  
In particular, \cite{Bhattacharyya:2008mz} give the stress tensor on the $d$-dimensional boundary metric $g_{\mu\nu}$ in a very elegant (and manifestly Weyl-covariant) form:
\begin{equation}
\begin{split}
T_{\mu\nu} &= p \, \left(g_{\mu\nu}+d  \, u_\mu  \, u_\nu \right)-2 \, \eta \, \sigma_{\mu\nu}\\
&-2 \, \eta  \, \tau_\omega \left[u^{\lambda} \, \mathcal{D}_{\lambda}\sigma_{\mu \nu}+\omega_{\mu}{}^{\lambda} \, \sigma_{\lambda \nu}+\omega_\nu{}^\lambda \,  \sigma_{\mu\lambda} \right]\\
&+2 \, \eta  \, b \, \left[u^{\lambda} \, \mathcal{D}_{\lambda}\sigma_{\mu \nu}+\sigma_{\mu}{}^{\lambda} \, \sigma_{\lambda \nu} -\frac{\sigma_{\alpha \beta} \, \sigma^{\alpha \beta}}{d-1} \, P_{\mu \nu}+ C_{\mu\alpha\nu\beta} \, u^\alpha  \, u^\beta \right]
\label{ddim2T}
\end{split}
\end{equation}
with
\begin{equation}
\begin{split}
b\equiv \frac{d}{4\pi T}\qquad;&\qquad p=\frac{1}{16\pi G_{\text{AdS}}b^d}\qquad;\\
\eta = \frac{s}{4\pi}=\frac{1}{16\pi G_{\text{AdS}}b^{d-1}}\qquad \text{and}& \qquad \tau_{\omega} =  b \int_{1}^{\infty}\frac{y^{d-2}-1}{y(y^{d}-1)}dy 
\end{split}
\end{equation}
where the pressure $p$ depends on the temperature $T$ and the viscosity of the fluid $\eta$ depends on the entropy density $s$ in the usual manner, and $\tau_{\omega}$ 
denotes a particular second-order transport coefficient of the fluid, and
$\sigma_{\mu\nu}$ and $\omega_{\mu\nu}$ are the shear and the vorticity as before. 
$C_{\mu\nu\alpha\beta}$ is the Weyl tensor for the metric $g_{\mu\nu}$ in which the fluid lives and $\mathcal{D}_{\lambda}$ is the Weyl-covariant derivative.

One rather intriguing aspect is the striking difference between the phenomenology of turbulent flows in 3+1 and 2+1 dimensions, as pointed out in \cite{VanRaamsdonk:2008fp}. In the 3+1 dimensional turbulent energy cascade, large scale eddies give rise to smaller scale eddies, eventually transferring energy down to scales where viscosity becomes important and energy is dissipated. In contrast, the 2+1 dimensional turbulent flows are characterized by an ``inverse cascade,'' in which smaller scale eddies merge into large scale eddies, creating large long-lived vortical structures. If these qualitative differences extend to relativistic fluids, they would suggest a profound difference in gravitational dynamics between four and five dimensional gravity. In particular, we might predict that black holes in AdS$_4$ would take much longer to equilibrate owing to the fact that the fluctuations on the horizon could coalesce in macroscopic structures.  From the gravitational standpoint, this would certainly seem very surprising and counter-intuitive.

\item
AdS/CFT, and in particular the fluid/gravity correspondence, extends to bulk spacetimes which are not just asymptotically AdS but also those which are only {\it locally} asymptotically AdS.  In particular, those which asymptote to
\begin{equation}
ds^2 = \frac{1}{z^2} \, \left[ dz^2 + ds_{\rm bdy}^2 \right]
\label{ALAdS}
\end{equation}	
where the 4-dimensional boundary metric $ds_{\rm bdy}^2$ can be any desired slowly-varying metric.
Hence we can consider fluids on curved (and not only conformally flat) manifolds, rather than just the Minkowski spacetime $\mathbf{R}^{d-1,1}$, as has been initiated in \cite{Bhattacharyya:2008ji} and carried out to arbitrary dimensions in \cite{Bhattacharyya:2008mz}.   It should still remain true that such solutions\footnote{
For a general (non-slowly varying) boundary metric $ds_{\rm bdy}^2$ we may not know the exact bulk metric corresponding to even the global thermal equilibrium.  For a discussion when the boundary metric describes a black hole with various asymptotics,
see \cite{Hubeny:2009ru,Hubeny:2009kz,Hubeny:2009rc}; 
in the high temperature regime the conjectured bulk solution, christened `black funnel', would be a bulk black hole whose horizon stretches all the way out to the boundary.
} give a universal subsector of CFTs on this curved background, with the dynamics given simply by the covariant form of the generalized Navier-Stokes equations on $ds_{\rm bdy}^2$.  
As a simple example, \cite{Bhattacharyya:2008ji} (in $d=4$) and \cite{Bhattacharyya:2008mz} (for general $d$) consider the explicitly-known case of rotating AdS black hole (which is asymptotically global AdS$_{d+1}$) which describes a rigid fluid flow on $\Sp^{d-1} \times \R$ (see also \cite{Bhattacharyya:2007vs}), confirming the general construction indicated above.

\item
In addition, one can include matter in the bulk.  This allows for richer dynamics, but typically at the expense of losing universality.  Early examples of such extensions include considering the dilaton (which corresponds to forcing of the fluid) in \cite{Bhattacharyya:2008ji}.
\item
More involved generalizations of the fluid/gravity correspondence with extra matter fields include Maxwell $U(1)$ field \cite{Erdmenger:2008rm,Banerjee:2008th}, multiple Maxwell fields and scalars, magnetic and dyonic charges, as well as more exotic models such as those relevant for holographic superfluids \cite{Sonner:2010yx} (see e.g.\ \cite{Rangamani:2009xk} for further references).
\item
Moreover, one can even extend the correspondence to non-conformal fluids \cite{Kanitscheider:2009as,David:2009np} as well as to non-relativistic fluids \cite{Rangamani:2008gi,Bhattacharyya:2008kq}, which allows us to make closer contact with familiar everyday systems.
\item 
Stringy ($1/\lambda$) and quantum ($1/N$) corrections to some of the transport coefficients have been computed in \cite{Buchel:2008ac,Dutta:2008gf,Buchel:2008ae,Buchel:2008kd,Buchel:2008bz,Bigazzi:2009tc}.

\end{itemize}

Nevertheless, many future directions and puzzles remain, as well  as the need for further generalizations.  For example, of particular current interest is to understand the fluid/gravity correspondence for extremal fluids  which are presently attracting much attention.  Also, to mimic many of the familiar aspects of fluid flows, we need to understand how to confine the fluid within walls in the gravity dual.  Still more ambitiously, to understand the rich phenomena rooted in quantum processes, we would like to get a better handle on finite-$N$ effects.

To summarize, one of the most intriguing features of the fluid/gravity correspondence is that it provides us with a window into the {\it generic} behavior of gravity in a nonlinear regime, mapping long-wavelength (but arbitrary amplitude) perturbations of AdS black holes to the more familiar physics of fluid dynamics.  Apart from the obvious conceptual advantages, one has a tremendous computational simplification for numerical studies of gravitational solutions since the fluid dynamics  lives in one lower dimension.

The bulk spacetime solutions discussed here describe a generic (time-dependent and non-uniform) planar black hole.  Each of these solutions  (with regular fluid data $T(x)$ and $u_\mu(x)$) has its singularities hidden from the boundary by a regular event horizon.  In this sense, all gravitational solutions dual to regular solutions of fluid dynamics uphold the cosmic censorship conjecture.  At the technical level, the key to the fluid/gravity  construction lies in utilizing the long-wavelength regime of fluid dynamics.  This allows effectively a linearization of Einstein's equations in that one ends up solving linear ordinary differential equations to solve for the  bulk metric, and corresponding boundary stress tensor, to any order in a boundary derivative expansion in a completely procedural manner.  Furthermore, having obtained the bulk geometry, this expansion likewise enables us to determine the radial position of its event horizon, remarkably expressed  locally in terms of the fluid dynamical data.  One can then easily verify that the area of this event horizon is necessarily non-decreasing, in accordance with the Area Theorem.  The corresponding entropy current induced from the bulk solution is then guaranteed to satisfy the Second Law of thermodynamics.

The boundary fluid stress tensor contains new quantities of interest, namely the various transport coefficients which characterize the fluid.  At first order one recovers the previously-known value of viscosity and verifies that $\frac{\eta}{s}=\frac{1}{4\pi}$.  More importantly, the construction allows the extraction of second order fluid parameters $(\tau_\pi, \lambda_1, \lambda_2, \lambda_3)$ (see also \cite{Baier:2007ix}).  This has been of interest in QCD phenomenology, especially in understanding certain characteristic features of the quark-gluon plasma.

\section{Beyond long-wavelength approximation} 
\label{s:dynam}

While impressively powerful, the fluid/gravity correspondence as formulated is only valid in the long-wavelength regime.  Albeit useful to study gravitational duals of fluid flows, it falls short of describing more 
interesting aspects of dynamics, viz., the approach to local equilibrium and the behaviour of general fluctuations which are not necessarily of long-wavelength. Likewise, from the bulk point of view, the long-wavelength regime does pose a substantial restriction, in the sense that many `natural' bulk solutions do not fall in this regime.

The simplest example is the static spherically symmetric \SAdS{} black hole in global AdS, whose radius is larger, but not parametrically larger, than the AdS radius, $r_+\gtrsim \RAdS$.
Such large(ish) global \SAdS{} black hole does not fall within the long-wavelength regime because the boundary curvature is comparable to the thermal scale (for instance the term proportional to the Weyl tensor in \req{ddim2T} obviates the expansion). However, this is a regular solution, admissible from the point of field theory, describing simply an equilibrated plasma on the boundary sphere. Furthermore, one expects (assuming cosmic censorship) that this solution actually captures the end-point of a generic collapse scenario in AdS for initial data whose energies are in the appropriate range.   This then suggests that the dual state in the CFT is not described by a fluid stress tensor in spite of being globally equilibrated and exhibiting no spatial variations!  

One would therefore like to understand the generic situation where perturbations take us away from hydrodynamic equilibrium, involving excitations of high frequency modes. A related question of interest concerns the evolution of a highly excited state in a quantum system. Consider preparing a quantum state, with a large vacuum expectation value for the Hamiltonian operator, which is not an energy eigenstate, as can be done for instance by injecting energy into the system. One expects this state to explore the available phase space and evolve dynamically to a thermal density matrix, in the sense that the late time observables are thermal expectation values. In such situations one would like to track the evolution of such a state.\footnote{This concept is also of interest in condensed matter systems, where it is referred to as quantum quench. See \cite{Sengupta:2004eu,Calabrese:2006rx,Calabrese:2007fk,Calabrese:2007rg} and references therein for details on this subject. We should also note that there has recently been some progress in holographic modelling of quantum quench behaviour in \cite{Das:2010yw}.}
 From the bulk standpoint this evolution should be associated with the formation of a black hole, and questions like the approach to equilibrium and validity of hydrodynamics can be addressed once the solution is understood. 
To summarize our motivation, we would like to move away from the axes in \fig{fig:regimes} and explore the region of arbitrary amplitudes and momenta. 

In this section we will review various works that have addressed these issues. While most of the techniques for investigating the general evolution involve numerical simulation (we are in the non-linear regime with no control parameter generically), there are some interesting examples where analytic techniques can be brought to bear. We will begin by describing simple toy models which allow detailed exploration of time dependence and then discuss some concrete attempts to study black hole formation in the AdS/CFT context. Towards the end we revert back to the story of equilibrium dynamics, but for field theories in curved spacetime.

\subsection{CFTs with strong time dependence}
\label{ss:fast}

We have seen above that even innocuous static configurations can lie outside the long-wavelength regime, by virtue of living on a compact space.  However, of greater interest in the present setting are configurations exhibiting explicit time-dependence, which are typically outside the long-wavelength regime.  This is because a typical dynamical process involving a black hole (such as its formation or coalescence, quasinormal ringing, etc.) occurs on time scales set by the black hole size, which in the regime of interest is given by the thermal scale, rather than being parametrically longer.  Nevertheless, there are interesting and important physical processes which may be at least partially understood from the fluid/gravity framework.  
In the following discussion, we will first consider systems with `mock' time-dependence given by simply considering a static configuration in different coordinates, and then discuss several configurations which are genuinely time-evolving.  We will see that even the first, seemingly rather trivial, class offers intriguing surprises.

\subsubsection{Mock time-dependence with evolving event horizon}
\label{ss:confsol}

\begin{figure}
\begin{center}
\includegraphics[width=2.6in]{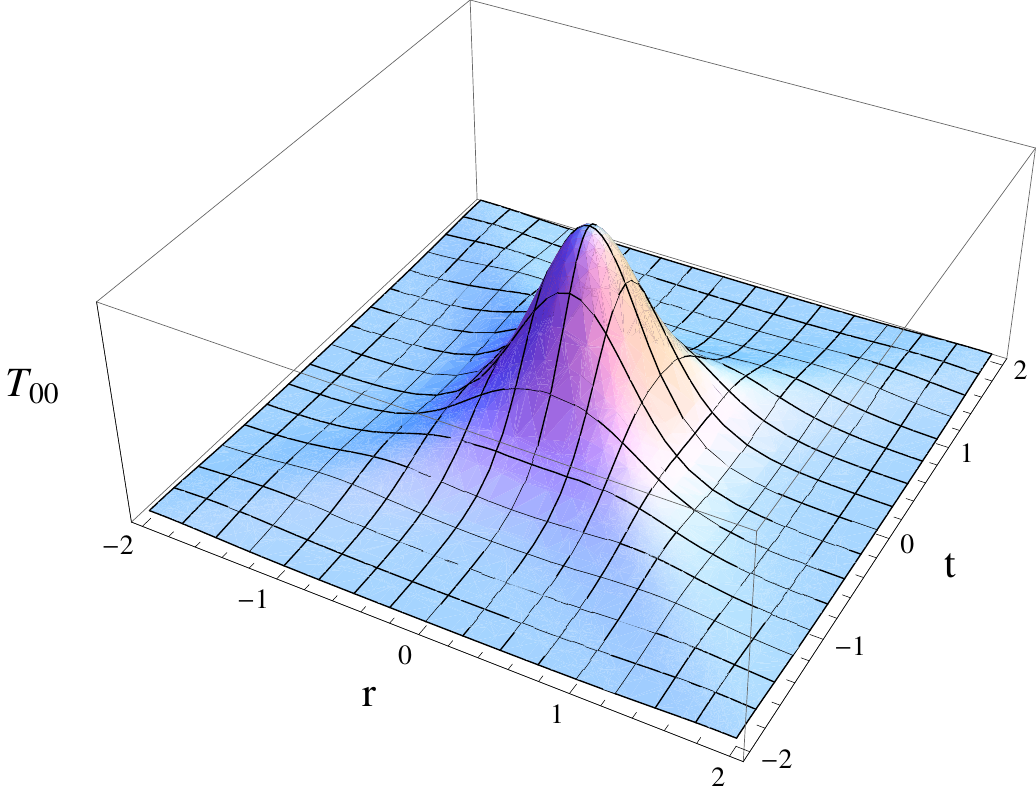}
\hspace{2cm}
\includegraphics[width=1.5in]{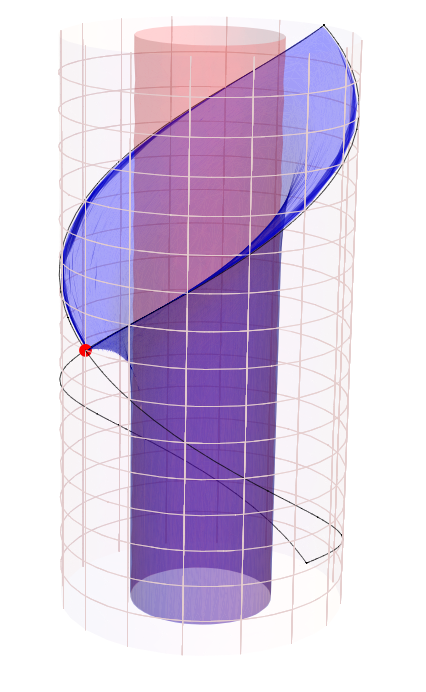}
\caption{Conformal Soliton energy density  (left) and dual geometry (right). 
On left, $T_{00}(t,r)$ is plotted as a function of boundary time $t$ and boundary radial variable $r$.  On right, the outer cylinder represents the AdS boundary, the inner (red) cylinder corresponds to the global \SAdS{} event horizon, which coincides with the apparent horizon of the conformal soliton, whereas the flared (blue) surface is the actual event horizon.}
\label{fig:CShorizon}
\end{center}
\end{figure}

One of the simplest examples of a secretly static configuration which nevertheless provides an instructive toy model for a time-dependent scenario is given by the conformal soliton geometry originally introduced in \cite{Friess:2006kw}, and then studied extensively in \cite{Figueras:2009iu}.   
The spacetime in question is simply a patch of the well-known \SAdS{} black hole; so
the explicit metric is known exactly, and admits a high degree of
symmetry.  Nevertheless, if we work in a coordinate system
such that the field theory lives on flat space $\R^{3,1}$ rather than the Einstein Static Universe $\Sp^3\times \R^1$, which corresponds to considering only a `Poincar\'e patch' of the \SAdS{} black
hole, the time translation symmetry is no longer manifest, and the
solution looks highly dynamical.  Pictorially, it describes a
black hole entering through the past Poincar\'e horizon and exiting
through the future Poincar\'e horizon, with its closest approach to
the boundary occurring at $t=0$ (Poincar\'e time).  In the boundary CFT, this corresponds to
a finite energy lump which collapses and re-expands in a
time-symmetric fashion (cf.\ the left panel of \fig{fig:CShorizon}). Here the fluid dynamical approximation is not
valid at all times, but because this fluid flow is conformal to a
stationary fluid on the Einstein static universe, the stress tensor is
shear-free; that is, there is no dissipation in this fluid flow.

Nevertheless, the event horizon in this case is rather interesting. Since our coordinate patch includes only part of the future infinity of the global \SAdS{}, the actual event horizon
for the conformal soliton lies outside the global event horizon of
\SAdS{}. 
For orientation, the event horizon for both global \SAdS{} and the conformal soliton is plotted in right panel of \fig{fig:CShorizon}.
In fact, \cite{Figueras:2009iu} observed that the area of this conformal soliton event horizon diverges at late times.
This surprising result led to a puzzle: if we associate the entropy
of the corresponding CFT conformal soliton state to the area of the conformal soliton
event horizon, as is usually assumed to be the case, then we find that
this entropy likewise diverges at late times.  But the conformal
soliton describes a shear-free flow, with no entropy production
whatsoever.  Said differently, it is easy to see \cite{Figueras:2009iu} that the conformal transformation from the
CFT on $\Sp^3 \times \R^1$ to the CFT on $\R^{3,1}$
leaves the entropy invariant.  But the former describes a perfect fluid
in global thermodynamic equilibrium: its entropy is finite and constant in time. 

Already prior to \cite{Figueras:2009iu}, it has been argued in several different contexts
\cite{Hubeny:2007xt,Chesler:2008hg} that it may be more appropriate to
associate the entropy of the CFT configuration to the area of the
apparent horizon, rather than the event horizon, in the bulk dual.  In fact, as explained in \cite{Figueras:2009iu},  the apparent horizon in Poincar\'e slicing
coincides with the global \SAdS{} event horizon, whose area is indeed
constant and given by the expected value.  Since the event horizon and the
apparent horizon behave radically differently (cf.\ \fig{fig:CShorizon}), this conformal soliton geometry provides a
good testing ground for studying the distinction between the two horizons and the role they play for the associated CFT dual.
We see that in this case the CFT entropy is clearly more naturally associated with the
apparent horizon rather than the event horizon.\footnote{
We will revisit the intriguing question of whether this can be taken as a general tenet in \sec{s:concl}.}

The large disparity between locations of the event and apparent horizons might seem enigmatic in light of \cite{Bhattacharyya:2008xc}, where it was argued that within the fluid dynamic regime, the event horizon and the apparent
horizon track each other closely.
The essential difference between these two contexts is that in \cite{Bhattacharyya:2008xc}, it was assumed that the geometry would settle down at late
times to a stationary finite-temperature black hole.   This is not the case for the conformal soliton geometry. It is precisely these late time boundary conditions that force the apparent
horizon and the event horizon to behave very differently.

This observation might prompt one to expect that if the spacetime does not settle down at late times, then the entropy will not be given by the event horizon area.  However, we will now briefly mention another example, this time with genuine time dependence, where this is not the case.
In particular, let us consider the holographic dual to Bjorken flow \cite{Janik:2005zt,Janik:2006gp}, also studied in \cite{Figueras:2009iu}.
Since Bjorken flow plays a
central role in understanding the post-thermalization evolution of the
QGP produced in heavy-ion collisions, this is a more broadly useful example to study. The basic physical picture
developed in \cite{Bjorken:1982qr} is that, assuming local
thermal equilibrium, in the central rapidity
region of ultra-relativistic collisions of heavy ions one can model the flow of the plasma via
quasi-ideal hydrodynamics. In the hydrodynamic description, it is
assumed that the fluid evolution respects the boost symmetry along the
collision axis. This implies a boost invariant expansion of the fluid,
consistent with the observed distribution of the particles in the
collision process.  

In \cite{Janik:2005zt,Janik:2006gp}, the corresponding bulk geometry was constructed as a
perturbation expansion in the boundary time coordinate which is valid
at late times. By demanding regularity of the solution at leading
orders, the authors were able to derive the transport properties of
the plasma, in particular the shear viscosity $\eta$ which saturates
the bound $\eta/s \ge 1/4\pi$ \cite{Kovtun:2004de} mentioned in \sec{s:bhquasi}. The study
of the gravitational dual at higher orders was undertaken to derive
the relaxation time of the plasma in
\cite{Heller:2007qt,Baier:2007ix}. Although initially the regularity of the dual
spacetime was brought into question as subleading singularities were
encountered \cite{Heller:2007qt, Benincasa:2007tp}, this issue was subsequently
addressed in \cite{Heller:2008mb,Kinoshita:2008dq} where the
authors used the framework of the fluid/gravity correspondence
\cite{Bhattacharyya:2008jc} to argue that the spacetime was indeed
regular.  Explicit confirmation was provided in \cite{Figueras:2009iu} by constructing the global event horizon for these geometries, finding that it is regular and
closely tracks the location of the apparent horizon, as previously computed in \cite{Kinoshita:2008dq} and verified in \cite{Figueras:2009iu}.

\subsubsection{Black hole formation in AdS and approach to equilibrium}
\label{ss:bhform}

Having discussed in \sec{ss:confsol}  a simple case where one engineers apparent time dependence by a non-trivial coordinate change in a static solution, we now turn to recent discussions of explicit time dependent phenomena in AdS/CFT. These examples provide a clean way to understand the deviations from equilibrium and approach to the same from an arbitrary initial state. 

One way to study time-dependent phenomena in AdS/CFT is to start with a system in the vacuum state and inject in energy at some suitably chosen moment, say $t =0$. At early times $t<0$ the system is holographically described by the pure AdS spacetime, while after the perturbation one needs to solve the bulk Einstein's equations in order to infer the nature of the evolving geometry.  The injection of energy can be achieved by perturbing the boundary Hamiltonian with an explicit insertion of a  source at $t=0$. In the bulk description, the source corresponds to a change of boundary conditions for the corresponding field; in effect one has to turn on non-normalizable modes for the bulk fields. Equivalently, one can consider sending in a null shell at $t=0$ from the boundary, which deforms the spacetime from pure AdS in the interior. In either case we expect to seed a black hole collapse; the energy contained in the deformation (or in the null shell) will cause a black hole to form in the interior of the spacetime, which at late times will asymptote to a new steady state solution.\footnote{
One can of course construct such geometries by patching together spacetimes across thin domain walls using the Israel junction conditions, but since we are interested in real dynamics we need to actually solve the bulk equations of motion.
}  A schematic Penrose diagram of such a process is sketched in \fig{fig:collapse_bcs}. 

In \cite{Chesler:2008hg} the first non-trivial such calculation was carried out.\footnote{In fact earlier analysis of equilibration in strongly coupled systems was considered in
\cite{Bak:2007qw} building on the construction of time dependent Janus type geoemtries in \cite{Bak:2007jm}.}
  The idea was to inject energy into the boundary field theory by changing the boundary metric. Since the boundary metric couples to the energy-momentum tensor, we are effectively pumping energy into the system directly. The set-up used by \cite{Chesler:2008hg} was to consider a background for the field theory which asymptotes to the flat Minkowski background in the far past and far future, but with different length scales. In particular, they consider explicitly time dependent background of the form:
\begin{equation}
ds^2_\text{bdy}  = -dt^2  + e^{B_0(t)} \, d{\bf x}_\perp^2 + e^{-2\,B_0(t)} \, dx_\parallel^2 \ .
\label{cybdy1}
\end{equation}	
With a suitable choice of $B_0(t)$, e.g.\ approximating a step function obeying the early and late time boundary conditions, one can inject energy into the system as a consequence of the time-varying spatial length scales. Physically the change in length scales corresponds to anisotropizing the pressure in the various directions.

Since the deformation is purely in the metric degrees of freedom, the problem can be treated within the consistent truncation of pure gravity in \AdS{d+1}, i.e., by solving \req{ein} for the bulk metric $G_{MN}$ subject to the boundary condition of approaching \req{cybdy1} asymptotically. The numerical solution of this problem  clearly illustrates the process of black hole formation; in particular, \cite{Chesler:2008hg} were able to demonstrate the presence of an event horizon by examining the causal past of the \AdS{} boundary. Moreover, the computation of the boundary energy momentum tensor using the holographic recipe 
\req{bdysten} showed clearly that the pressure in the system starts out being anisotropic between the longitudinal ($x_\parallel$) and transverse (${\bf x}_\perp$) directions, but evolves towards an isotropic equation of state characterizing eventual approach to equilibrium.
In \cite{Chesler:2009cy}, the phenomenologically interesting boost invariant Bjorken flow \cite{Bjorken:1982qr} was also modeled in the bulk by a numerical solution. This calculation provides a useful way to make contact with the hydrodynamic description of the system at late time for which approximate solutions in the fluid/gravity sense were constructed in \cite{Janik:2005zt, Janik:2006gp} as described at the end of \sec{ss:confsol}.\footnote{
These numerical techniques have also been adapted to study the evolution of perturbations on unstable phases of the field theory, see \cite{Murata:2010dx}.} Numerical investigations of the early time behaviour relevant for  the Bjorken flow have also independently been considered in \cite{Beuf:2009cx}.

While the constructions described above involve numerically integrating Einstein's equations, it is nevertheless possible to understand aspects of black hole formation and thermalization in  \AdS{} spacetimes analytically. This was achieved in \cite{Bhattacharyya:2009uu} by considering a perturbative solution in the amplitude of the fluctuation.  The basic idea is to start with the initial data similar to the one described above, but to take the injection of energy to be of small amplitude to allow for a perturbation expansion in amplitude.\footnote{
In addition to working in pure gravity, \cite{Bhattacharyya:2009uu} also considered the somewhat simpler context of a scalar collapse in \AdS{}.
} The key ingredient in the analytic construction was the observation that the perturbation series in amplitude needs to be resummed, for the naive expansion fails to converge.  This is perhaps not so surprising given that a black hole, no matter how small, is necessarily a large perturbation on causally-trivial constant curvature spacetime.  The resummed perturbation series naturally leads to a change in the background reflecting the process of black hole formation. Assuming that one started with pure \AdS{} spacetime and injected energy (or sent in a null scalar pulse) at $t=0$, one finds that the resummation of the perturbation series shifts the background towards a Vaidya type solution, which describes gravitational collapse. 

The analytic results of \cite{Bhattacharyya:2009uu} illustrate two key features of gravitational collapse in \AdS{} spacetimes:
\begin{itemize}
\item In global \AdS{d+1} spacetimes there is a threshold for black hole formation if one examines the system on a time-scale of the order of the AdS scale after the perturbation. This is similar to the gravitational collapse in asymptotically flat spacetimes where the space of initial data splits into distinct domains separated by a co-dimension one critical surface which captures the threshold configurations for black hole collapse.  
\item Arbitrarily small energy densities lead to black hole formation when the boundary is a copy of Minkowski space; in this context there is no threshold for black hole formation. 
\end{itemize}

\subsection{CFT on non-trivially curved backgrounds}
\label{ss:funnel}

In \sec{ss:bhform} we indicated how one might perturb a system in equilibrium by deforming the spacetime background in which it lives. While this is interesting in order to understand deviations from equilibrium, examining the behaviour of quantum fields in curved spacetime brings another dimension to the problem. As described towards the end of \sec{s:regimes} the AdS/CFT correspondence provides a tool to investigate such issues as well. In general one needs to construct all bulk manifolds whose boundary is the background on which the field theory dynamics takes place. 

\begin{figure}
\begin{center}
\includegraphics[width=5in]{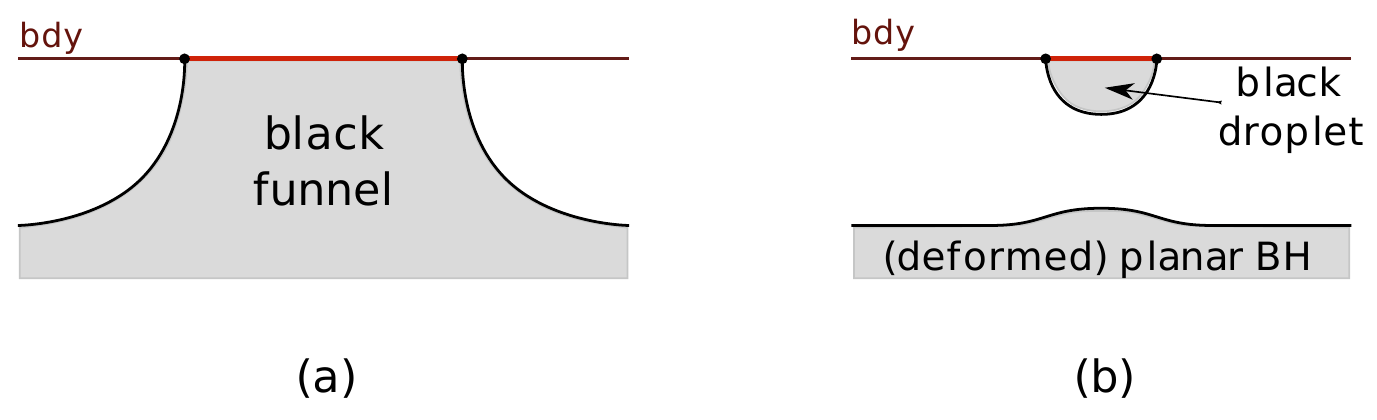}
\caption{A sketch of our two novel classes of solutions dual to a deconfined phase of a strongly coupled field theory on a black hole background: {\bf (a):} black funnel and  {\bf (b):} black droplet above a deformed planar black hole. The shaded part indicates regions which are behind horizons in the bulk spacetime, and the thick (red) line indicates the black hole of the boundary metric.}
\label{f:fundrop}
\end{center}
\end{figure}

Moreover, from the standpoint of our general discussion,  static configurations might have explicit non-long-wavelength feature resulting from the boundary background geometry the CFT lives on.  For example, consider the field theory on a Schwarzschild black hole background.  Far away, the equilibrium state is simply a thermal state at a temperature given by the black hole's temperature, which in turn determines the microscopic scale $\ell_{\rm mfp} \sim 1/T_{\rm BH}$.  But the curvature of the background geometry is of that same scale near the black hole.  
From the dynamical point of view, the fluid near the black hole wants to fall in on a timescale determined by the black hole, i.e.\ also $\ell_{\rm mfp}$.  Considering such system in equilibrium with the outgoing Hawking quanta leads to an interesting bulk dual:  \cite{Hubeny:2009ru} suggested that the dual geometry exhibits new (yet to be found) solutions, dubbed black funnels and black droplets.\footnote{
As a follow-up, evidence for lower-dimensional realization of these was analyzed in \cite{Hubeny:2009kz} and further discussion of field theory on AdS black hole background was presented in \cite{Hubeny:2009rc}.
}  For ease of orientation, these are sketched in \fig{f:fundrop}.
Far away from the black hole in the boundary directions, the bulk configuration is well-described by the fluid/gravity framework, approaching the planar black hole at the requisite temperature (namely the local temperature of the boundary field theory heated up by the boundary black hole).  However the event horizon of the background spacetime which the field theory lives on must be continued into the bulk on the bulk event horizon.  Near the boundary, such an extension looks like the AdS black string, but in the bulk this black string is deformed.  If it joins with the bulk planar black hole and forms a single connected horizon, the solution is called a black funnel; on the other hand, if it caps off and only slightly deforms the disconnected planar black hole horizon, it is called a black droplet.  Depending on which geometry is realized, the physics (such as the response of the system to perturbations) differs significantly \cite{Hubeny:2009ru}.  As the parameters of the background are varied, we furthermore expect a phase transition to occur.
Such rich physics would of course not be apparent had one attempted to use the fluid/gravity framework with thermal fluid stress tensor everywhere (for example we could never see the droplets this way), but then one would be alerted to being well out of its regime of validity by the description breaking down.

These examples serve to illustrate the rich physics exhibited by systems which are not in local thermodynamic equilibrium but which are nevertheless accessible for exploration within AdS/CFT. 

\section{Diagnostics of dynamics}
\label{s:diagnostics}

Let us pause to take stock of the discussion so far.
In \sec{s:linresp} we have discussed how to extract certain properties, such as transport coefficients of the CFT by considering infinitesimal deviations from equilibrium, which is fortuitously captured by probes in the equilibrium configuration itself.
We then went on to consider genuinely time-dependent configurations in \sec{s:flugra}, and explicitly constructed the bulk spacetime and the corresponding CFT fluid stress tensor to second order.  While the deviations from global equilibrium can be large, the spatial and temporal dependence had to be sufficiently slow for the boundary derivative expansion to be valid.  This posed a rather severe restriction, as many processes of interest happen on faster time scales, comparable with the thermal scale.
Hence in \sec{s:dynam} we tried to venture beyond this long-wavelength regime.  Except for toy models with `mock' time-dependence, we have much less handle on such systems, but these strongly time-dependent cases nevertheless yield more insight about {\it generic} physical processes.

In such genuinely time-dependent scenarios, it is often the case that we have a good idea of what the bulk spacetime looks like -- typically we know the explicit metric, at least approximately, at least in some region.  However, having handle on the bulk does not mean that we know much about the CFT dual; actually {\it constructing the CFT state} dual to the given bulk configuration is well beyond our means at any rate, though we can use the gravity side calculation to read off various observables in that state.  For example, using the asymptotic fall-off of the metric, we can determine the stress tensor expectation value in the dual CFT state.  Conversely, we can turn the logic around and ask: which CFT data should we use to learn as much as possible about the bulk spacetime?

The purpose of this section is to review some of the CFT `observables' which reveal the salient features of the bulk geometry, focusing on probes which we can use in time-dependent cases.  The philosophy here is to suppose that we have at our disposal all that we wish to know about the CFT state, and ask where should we look to see a particular bulk feature.  As already indicated in the Introduction and reviewed in \sec{s:AdSCFT}, this is not an easy question, since bulk locality is not manifest, nor indeed well-understood, in the dual CFT.  On the other hand, this was part of the motivation of the works reviewed below: by determining the signature of specific local bulk features in the CFT dual, we gain some insight into how holography encodes bulk locality, and in turn into the emergence of spacetime (see \cite{Heemskerk:2009pn,Heemskerk:2010ty} for recent progress). 

In classical general relativity, concepts such as causal structure, event horizons, singularities, etc., play an important role in understanding the spacetime geometry. In the semi-classical approximation these concepts are useful in understanding the dynamics of quantum fields in curved backgrounds. Given the AdS/CFT correspondence, it is therefore interesting to understand the field theoretic encoding of these geometric features. From the central role played by geometry in classical general relativity, one naively expects them to have a well defined representation in the field theory.
For example, consider a time dependent process of black hole collapse in the bulk.  Since the formation of the event horizon is a sharply-localized spacetime event (its teleological nature notwithstanding), we would expect that this will manifest itself in some correspondingly sharp feature in the gauge theory.  This expectation was indeed verified in \cite{Hubeny:2006yu}, as we explain below.  
Of course, if we could extract the full spacetime metric from the CFT data, we could then reconstruct these salient features of the geometry.  However, that is not a very direct route, since extracting the actual spacetime metric is harder than extracting its geometrically interesting features.  

One might naively think, using the ideas of \sec{s:flugra}, that given the stress tensor expectation value everywhere, one can nevertheless reconstruct the entire bulk metric.  This is not correct.  The stress tensor expectation value only knows about the asymptotic behaviour of the metric.   In case of holographic RG \cite{deHaro:2000xn}, one can write the metric in a radial series expansion around the boundary, but there is no guarantee that this series will converge inside the bulk.  Indeed as explained in \cite{Bhattacharyya:2008jc}, a general (non-fluid) conformal stress tensor will lead to naked singularities in the bulk.   On the other hand, within the fluid/gravity correspondence framework reviewed in \sec{s:flugra}, we can reconstruct the spacetime exactly in the radial direction well inside the (regular) event horizon -- but only at the expense of confining ourselves to the long-wavelength regime.  In other words, in such a class of spacetimes, the asymptotic behaviour {\it determines} the geometry in the bulk as well, in the `tubewise' manner described above.  While it is impressive that local expectation values carry so much information about the bulk\footnote{
This is intimately tied to the AdS asymptotics, which in a sense `magnify' the signals to be discernible from the boundary. A very early example which illustrates that far more information about the bulk is contained in the local CFT expectation values than naively expected from the UV/IR correspondence is given in  \cite{Horowitz:2000fm}, where the authors show that even such detailed information as the size of a small ($\ll \RAdS$) object can be read-off from local CFT expectation values.
}, it is nevertheless clear that we will need at least bi-local observables, such as correlation functions, to learn about the full bulk geometry.

\subsection{Extracting bulk metric from CFT correlators}
\label{s:correls}

The conventional lore concerning the UV/IR relation, which maps local regions in the interior of the bulk to non-local objects in the CFT, naively appears to preclude extracting much useful information about bulk geometry, especially any precise signal about the causal structure in the neighbourhood of the event horizon.
The initial clue on how the field theoretic observables transcend the classical barrier of the event
horizon to encode information of behind-the-horizon physics\footnote{
One of the early motivations that this should be possible in the first place, despite bulk causal obstructions, was presented in the gedanken-experiment of \cite{Hubeny:2002dg}.
}
 emerged from the progress made in identifying the CFT signature of the black hole singularity
\cite{Fidkowski:2003nf,Festuccia:2005pi}, which built on the prior work of \cite{Balasubramanian:1999zv,Louko:2000tp,Kraus:2002iv}.  
Schematically, the idea was to use the
intrinsic non-locality of the boundary correlation functions to
identify signals of the bulk curvature singularity.

\begin{figure}
\begin{center}
\includegraphics[width=2.5in]{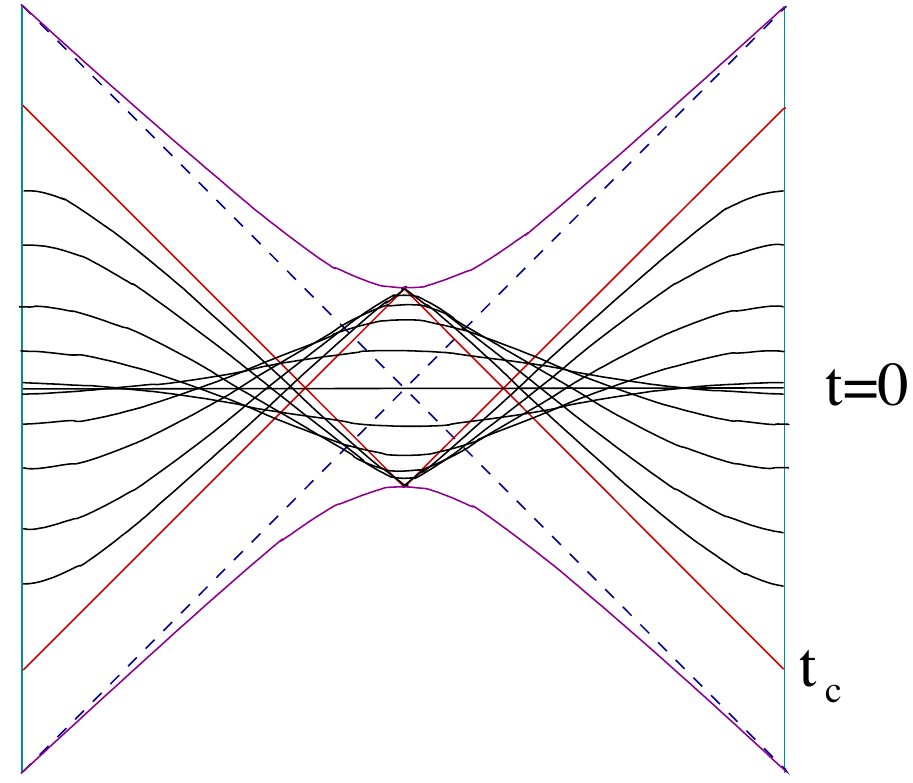}
\hspace{1cm}
\includegraphics[width=2.9in]{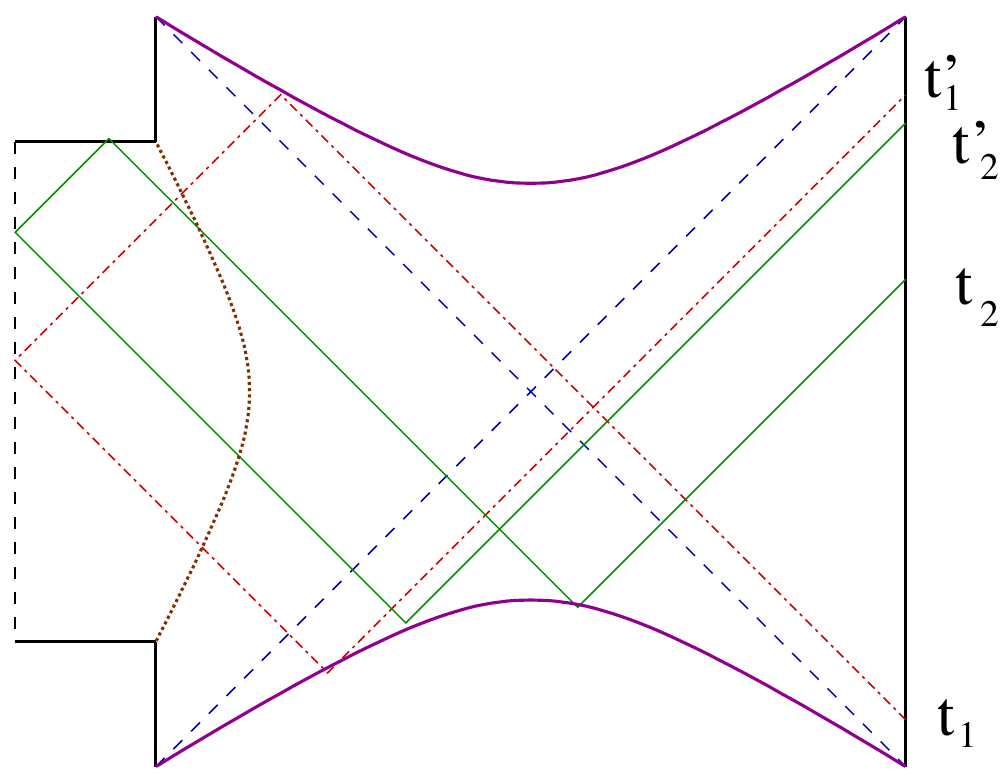}
\caption{Illustration of mapping out bulk causal structure by analytically continued correlators, as discussed in \cite{Fidkowski:2003nf} for the \SAdS{5} spacetime (left panel) and in \cite{Freivogel:2005qh} for a more complicated geometry describing deSitter universe behind the horizon of a \SAdS{} black hole (right panel).  In both Penrose diagrams, the vertical solid lines represent the AdS boundaries, the top and bottom (purple) curves the curvature singularities, the dashed diagonal lines the horizons; on right the horizontal lines are de Sitter boundaries, the vertical dashed line the de Sitter origin, and the vertical (brown) curve a domain wall separating the dS and SAdS regions.  All other curves correspond to the geodesics discussed in the text.
}
\label{fig:correls}
\end{center}
\end{figure}

The basic strategy
was to look at the bulk Green's functions in a saddle
approximation where they are dominated by geodesics and  thereby use simple
bulk computations to extract the behaviour of the boundary correlation
functions. 
When spacelike geodesics which get repelled by the curvature singularity become almost null, they provide a large contribution to the correlation function, akin to light-cone singularities, in the appropriate regime.  At the same time, such geodesics penetrate arbitrarily close to the curvature singularity, where their contribution is most significant.
 Hence the black hole singularity
is manifested by a particular light-cone-like
singularity in the field theory correlation function.\footnote{
The identification of the singularity in the strict large $N$ limit is
easiest when formulated in terms of momentum space correlators
\cite{Festuccia:2005pi} as opposed to direct computation in
position space where the singularity is not visible in the primary
sheet of the correlator \cite{Fidkowski:2003nf}. In particular, it
was confirmed in \cite{Festuccia:2005pi} that the signatures of the
singularities disappear for finite rank of the boundary theory gauge
group, implying that the singularities are resolved in quantum gravity.
} 
The left panel of \fig{fig:correls} sketches the relevant spacelike geodesics for the original \SAdS{5} spacetime; as a critical time $t_c$ is approached by the geodesic endpoints (which we can think of as the correlator insertion points), the geodesic approaches two null geodesics which meet at the singularity.
Moreover, it has been proposed in \cite{Freivogel:2005qh} that this technique
can be exploited to extract physics from further beyond the event horizon,
and in particular used to investigate aspects of inflationary
geometry within AdS/CFT. 
On the right panel, such almost-null geodesics are indicated for the far more complicated spacetime considered in \cite{Freivogel:2005qh} containing a de Sitter universe behind the horizon of a \SAdS{} black hole.  Intriguingly, unlike for the previous case of just the \SAdS{} geometry, in these more complicated cases the relevant correlator insertion points appear on the same boundary.

As evident, these correlators are therefore highly sensitive to certain bulk information which might otherwise be completely inaccessible to a local bulk observer. 
In fact, even such subtle differences as the fuzzball picture of the black hole \cite{Mathur:2005zp,Skenderis:2008qn} and a genuine eternal black hole geometry would produce glaringly different signatures in these correlators \cite{Balasubramanian:2007qv}.  For instance, as discussed in \cite{Maldacena:2001kr} the fact that the eternal black hole corresponds to a density matrix can be used to argue that the correlation functions computed from this geometry should be periodic in imaginary time. This is of course well known from the discussions of computing thermal correlation functions \cite{Son:2002sd,Herzog:2002pc,Skenderis:2008dg}. However, any single horizon-free geometry will give rise to correlation functions that do not display this periodicity in imaginary time. 

This geometric picture can therefore be used to distinguish black hole spacetimes from the putative horizon-free microstate geometries. To understand this issue further recall that while the computation of the Green's functions in the black hole geometry leads to field theory observables in the canonical (or more generally grand canonical) ensemble, the computation in a single microstate geometry leads to a pure state correlator. One can of course consider an ensemble of pure states and indeed one would then recover the thermal answer via appropriate ensemble averaging -- this then corresponds to a micro-canonical ensemble computation, which by ensemble equivalence should agree with the thermal result, as was convincingly argued in \cite{Balasubramanian:2005mg, Balasubramanian:2005qu}. However, as demonstrated in \cite{Balasubramanian:2007qv}, by examining the detailed analytic structure of the correlation functions, one can amplify the small differences between the micro-canonical and canonical computations and use this as a distinction to tell apart spacetimes with and without horizons.

Given the remarkable level of detailed bulk information encoded in such CFT correlators, one might wonder how far-reaching this tool is for `decoding the hologram' in general situations.
The main shortcoming of this method is that it is most useful for static\footnote{
Following conventional terminology, by {\it static}  we mean admitting a Killing field which is asymptotically timelike, but not necessarily globally static.  In particular, this includes nontrivial geometries such as black holes or de Sitter, which have regions of strong time-dependence.  Indeed, it is precisely due to these regions that the CFT signals proposed in \cite{Fidkowski:2003nf, Freivogel:2005qh} are present.
} geometries, where we can determine the correct physics using the Euclidean continuation.
Nevertheless, this limitation can be overcome, as demonstrated by \cite{Hubeny:2006yu}, where the authors consider a manifestly time-dependent spacetime describing gravitational collapse and argue that the horizon formation event can be detected in the boundary theory by examining the structure of singularities of generic Lorentzian correlators.  The basic idea is that
CFT correlators will exhibit light-cone singularities when the operator insertion points are connected by a strictly null geodesic.  Such a geodesic may be confined to the boundary (in which case the corresponding light-cone singularity is a familiar feature of the field theory), but more interestingly it may also penetrate into the bulk spacetime. In the latter case, the
connection implies that CFT correlators in excited states have
additional Lorentzian singularities {\it inside}\footnote{
As proved by \cite{Gao:2000ga}, for certain wide class of spacetimes (including the present case of interest) with timelike conformal boundary, any Òfastest null geodesicÓ connecting two points on the boundary must lie entirely within the boundary.
The vacuum state of the CFT plays a distinguished role, in that all null geodesics through pure AdS bulk take equal time to reach the antipodal point as the boundary null geodesics.
} the boundary light cone,
which in \cite{Hubeny:2006yu} were christened {\it bulk-cone singularities}.   

\begin{figure}
\begin{center}
\includegraphics[width=1.8in]{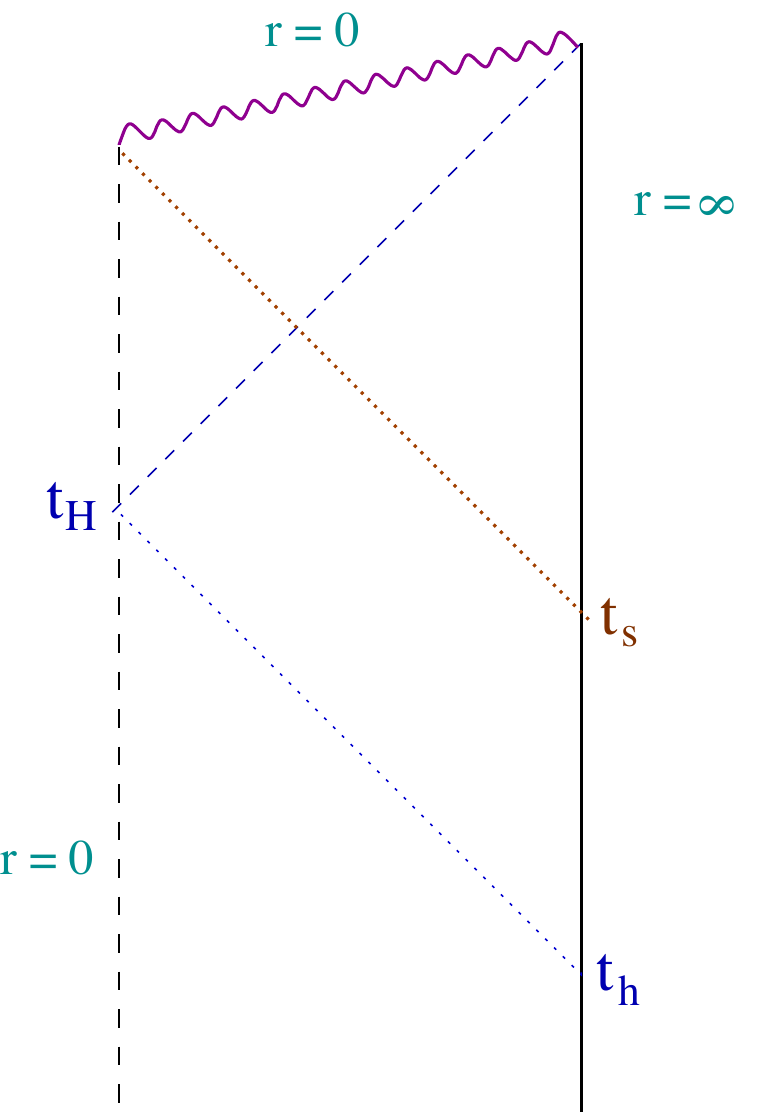}
\caption{Penrose diagram for collapse in AdS.  The vertical solid line represents the AdS boundary, the vertical dashed line the AdS origin, and the upper wavy curve the curvature singularity, produced by a null shell which originates on the boundary at time $t_s$ (CFT dual of this process is discussed in \sec{ss:bhform}), and collapses to zero area.  In the bulk, event horizon (dashed diagonal blue line) nucleates at bulk time $t_H$, whose signal can be seen by bulk-cone singularity with one insertion at $t_h$ and the other at $t= \infty$.}
\label{fig:collapse_bcs}
\end{center}
\end{figure}

This remarkable result has important and useful applications.
Because the endpoints of null geodesics through the bulk depend on the bulk geometry, the locus of the bulk-cone singularities changes as the bulk geometry changes.  In \cite{Hubeny:2006yu} the nature of these changes was explored for AdS deformed by a presence of a radiation ``star", a collapsing shell, and an eternal black hole; in each case, the pattern of bulk-cone singularities exhibits qualitatively distinct features.  In particular, in the collapse scenario, \cite{Hubeny:2006yu}  demonstrate that a sharp horizon-formation time can indeed be extracted from the pattern of singularities; cf.\ \fig{fig:collapse_bcs}: Consistently with the teleological nature of the event horizon, the relevant bulk cone singularity occurs for characteristic insertion points, one at infinite time $t=\infty$, the other null-separated from the location of the horizon formation event, $t=t_h$.  This is because the corresponding null geodesic (drawn as diagonal blue dotted/dashed line from $t_h$ in \fig{fig:collapse_bcs}) is the latest null geodesic to reach the boundary.  Note that the time $t_h$ precedes the formation of the shell, $t_s$, while $t_H$ and $t_s$ are spacelike-separated.  Moreover, as the black hole is about to form, the separation $(t_o-t_i)$ between the insertion points $(t_i,t_o)$ of the bulk-cone singularities grows without bound  as $t_i \to t_h^-$ ; in fact, the exact evolution of this separation should carry information about criticality in black hole collapse, such as that present in context of Choptuik scaling \cite{Choptuik:1992jv}.

Given that distinct geometries typically lead to distinct endpoints of null geodesics through these geometries, one might well ask whether one can invert this relation to actually {\it extract} the bulk metric from the locus of the bulk-cone singularities.  
At first sight this appears a daunting task since each geodesic passes through a 1-dimensional set of spacetime events, so there is no natural map between a single geodesic and a single spacetime event.  On the other hand, each spacetime event in $(d+1)$-dimensional bulk is pierced by a $(d-1)$-dimensional set of null geodesics, each specified by its direction at that event.  Moreover, by continuity nearby geodesics pass through nearby events; but since the separation between the nearby geodesics is typically larger in some regions than in others, such regions dominate the corresponding changes in the bulk-cone singularities. This motivates the expectation that if one takes the combined information carried by an appropriate set of geodesics, one can recover information about the metric in a localized region.
However, to what extent one can carry out this metric extraction in general remains an interesting open question.

The simplest setting in which to investigate this question is to focus on static, spherically symmetric spacetimes.  The most general such spacetime can be written in the form
\begin{equation}
ds^2 = -f(r) \, dt^2 + h(r) \, dr^2 + r^2 \, d\Omega^2 
\label{genstatssmet}
\end{equation}	
where $f(r)$ and $h(r)$ are arbitrary\footnote{
Here we ignore the bulk field equation, or equivalently assume no knowledge of the bulk matter content, and ask whether the metric can be recovered solely from the restricted set of boundary data given by the bulk-cone singularities.
} functions of one variable with the prescribed boundary conditions.  Spherical symmetry allows us to consider only the equatorial plane, so that the geodesics can be described in 3-dimensional $(t,r,\varphi)$ space, and from each boundary point $(t,\varphi)$ at $r=\infty$ there emanates a 1-parameter family of null geodesics, specified by the `angular momentum per energy' $\ell$.  On the boundary, time translation and rotational invariance imply that the bulk-cone singularity locus is specified by $(\Delta t(\ell), \Delta \varphi(\ell))$, i.e.\ one function's worth of data $\Delta t(\Delta \varphi)$.  It is then natural to expect that without additional information, this single function  does not suffice to recover two independent functions $f(r)$ and $h(r)$.  Indeed, from the bulk perspective, this is consistent with the fact that null geodesics are insensitive to conformal rescaling of the metric.  Nevertheless, the above counting suggests that once we fix the conformal factor, we should be able to recover the metric in at least some region.  Explicit reconstruction was carried out in \cite{Hammersley:2006cp, Bilson:2008ab} for spacetimes of the form \req{genstatssmet} with $h(r) = 1/f(r)$, using numerical and analytical means respectively.  Considering various ad-hoc examples of $f(r)$, the authors demonstrated that using only the corresponding bulk-cone singularities, $f(r)$ can be recovered for all $r>r_{\rm min}$, where either $r_{\rm min} = 0$ or it corresponds to a null circular orbit.  Furthermore, if  the available boundary data includes the endpoints of not only null, but also spacelike geodesics at  constant $t$, these works demonstrated extraction of both $f(r)$ and $h(r)$ in \req{genstatssmet}.  

These examples illustrate that by using the non-local nature of correlation functions one can obtain information about the details of the bulk spacetime.  In fact, this is already a vast improvement to the previously-mentioned avenues to decoding the bulk; in particular, it is not reliant on long-wavelength regime.   Moreover, in contrast to the previous discussions of probing the black hole singularity in AdS/CFT \cite{Fidkowski:2003nf,Festuccia:2005pi}, the use of bulk-cone singularities does not rely on analytic properties of the CFT correlators.  
Nevertheless, as indicated above, bulk cone singularities have several shortcomings of their own, primarily related to general properties of null geodesics.  First, they are insensitive to conformal rescaling of the bulk geometry; hence they cannot determine the metric fully.   Second, since null geodesics are causal, bulk cone singularities can never be used to probe the geometry inside an event horizon.  In fact, using the methods mentioned above for recovering $f(r) = 1/h(r)$ in \req{genstatssmet}, one cannot even probe the geometry beyond a null circular orbit (which is separated from the horizon by a factor of ${\cal O} (r_+)$).  Finally, another limitation of using the bulk-cone singularities to extract information about the bulk geometry arises from the fact that it is well-tailored mainly to asymptotically {\it globally} AdS spacetime, i.e.\ the field theory living on a compact space.  When the field theory lives on e.g.\ $R^{3,1}$, there may be no bulk-cone singularities, as demonstrated in \cite{Hubeny:2006yu} for any state respecting the same symmetries.

\subsection{Detour: surfaces of different dimensionality as probes of geometry}
\label{s:surfprobe}

So far, we have seen that although CFT correlators, in particular the locus of bulk-cone singularities, may encode a large amount of information about the bulk geometry and for certain class of states recover the bulk metric entirely, they nevertheless do not suffice in general to recover the full metric.  We will therefore take a brief detour to motivate our expectations as to which CFT observables would be likely to provide better probes of the bulk geometry.

Typically, geometrically defined objects such as geodesics, extremal surfaces, etc., which are anchored on the boundary, correspond to some probe (or observable) in the dual CFT.  For example, we have already seen the relation between geodesics and correlators in \sec{s:correls}, we have mentioned the well-known relation between the area of two-dimensional minimal surface describing a string world-sheet and the Wilson-Maldecena loop \cite{Maldacena:1998im} in \sec{s:projectiles}, and in \sec{s:entent} below we will use the relation between co-dimension two extremal surface and entanglement entropy \cite{Ryu:2006bv,Ryu:2006ef}.
One may therefore ask, purely as a geometrical question in the bulk, how far into the bulk can such surfaces penetrate.  The motivation behind this question is the tacit assumption that we can extract the information about the geometry to wherever the bulk dual of our CFT probes can reach.  This is in turn motivated by our experience with geodesics mentioned above, as well as with certain higher-dimensional surfaces discussed in \sec{s:entent} below.  Conversely, it is clear that if the bulk dual of a given CFT probe does not pass through a specific bulk region, then it does not carry any information about the spacetime geometry in that region.

Let us for simplicity continue considering only static spherically symmetric spacetimes of the form \req{genstatssmet}  discussed above.  Geodesics in such a spacetime can be described in terms of effective potential describing the radial motion and conserved quantities $E$ and $L$ conjugate to the temporal and rotational Killing fields.  The effective potential is given by
\begin{equation}
\dot{r}^2 + V_{\rm eff}(r) = 0 \ , \qquad \qquad
V_{\rm eff}(r) = \frac{1}{h(r)} \, \left[ -\kappa - \frac{E^2}{f(r)} + \frac{L^2}{r^2} \right]
\label{Veffgen}
\end{equation}	
where $\kappa = -1,0,1$ for timelike, null, and spacelike geodesics, respectively.
We are most interested in the radial turning point of the geodesic, denoted by $r_{\rm min}$, which corresponds to the largest root of $V_{\rm eff}$.   In particular, we wish to ask, given generic $f(r)$ and $h(r)$ describing our asymptotically AdS spacetime, how small can we make $r_{\rm min}$ by adjusting $\kappa$, $E$, and $L$?
It is clear from the form of the potential that for fixed $E$ and $L$, null geodesics have higher effective potential (outside the horizon) than spacelike ones, which in turn means that the corresponding $r_{\rm min}$ will be larger for a null geodesic than for the corresponding spacelike one.   This quick argument is further substantiated by actually determining where the turning point can occur.  This is carried out in \cite{Hubeny:2010aa} ; here we will only summarize the main points:
\begin{itemize}
\item 
For causally trivial spacetimes without an event horizon, one can probe the entire spacetime, both with spacelike and with null geodesics.  In particular, for radial geodesics, $r_{\rm min} = 0$.

However, here it is interesting to ask a more refined question: suppose we have access only to a certain region on the boundary, in particular we can only use geodesics with both endpoints in that region; out of this constrained set of geodesics, which one probes the deepest?  While the answer depends on the details of the geometry, in AdS we can explicitly confirm that for a fixed angular span of the endpoints, $r_{\rm min}$ is minimized for $E=0$ (which necessarily implies spacelike geodesic) and grows monotonically with $E^2$.  In fact, null geodesics probe the bulk least, as in the limit of $E \to \infty$ the spacelike geodesic approaches a null one.
\item
For spacetimes with an event horizon, null geodesics can probe only down to a finite distance outside the horizon, whereas specific spacelike geodesics (namely with $E=0$ and $L$ approaching the horizon size from above) can probe arbitrarily close to the horizon.\footnote{
More specifically, in all cases the geodesics can only probe to depth given by circular orbit; for null geodesics this location is fully specified by the geometry, whereas for spacelike geodesics, by adjusting $L$ we can bring this orbit arbitrarily close to the horizon.
}  However, no geodesic with both endpoints anchored to the boundary can probe past the horizon.
\end{itemize}
It is important to keep in mind that the above statements pertain to static spacetimes of the form \req{genstatssmet}, and do not necessarily hold for rapidly evolving spacetimes.  In particular, if one collapses a black hole sufficiently quickly, spacelike geodesics {\it can} penetrate inside the event horizon.  This can be easily seen by the simple gedanken experiment of \cite{Hubeny:2002dg}, whose main point was to argue that precisely due to its teleological nature, the event horizon cannot pose a limitation to how far such bulk probes can reach.
The idea, sketched in \fig{fig:precursors}, is simply the following:
\begin{figure}
\begin{center}
\includegraphics[width=1.5in]{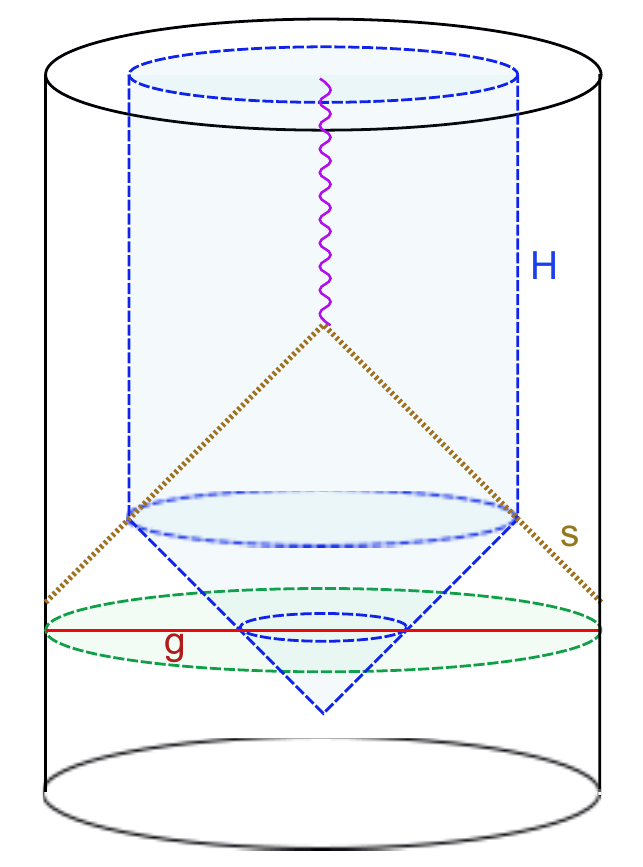}
\caption{Gedanken experiment of \cite{Hubeny:2002dg} demonstrating that spacelike geodesics $g$ (or minimal surfaces) which are anchored on the AdS boundary can pass through an event horizon $H$ of a dynamically evolving spacetime, here exemplified by a null shell $s$ collapsing to form the black hole (shaded blue).
The corresponding Penrose diagram is sketched in \fig{fig:collapse_bcs}.}
\label{fig:precursors}
\end{center}
\end{figure}
 Suppose we collapse null shell with sufficiently high energy to create a very large black hole.  Inside the shell, the spacetime is that of pure AdS, so the spacelike geodesics just to the past of the shell creation probe the full spatial slice of the bulk; yet the event horizon from the black hole formed by the shell extends though this pure AdS region, and therefore gets penetrated by these unsuspecting geodesics (such as $g$ in \fig{fig:precursors}).

Having classified the behaviour of geodesic probes, let us now turn to higher-dimensional surfaces. The motivation for considering such surfaces is that one has access to observables in the field theory whose holographic avatars involve such surfaces. The fully story for these higher dimensional surfaces is rather more involved, so we will mostly draw lessons from cases with sufficiently high degree of symmetry.   In particular, we will keep the bulk spacetime static and we will consider a bulk minimal surface at constant time anchored on two parallel $(n-1)$-dimensional planes or an $(n-1)$-sphere on the boundary, with the boundary metric that of $d$-dimensional Minkowski spacetime.  To warm up we'll start with the bulk spacetime being that of pure AdS in Poincare coordinates, and then move on to a more general class of asymptotically AdS bulk spacetimes.  While some of the results mentioned below are well-known and have been derived previously (see e.g.\ \cite{Ryu:2006ef, Hubeny:2007xt}), the following presentation is based on \cite{Hubeny:2010aa} which contains a more systematic analysis.
 
 Let us start with pure AdS$_{d+1}$ in Poincare coordinates, 
\begin{equation}
ds^2 = \frac{1}{z^2} \left(-dt^2 + dx^2 + d\vec{y}^{\, 2} + dz^2\right) ,
\label{}
\end{equation}	
where $z$ is the bulk radial coordinate with $z=0$ corresponding to the AdS boundary.
 One of the simplest cases to consider is an $n$-dimensional minimal surface anchored on the boundary of a ``strip" of width $\Delta x$, with infinite extent in the other $n-1$ directions. The cross section $z(x)$ of the corresponding minimal surface is obtained by minimizing its area. 
 Since the area depends on $n$, the cross-section $z(x)$ likewise depends on $n$, despite the translational invariance in $n-1$ directions.  Although one can easily obtain the cross section analytically, here we content ourselves with indicating its behaviour graphically.  
\begin{figure}
\begin{center}
\includegraphics[width=3in]{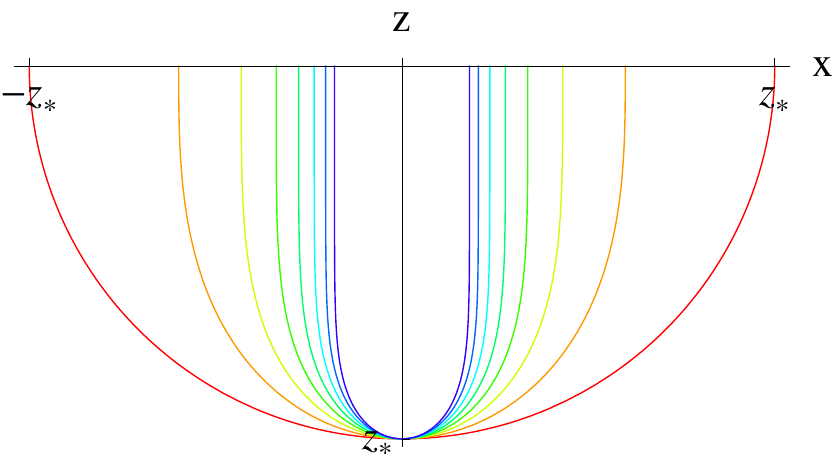}
\hspace{1.5cm}
\includegraphics[width=2.4in]{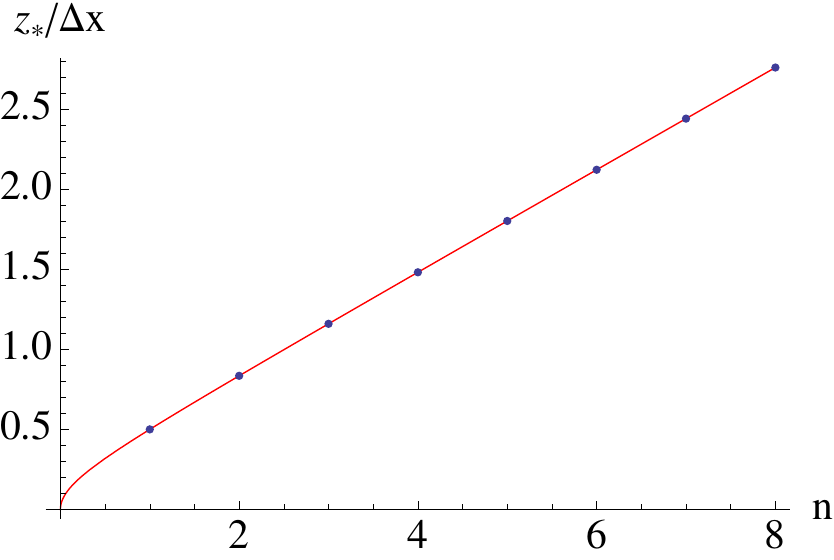}
\caption{
Left: cross-section of $n$-dimensional minimal surfaces in Poincare AdS$_{d+1}$, with varying dimensionality $n=1,2,\ldots,8$: the outermost (red) curve corresponds to $n=1$ while the innermost (purple) curve to $n=8$.  Note that $d$ (as long as it's high enough to accommodate the surface) does not enter.  Right: corresponding ratio of maximal bulk radial extent $z_{\ast}$ to its boundary size $\Delta x$.
}
\label{fig:minsurfAdS}
\end{center}
\end{figure}
The left panel of \fig{fig:minsurfAdS} shows the cross section of an $n$-dimensional minimal surface for $n=1,2,\ldots,8$, with the outermost curve representing $n=1$, i.e.\ a spacelike geodesic, which has semi-circular shape.  We see that as we increase the dimensionality, for a fixed depth $z_{\ast}$ to which such a surface reaches, the strip width $\Delta x$ decreases.  Conversely, keeping $\Delta x$ fixed, higher dimensional surfaces reach deeper into the bulk.  This is summarized in the right panel of  \fig{fig:minsurfAdS} where the ratio of depth $z_{\ast}$ probed  to boundary extent $\Delta x$ is plotted as a function of $n$; we see that it increases approximately linearly.\footnote{
The actual relation is somewhat more complicated, given by 
$z_{\ast} = \frac{\Delta x}{2} \, \frac{n+1}{\sqrt{\pi}} \, 
\frac{\Gamma \left( \frac{2n+1}{2n} \right) }{\Gamma \left( \frac{3n+1}{2n} \right) }$.
}

One might conclude from this observation that higher-dimensional surfaces are therefore better probes of the bulk geometry.  This will indeed partially motivate \sec{s:entent}; but the indication given above is a bit too glib.  In particular, we have only considered the extent in the $x$ direction and ignored the fact that the higher dimensional surfaces utilize extra directions of infinite extent.  A slightly better comparison would  therefore involve a boundary region of finite extent in all directions.  The most natural such region is a $(d-1)$-ball (with some specified radius $R_0$) in $\R^{d-1,1}$.  Let us therefore compare minimal surfaces anchored on $\Sp^{n-1}$ of radius $R_0$, for different values of $n$.  Although the equation of motion for such a surface depends on $n$, this equation is satisfied by a very simple solution\footnote{
Here $\bar{r}$ denotes the boundary radial variable, such that the corresponding boundary metric takes the form
$ds^2 = \frac{1}{z^2} \left[ -dt^2 + d\bar{r}^2 +\bar{r}^2 \, d\Omega_{n-1}^2+ d\vec{y}^{\, 2} + dz^2\right] $,
and by $(n-1)$-spherical symmetry, the minimal surface will be described by a single function $z(\bar{r})$.
} $z(\bar{r}) = \sqrt{R_0^2 - \bar{r}^2}$ for all $n$ -- that is, the minimal surface is simply an $n$-hemisphere, whose bulk extent $z_{\ast}= R_0$ is independent of $n$.  However, it is clear that this can happen only due to a special cancellation in the bulk geometry; for general geometries the minimal surface profile $z(r)$ {\it would} depend on $n$.   The interesting question, then, becomes, for a fixed boundary region specified by $R_0$, how does the depth probed $z_{\ast}$ vary with $n$?

This question can be studied numerically for various spacetimes.  Perhaps not surprisingly, it turns out that for ``sensible" spacetimes, such as the planar Schwarzschild-AdS black hole, similar result as in \fig{fig:minsurfAdS} holds: the higher-dimensional surfaces again probe deeper.  However, one can easily write a bulk spacetime metric violating energy conditions for which the opposite effect takes place: lower-dimensional surfaces probe deeper.  Nevertheless, since such cases are unphysical, we will take the black hole case as the prototypical one.
The basic rule of thumb is then very simple and consistent with naive expectations based on the UV/IR duality:
If one has access to only a certain region ${\cal A}$ on the boundary, and therefore is constrained to probes contained entirely within that region, then one can probe maximal amount of the bulk geometry by utilizing the full region ${\cal A}$.  In particular, if ${\cal A}$ is $d-1$ dimensional, and thus covers some open set in the boundary space, then one should use probes dual to the full $d-1$-dimensional surfaces in the bulk for probing the bulk geometry.

Finally, ere ending this section, let us revisit the issue of penetrating the horizon of a static bulk spacetime in the context of these higher-dimensional minimal surfaces.  We have seen thtat geodesics with both endpoints anchored on the boundary cannot reach past a horizon, based on quite general arguments independent of the details of the bulk geometry.  But we have also seen that higher-dimensional surfaces can probe deeper then geodesics; so one might naturally wonder whether such surfaces could in fact be used to probe past the horizon.  The answer turns out to be negative, and similarly robust as for the geodesic case: no minimal surface anchored on the boundary can reach past the horizon.  Therefore, in this sense, if we have access to the full boundary, for static spacetimes, spacelike geodesics are as good probes as higher-dimensional extremal surfaces, since in each case, the horizon is the limit to how deep these types of probes reach.  Although we have so far focused on static geometry, shortly in \sec{s:entent} we will see an example of the behaviour of extremal surfaces for dynamically evolving geometry describing a black hole collapse.

\subsection{Entanglement entropy as probe of bulk dynamics}
\label{s:entent}

Motivated by the discussion in \sec{s:surfprobe}, we are naturally led to consider spacelike extremal hypersurfaces anchored on the boundary with highest dimensionality.  
Ryu and Takayanagi \cite{Ryu:2006bv,Ryu:2006ef} proposed that in static geometries, the area of a bulk minimal co-dimension two spacelike surface (at a constant time) anchored on the boundary $\partial {\cal A}$ or certain region ${\cal A}$ on the boundary of AdS gives the entanglement entropy of the region ${\cal A}$ in the dual CFT state.\footnote{
Justification of this remarkable statement was subsequently provided in \cite{Fursaev:2006ih} using replica trick to evaluate entanglement Renyi entropies; although an error in this derivation was recently pointed out by \cite{Headrick:2010zt}, the correction nevertheless gives further evidence for the correspondence.}
 This was generalized in \cite{Hubeny:2007xt} to general time-dependent configurations, where the entanglement entropy is given by the area of an extremal surface, which is in fact related to  light-sheet constructions of the covariant entropy bound \cite{Bousso:1999xy} in the bulk spacetime.
It is gratifying to note that such class of special highest-dimesionality surfaces are related to an appropriately special notion on the dual side: entanglement entropy is an important concept in field theory systems, with applications to condensed matter systems, quantum information, \etc. Further, given its non-extensive nature \ie, area scaling  \cite{Srednicki:1993im,Kabat:1994vj}, one is tempted,  in the context of holography, to  think of it as providing a measure for the effective degrees of freedom associated with a given region.  This is indeed supported by the proposal of \cite{Ryu:2006bv,Ryu:2006ef} and its covariant extension  \cite{Hubeny:2007xt}.\footnote{Further evidence of the entanglement entropy providing a measure of the active degrees of freedom is provided in \cite{Emparan:2006ni} in the context of induced gravity (brane-world) models. See also \cite{VanRaamsdonk:2009ar} for recent attempts to reverse-engineer spacetime geometry from knowledge of entanglement entropy.}

To be slightly more specific, let us recall the definition of entanglement entropy and the holographic prescription for computing it. Consider a quantum field theory on $\partial{\cal M} = {\cal N} \times {{\rm \bf R}}_t$ and focus on a region ${\cal A}\subset {\cal N}$. Given a density matrix $\rho$ (or a pure state) on ${\cal N}$ we define the reduced density matrix $\rho_{\cal A} = {\rm Tr}_{{\cal N} \, \backslash {\cal A}} ( \rho)$. The entanglement entropy is then simply the von Neumann entropy associated with $\rho_{{\cal A}}$: $S_{{\cal A}} = - {\rm Tr} \left( \rho_{{\cal A}} \, \log \, \rho_{\cal A} \right)$. 
The covariant holographic prescription \cite{Hubeny:2007xt} for computing this  is to consider the problem of finding a co-dimension two extremal surface ${\cal W}_{\cal A}$ in the bulk geometry  ${\cal M}$ (dual to the state in question), which ends on the boundary of the chosen region ${\cal A}$ at the spacetime boundary $\partial{\cal M}$. $S_{\cal A}$ is then given by the area of ${\cal W}_{\cal A}$ in Planck units \ie, $S_{\cal A} = {{\rm Area }\left({\cal W}_{\cal A}\right) \over 4 G_N}$. In the situation where the bulk geometry is static (and hence the dual state is time-invariant) the problem reduces to the simpler one of finding minimal area surfaces \cite{Ryu:2006bv,Ryu:2006ef}. An excellent summary of the developments in understanding this holographic entanglement entropy can be found in \cite{Nishioka:2009un}.

\begin{figure}[htpb]
\begin{center}
\includegraphics[width=4in]{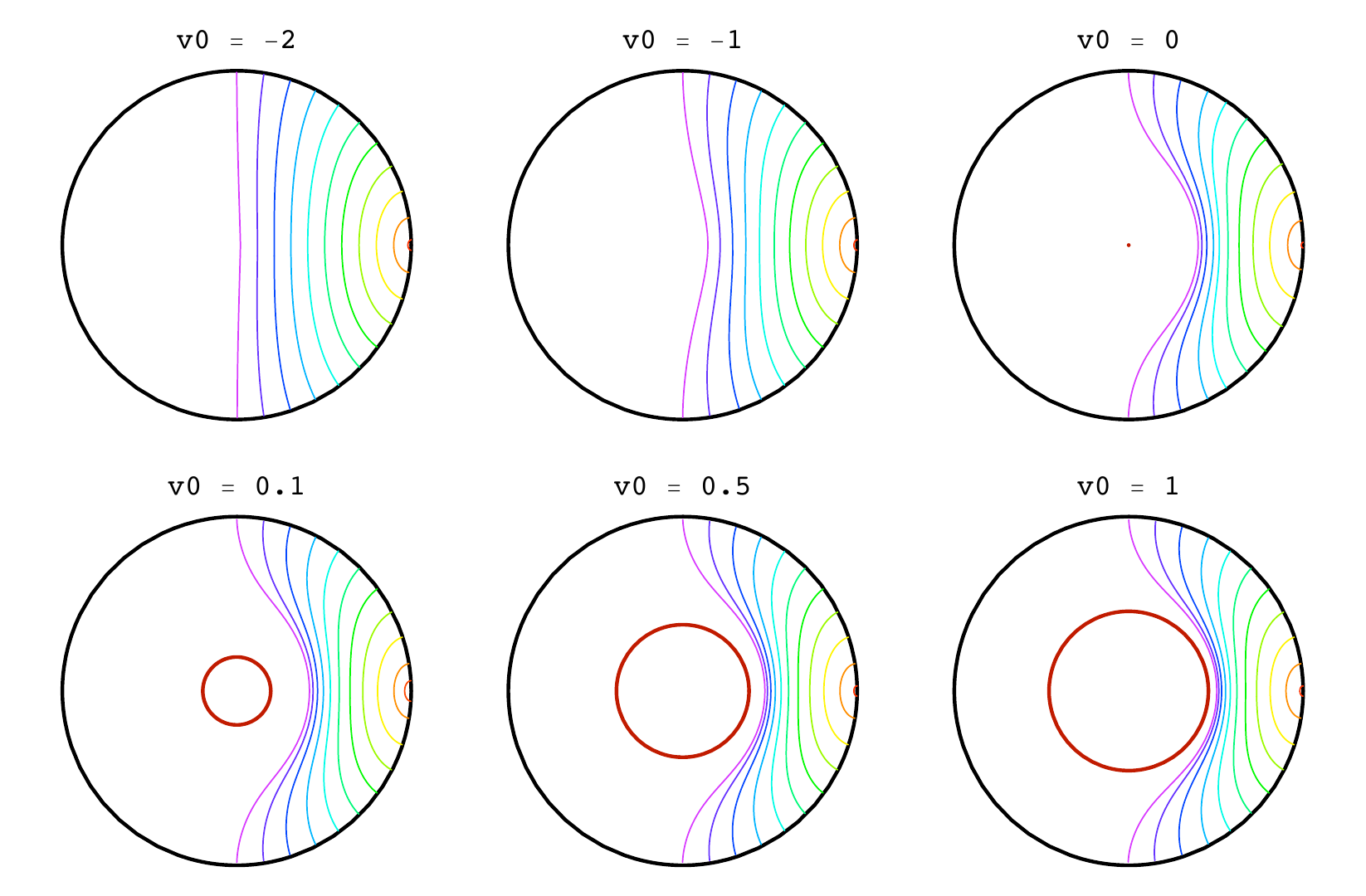}
\caption{Spatial projection of extremal surfaces in Vaidya-AdS (collapsing black hole) geometry, calculated at various times in the black hole formation process.  The outer (black) circles correspond to the AdS boundary at time given by the $v0$ indicated; the inner (red) circles correspond to the event horizon cross-section at corresponding time.  The remaining curves correspond to the extremal surfaces anchored on the boundary, colour-coded by the size of the boundary region they encompass.
At early times $v0$ before the black hole forms, these extremal surfaces coincide with minimal surfaces in pure AdS.  We can see that as the black hole grows, these surfaces are `repelled' to remain outside the horizon.}
\label{fig:vaidya_bcs}
\end{center}
\end{figure}

Let us now apply this prescription to the interesting case of black hole collapse. In \cite{Hubeny:2007xt} a setting rather similar to the ones discussed earlier in \sec{ss:bhform} was considered to ascertain the time evolution of entanglement entropy. In particular, these authors considered the Vaidya solution in global AdS,
\begin{equation}
ds^2=-\left(r^2+1-\frac{m(v)}{r^{d-2}}\right)\, dv^2+2\, dv \, dr+r^2\,
d\Omega_{d-1}^2 \ .
\label{}
\end{equation}	
with $m(v) = \tanh v$ smoothly interpolating between 0 (corresponding to pure AdS) and 1 (corresponding to \SAdS{}), in 3 bulk dimensions.  The corresponding extremal surfaces (which in 2+1 dimensions are simply spacelike geodesics) are plotted in \fig{fig:vaidya_bcs}.  At early times, before a black hole forms, these minimal surfaces probe the entire bulk, $r \in (0,\infty)$.  However, as the black hole forms and grows, these surfaces are repelled by the horizon, so that the region of spacetime they are sensitive to is only $r \in (r_+(v),\infty)$.

\begin{figure}[htbp]
\begin{center}
\includegraphics[width=4in]{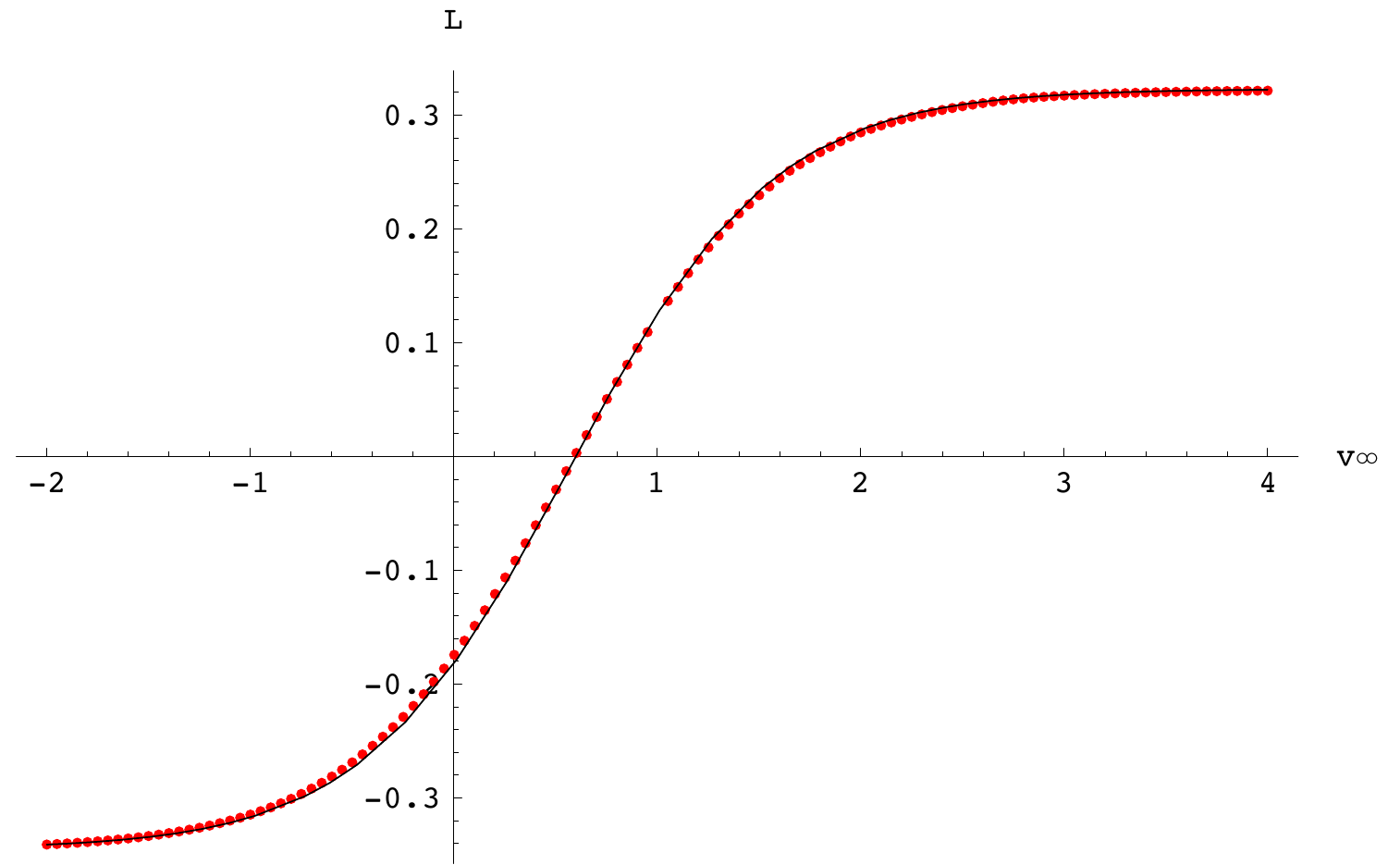}
\caption{ Regularised proper length $L_{reg}$, or equivalently, the entanglement entropy $\CS_\CA$,  as a function of the
boundary $v_{\infty}$, for the particular region (an arc of fixed length along the spatial direction) in the Vaidya spacetime (red dots) and the corresponding analytic 
prediction in BTZ (black curve). } \label{f:LvarfitVad}
\end{center}
\end{figure}

Having obtained the extremal surfaces  for the Vaidya-AdS geometry, we can compute the entanglement entropy of the region $\CA$ using the area of these surfaces. In the case of Vaidya-AdS$_3$ spacetimes one can in fact carry out this computation explicitly, since the extremal surfaces are simply spacelike geodesics in the geometry. The entanglement entropy is then just the regularized length $L_{reg}$ of these geodesics. 
In fact, the calculation can be done analytically if the variation of the mass is slow, analogously to the discussion in \sec{s:flugeom}. The numerical evaluation of the time-dependent entanglement entropy in this case and its comparison to the analytic approximation is shown in \fig{f:LvarfitVad}.
%
For matter fields in the bulk satisfying the null energy conditions, and for a fixed boundary region, one can can show that the
entanglement entropy increases (at least  in the adiabatic approximation). Hence assuming that the matter undergoing collapse to form the black hole satisfies the null energy condition, it is evident from the adiabatic approximation that the entanglement entropy $\CS_\CA(v) \propto L_{reg}(v)$ increases in the process of a gravitational collapse.

As suggested by our discussion above, it is clear geometrically that the entanglement entropy being described by an extremal surface should be sensitive to the causal structure of the spacetime. This is borne out explicitly in the computations of \cite{Hubeny:2007xt} which provide a strong coupling counterpart to the field theoretic results on temporal evolution of entanglement entropy \cite{Calabrese:2005in}. 

In summary, we have seen that there exists a rich variety of bulk probes, with different merits.  The bulk-cone singularities discussed in \sec{s:correls} probe smallest region of the bulk, but are quite robust and presumably easiest to extract as CFT `data'.  Spacelike geodesics or higher-dimensional minimal surfaces probe deeper, and as we've explicitly seen in \sec{s:surfprobe}, can be used even when the boundary lives on a non-compact flat spacetime.  Moreover, unlike bulk-cone singularities, these probes are all sensitive to the conformal factor of the bulk spacetime, but on the other hand, for static geometries, any surface at constant time is completely insensitive to the time component of the metric.    Nevertheless, by considering finite-energy spacelike geodesics, or extremal surfaces which are not localized to constant time, we can probe the temporal aspect of the geometry as well.  Although part of our discussion focused on static geometries because these allowed us to make more general statements and were easier to analyze, the most important point of this section in the present context  is that these probes can tell us much about physics far out of equilibrium.

\section{Conclusion}
\label{s:concl}

In this Review we have focussed on various aspects of non-equilibrium physics that can be accessed via holographic methods. The intriguing aspect of studying dynamical phenomena in AdS/CFT, apart from the intrinsic interest in developing computational techniques to facilitate investigation of strong coupling physics, is the natural mapping of these dynamics into the time dependent phenomena in a classical bulk spacetime. 

We have in the course of this Review touched upon various developments in the field, ranging from linear response regime, to the hydrodynamic regime, and finally to the fully non-linear regime. While aspects of the physics in various situations are by now well understood, it is clear that there are still open issues. In the following we will discuss some of the interesting open questions which could provide a fruitful avenue for further progress.

In \sec{s:probes} we have mentioned some of the seminal attempts to model probes moving through a thermal medium, which is of relevance to the QGP phenomenology. 
The thermal medium is provided by a black hole geometry in the gravitational dual, while a probe in the form of an external quark is modelled by a fundamental open string (ending on the quark) in this background.   The motion of the quark determines the trajectory of the string's world-sheet, which in turn backreacts on the bulk metric to induce a non-trivial stress tensor on the boundary.  The latter describes not only the bare quark, but also the cloud of partons around it, as well as the radiation produced by the quark's motion.  It is interesting that these apparently distinct manifestations of the quark's motion are unified by a single fundamental quantity, namely the string, in the bulk.  Conversely, given the bulk string, one might want to ask, which part of this string describes the quark and its surrounding halo, and which part the radiation?\footnote{\label{f57} It is in fact commonly assumed that the part inside the world-sheet horizon corresponds to the radiation, since this portion can no longer causally influence the quark, whereas the part above the horizon contributes to the halo.} In this context, it has been argued in \cite{Chernicoff:2008sa,Chernicoff:2009re,Chernicoff:2009ff} building on the formalism developed in \cite{Mikhailov:2003er} that it is indeed possible to separate the total four-momentum of the string into a part intrinsic to the boundary quark, and another that corresponds to the radiation. The results of these computations seem to differ from the commonly assumed picture quoted in footnote \ref{f57}, which has been advocated earlier in  \cite{Dominguez:2008vd,Xiao:2008nr,Beuf:2008ep}.\footnote{We thank Alberto Guijosa for alerting us to these references.}

The fluid/gravity correspondence was reviewed in \sec{s:flugra}. From the geometrical standpoint, several intriguing issues deserve further investigation.  Note that the fluid/gravity correspondence establishes a one-to-one map between long-wavelength gravitational solutions and solutions of fluid dynamics.  Naively, one might think of this as merely a low-energy effective description of the AdS/CFT correspondence.  However is it not true that any gravitational solution in AdS admits a fluid description.  In particular, it would be useful to understand the role of non-long-wavelength bulk classical or semiclassical solutions, such as small AdS black holes.  Likewise, there is room for more detailed bulk analysis, such as probing the allowed horizon topology and dynamics in more general situations, the nature of curvature singularity, or Cosmic Censorship.

More generally, the fluid/gravity correspondence offers new insight into the black hole membrane paradigm, \cite{Hubeny:2009zz}.  As described earlier, conventional membrane paradigm (cf., \cite{Thorne:1986iy}) provides a simple picture of black hole dynamics in terms of classical physics of fluid living on a ``membrane" (or stretched horizon) just outside the event horizon.   Taking a more general view of trying to encode the black hole dynamics by fluid dynamics localized on a membrane in the spacetime, the immediate natural question is:  where should such a membrane live?  Perhaps the most obvious candidate is the event horizon; but this is problematic due to its null nature, and more importantly, because it is globally defined so we can't fix its position without knowing the full future evolution of the spacetime.  Alternately, several (quasi)local notions of a black hole have been proposed, such as the so-called dynamical horizon, which however are spacelike surfaces inside the event horizon, and therefore do not admit the standard notion of evolution.  A more popular suggestion is the stretched horizon, which is the formulation given by the membrane paradigm \cite{Thorne:1986iy}.  However, there likewise remain ambiguities in localizing stretched horizon. Within the fluid/gravity correspondence, the full spacetime dynamics is mapped to the dynamics of a conformal fluid, which albeit reminiscent of the membrane paradigm, has one important twist: the membrane lives on the boundary of the spacetime (which is unambiguously defined and admits a fluid description with well-defined dynamics), and gives a perfect mirror of the bulk physics. This  ``membrane at the end of the universe" picture is a natural consequence of the holographic nature of the fluid/gravity correspondence. Recently this view has been countered in \cite{Eling:2009sj}, where the authors argue that the projection of Einstein's equations onto any null hypersurface leads in a long-wavelength approximation to hydrodynamical behaviour. While this is interesting, it does not address the objections given above and more importantly it seems to us somewhat unnatural that a feature common to all null hypersurfaces has something fundamental to do specifically with black hole horizons.\footnote{More recently \cite{Bredberg:2010ky} argue that a Wilson renormalization group  perspective allows one to relate the `membrane at the end of the universe' picture of the fluid-gravity correspondence with the conventional membrane paradigm story. In particular, the authors argue that as the location of the stretched horizon is brought in from the boundary to the actual horizon, one in fact sees an RG flow  from a relativistic conformal fluid to a non-relativistic incompressible fluid.}

As mentioned in \sec{s:intro}, one of the outstanding problems in fluid dynamics is understanding turbulence.  We might therefore ask how relevant is the fluid/gravity correspondence (appropriately generalized) to tackling such a problem.  We argue that the correspondence is indeed relevant and worth vigorously pursuing further (see \cite{VanRaamsdonk:2008fp,Bhattacharyya:2008kq,Eling:2009sj,Eling:2010vr} for thoughts on these issues). The important point is that the regime  where our derivative expansion is valid corresponds to  large Reynolds number (where the viscous terms are small relative to the leading order terms), so the phenomenology of turbulence should be directly relevant to the study of near-equilibrium AdS black hole dynamics. At least for the widely studied case of non-relativistic fluids described by the Navier-Stokes equations, turbulence has many striking qualitative features, including the tendency to dynamically break symmetries as Reynolds number is increased, the sharp onset of turbulence at critical Reynolds numbers, and an ``energy cascade'' in fully developed turbulence in which energy is transferred in a predictable way between eddies at different scales. Less is known about turbulence for the microscopically relativistic fluids relevant in our context, but it would be fascinating to understand the gravitational interpretation for those features that do generalize to the relativistic case.

Finally, one intriguing observation, made in \sec{ss:confsol}, is that in certain cases of large disparity between apparent and event horizon, such as in the conformal soliton setting analyzed in \cite{Figueras:2009iu}, the CFT dual entropy seems to be given by area of the apparent horizon rather than the event horizon as conventionally  expected.
Since this striking observation is rather surprising and, if true in general, would have far-reaching implications, let us end our discussion by probing this issue a bit deeper.  In particular, can it be true that the entropy is associated with the area of the apparent horizon?  The fact that CFT entropy is not determined by the area of the event horizon in general is perhaps not so surprising, given that the event horizon is a global construct -- its precise location behaves teleologically: it depends on the spacetime arbitrarily far in the future -- whereas CFT entropy at a given time is measured independently of what happens at a later time.\footnote{
Although some might argue that there is little distinction between the two cases, since  with certain analyticity assumptions on the spacetime the evolution of the entire spacetime is determined by the initial state (and the boundary conditions), we do not take this viewpoint:  there is a difference between simply measuring certain properties of a state at a given time, and evolving to that state from much earlier initial time -- the latter requires far more detailed information.
}
On the other hand, there are nevertheless serious problems with associating the area of an apparent horizon, or any other quasilocal\footnote{
For a good review of various notions of quasilocal horizons, see \cite{Booth:2005qc,Ashtekar:2004cn}.  Technically speaking,  an apparent horizon is defined as a co-dimension 2 surface, on a given leaf of foliation, corresponding to the outermost marginally trapped surface or the boundary of trapped points.  In fact, the set of apparent horizons on all leaves of a given foliation need not even form a smooth co-dimension 1 surface in the full spacetime.  In the present review, we will however stick to the more commonly used terminology in non-GR literature of apparent horizon signifying the full timelike co-dimension 1 surface.
}  horizon, to the dual CFT state entropy.  One difficulty is that  the entropy is expected to be smoothly varying in time, whereas the area of a given apparent horizon can jump discontinuously. 
A more problematic issue, however, is that the location of apparent horizons depends on a choice of spacetime foliation, whereas the CFT state at a given time carries no such information.
 This seems to imply that our bulk
prescription has more freedom or ambiguity in defining the entropy
than that afforded by the boundary theory. 

One possible resolution is that there is a preferred foliation of the bulk, such as a zero-mean-curvature slicing, on which one is supposed to evaluate the area.  However, we have no compelling justification for this option. A simpler resolution to this puzzle would be that in the regime where the concept of entropy is
meaningful, the horizon has to be evolving slowly enough that there is
negligible difference between the areas of all slices of horizon which
end on the same boundary time-slice.  This is essentially the same
picture as that advocated in \cite{Bhattacharyya:2008jc,Bhattacharyya:2008xc}, except that here we
use it for apparent horizon rather than the event horizon.
In effect the field theory on the boundary should achieve local equilibrium in order for entropy to constitute a meaningful observable.

However, intriguingly even this proposition cannot be the full story.
If there is no event horizon at early times, such as in the case of a black hole formed from a collapsed star, then we can find foliations without apparent horizon for {\it arbitrarily late times}.   
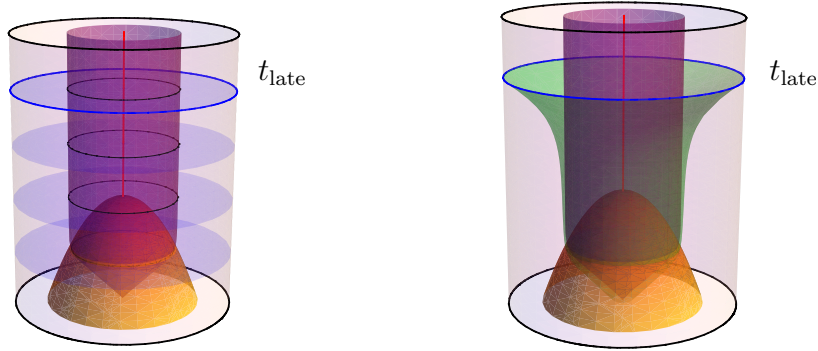
\begin{figure}
\begin{center}
\includegraphics[width=1.7in]{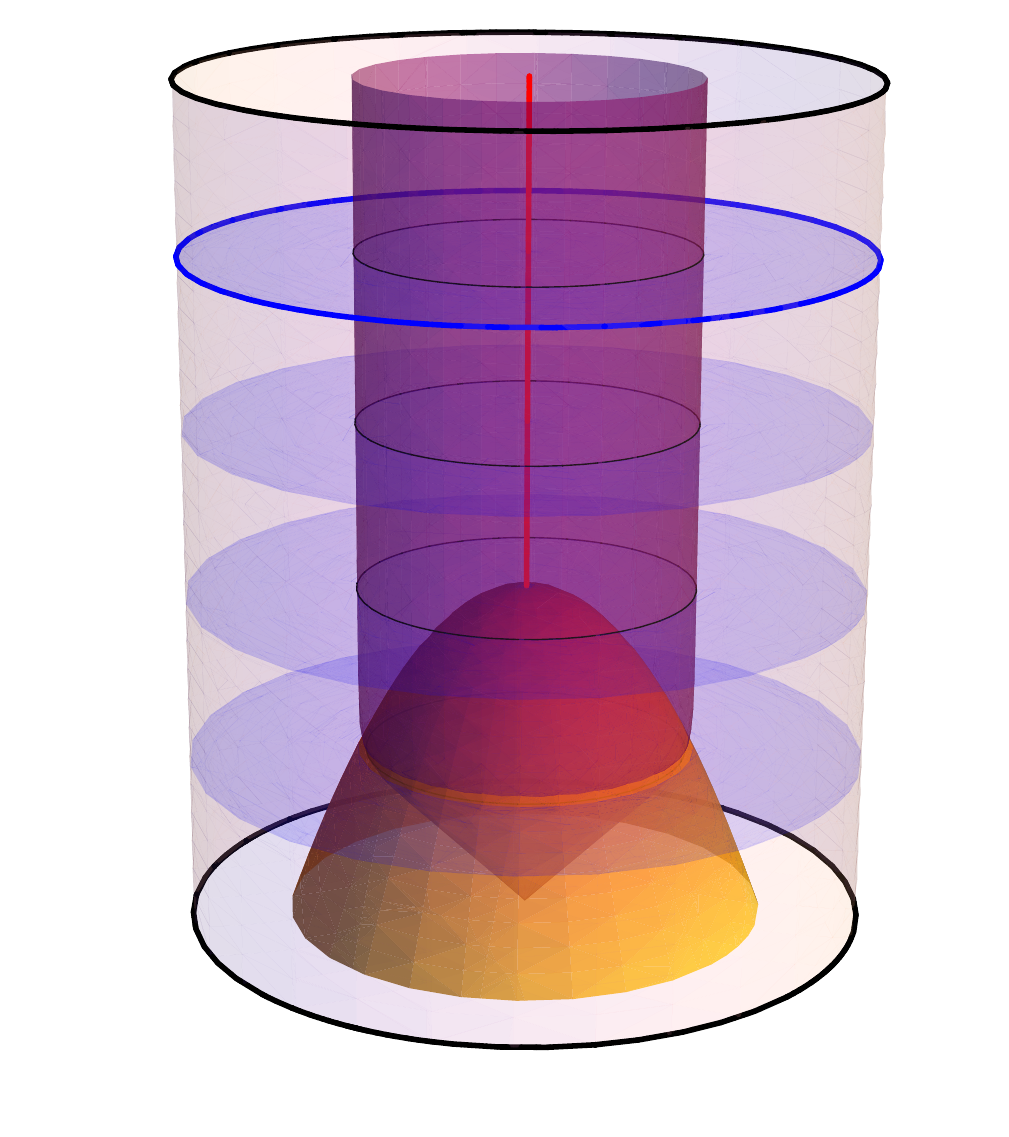}
\hspace{2cm}
\includegraphics[width=1.7in]{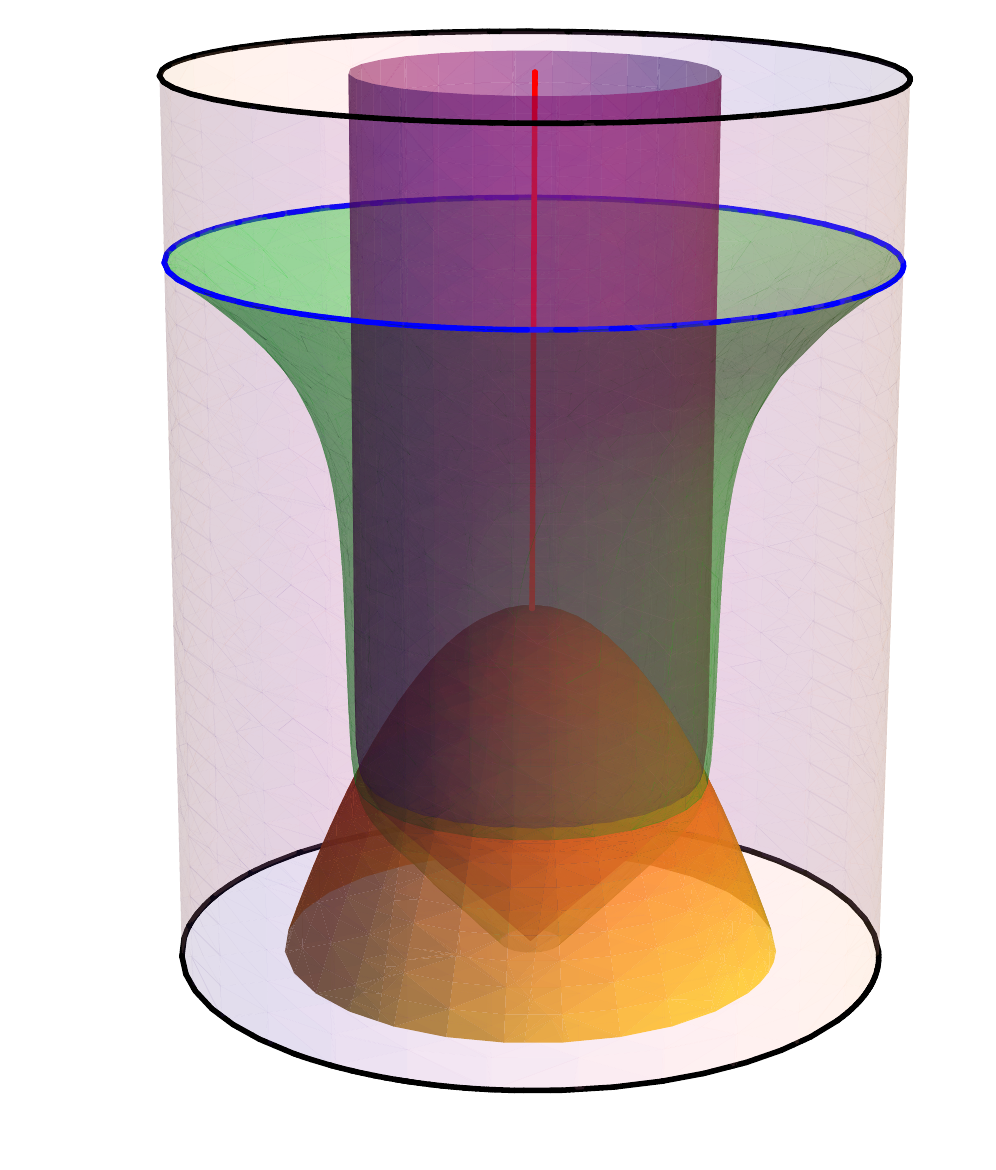}
\begin{picture}(0,0)
\setlength{\unitlength}{1cm}
\put(-7,3.9){$t_{\rm late}$}
\put(-0.2,3.9){$t_{\rm late}$}
\end{picture}
\caption{Possible spacelike foliations of a geometry a black hole collapse. The outer cylinder represents the AdS boundary; the bottom (orange) surface represents a collapsing star, the (red) vertical line the curvature singularity, the inner (purple) cylinder the event horizon.  Two distinct sets of foliations are sketched.  On {\bf left}, the corresponding apparent horizon has settled down at late time $t_{\rm late}$ (indicated in blue). On {\bf right}, the spacelike surface (green) at $t_{\rm late}$ admits no apparent horizon at all.}
\label{fig:foliations}
\end{center}
\end{figure}
A simple example of such a foliation is sketched in \fig{fig:foliations}. Suppose we collapse a black hole, and wait for very long time after the black hole has settled down, say time $t_{\rm late}$ indicated in \fig{fig:foliations}.   A `natural' foliation (sketched in the left panel) would give the expected results: by the time $t_{\rm late}$, the apparent horizon area has grown and settled down to a constant  which coincides with the event horizon area.  However, we can devise a foliation (sketched in the right panel) which, even at $t_{\rm late}$, admits no trapped surfaces and therefore has zero apparent horizon area.  The corresponding spacelike surface `dips just below' the event horizon formation; and since the event horizon itself is null, the requisite foliation surface can always be constructed so as to be spacelike.

It would be interesting to understand the full implications of this observation, since it bears on the discussion in \sec{s:intro} regarding the relation between boundary time and bulk time: which bulk region is most naturally associated with a given boundary time, such as $t_{\rm late}$?   Since the notion of entropy depends on amount of coarse-graining, perhaps this is the natural feature which the choice of bulk foliation reflects.

\subsection*{Acknowledgements}

It is a pleasure to thank our collaborators, Vijay Balasubramanian, Sayantani Bhattacharyya, Jan de Boer, Bartek Czech, Lukasz Fidkowski, Pau Figueras, Ben Freivogel, Gary Horowitz,  Matt Kleban, Klaus Larjo,  Hong Liu, R.\ Loganayagam,  Alex Maloney, Gautam Mandal, Don Marolf, Shiraz Minwalla, Takeshi Morita,  Rob Myers, Harvey Reall, Simon Ross, Steve Shenker, Masaki Shigemori,  Joan Simon, Dam Son,  Tadashi Takayanagi, Ethan Thompson and Mark Van Raamsdonk for wonderful collaborations on various aspects discussed in this review. We would also like to thank the Weizmann Institute and MIT for hospitality during the course of writing this review. We are partly supported by STFC Rolling grant and a STFC SPG grant.

\newpage

\providecommand{\href}[2]{#2}\begingroup\raggedright\endgroup

\end{document}